\documentclass[11pt]{article}
\usepackage{epsfig}
\usepackage{cite}
\usepackage{ifthen}
\usepackage[usenames,dvipsnames,svgnames,table]{xcolor}
\usepackage{multirow}
\usepackage{scalefnt}
\usepackage{hyperref}
\usepackage[normalem]{ulem}
\usepackage{amsmath,amssymb}
\newcommand{\abbrev}{\scalefont{.9}}
\newcommand{\PS}{{\abbrev PS}}
\newcommand{\order}[1]{{\cal O}(#1)}
\newcommand{\qrest}{Q_t}
\newcommand{\qresb}{Q_b}
\newcommand{\qresint}{Q_\text{int}}
\newcommand{\wrest}{w_t}
\newcommand{\wresb}{w_b}
\newcommand{\wresint}{w_\text{int}}

\newcommand{\cp}{{\abbrev CP}}
\newcommand{\dd}{\mathrm{d}}
\newcommand{\citere}[1]{Ref.\,\cite{#1}}
\newcommand{\citeres}[1]{Refs.\,\cite{#1}}
\newcommand{\hmw}{{\abbrev HMW}}
\newcommand{\bv}{{\abbrev BV}}
\newcommand{\AR}{{\abbrev AR}}
\newcommand{\mcnlo}{{\tt MC@NLO}}
\newcommand{\powheg}{{\tt POWHEG}}

\newcommand{\qsh}{\ensuremath{Q_\text{sh}}}

\newcommand{\qcd}{{\abbrev QCD}}
\newcommand{\scet}{{\abbrev SCET}}
\newcommand{\bsm}{{\abbrev BSM}}
\newcommand{\sm}{{\abbrev SM}}
\newcommand{\mssm}{{\abbrev MSSM}}
\newcommand{\thdm}{2{\abbrev HDM}}
\newcommand{\nnll}{{\abbrev NNLL}}
\newcommand{\nnlo}{{\abbrev NNLO}}
\newcommand{\nlo}{{\abbrev NLO}}
\newcommand{\nlops}{{\abbrev NLO+PS}}
\newcommand{\lo}{{\abbrev LO}}
\newcommand{\fnlo}{f\nlo}
\newcommand{\lhc}{{\abbrev LHC}}

\newcommand{\eqn}[1]{Eq.\,(\ref{#1})}
\newcommand{\fig}[1]{Figure\,\ref{#1}}

\newcommand{\mtop}{m_t}

\newcommand{\pt}{\ensuremath{p_{\bot}}}
\newcommand{\pth}{\pt}
\newcounter{notecount}
\newcommand{\als}{\ensuremath{\alpha_s}}

\newcommand{\llog}{{\abbrev LL}}
\newcommand{\nll}{{\abbrev NLL}}
\newcommand{\sct}[1]{Section~\ref{#1}}

\newcommand{\qres}{\ensuremath{Q_{\text{res}}}}

\newcommand{\plus}{+}
\newcommand{\born}{\ensuremath{B}}
\newcommand{\virtuals}{\ensuremath{\hat V_{\text{fin}}}}
\newcommand{\reals}{\ensuremath{R}}

\newcommand{\pdf}{{\abbrev PDF}}
\newcommand{\pdfs}{\ensuremath{\Gamma}}

\newcommand{\uv}{{\abbrev UV}}
\newcommand{\ir}{{\abbrev IR}}

\newcommand{\lowma}{low-$\ma$}
\newcommand{\largeint}{large-int}
\newcommand{\larget}{large-$t$}
\newcommand{\largeb}{large-$b$}
\newcommand{\largetop}{\larget{}}
\newcommand{\largebot}{\largeb{}}

\newcommand{\MC}{{\abbrev MC}}

\def\be{\begin{equation}}
\def\ee{\end{equation}}
\newcommand{\bea}{\begin{eqnarray}}
\newcommand{\eea}{\end{eqnarray}}
\newcommand{\bdm}{\begin{displaymath}}
\newcommand{\edm}{\end{displaymath}}
\long\def\symbolfootnote[#1]#2{\begingroup%
\def\thefootnote{\fnsymbol{footnote}}\footnote[#1]{#2}\endgroup}

\def\sq2{\sqrt{2}}

\def\x1g{x_{1}}

\def\moresushi{{\tt MoRe-SusHi}}
\def\sushi{{\tt SusHi}}

\newcommand{\as}{\alpha_s}

\newcommand{\smallH}{{\scriptscriptstyle H}} %
\newcommand{\smalla}{{\scriptscriptstyle A}} %

\newcommand{\mh}{m_h}
\newcommand{\mphi}{m_\phi}
\newcommand{\mH}{m_\smallH}
\newcommand{\ma}{m_\smalla}
\newcommand{\muF}{\mu_{\scriptscriptstyle F}}

\def\mt{m_t}

\def\mb{m_b}

\begin{document}

\begin{titlepage}


\mbox{}
\vspace*{-2cm}

{\flushright{
        \begin{minipage}{5cm}
          CERN-PH-TH-2015-242 \\
          KA-TP-20-2015 \\
          MCnet-15-21 \\
          DESY-15-155 \\
          TIF-UNIMI-2015-7\\
          WUB/15-06\\
          ZU-TH 31/15
        \end{minipage}        }

}
\renewcommand{\thefootnote}{\fnsymbol{footnote}}
\vskip 1.2cm
\begin{center}
{\LARGE\bf Resummation ambiguities in the Higgs transverse-\\[0.2cm]
momentum spectrum in the Standard Model and beyond}
\vskip 1.0cm
{\Large  E.~Bagnaschi$^{a}$, R.V.~Harlander$^{b}$, 
H.~Mantler$^{c,d,e}$,  \\ [7pt]
A.~Vicini$^{f}$, and M.~Wiesemann$^{g}$}
\vspace*{8mm}\\
\vspace*{2mm}{\sl ${}^a$ DESY, Notkestra\ss e 85, D–22607 Hamburg, Germany}\\
\vspace*{2mm}{\sl ${}^b$ Fachbereich C, Bergische Universit\"at Wuppertal,
  D-42097 Wuppertal, Germany}\\
\vspace*{2mm}{\sl ${}^c$ TH Division, Physics Department, CERN, CH-1211 Geneva 23, Switzerland}\\
\vspace*{2mm}{\sl ${}^d$ Institute for Theoretical Physics (ITP), Karlsruhe Institute of Technology,\\ Engesserstra{\ss}e 7, D-76128 Karlsruhe, Germany}\\
\vspace*{2mm}{\sl ${}^e$ Institute for Nuclear Physics (IKP), Karlsruhe Institute of Technology,\\ Hermann-von-Helmholtz-Platz 1, D-76344 Eggenstein-Leopoldshafen, Germany}\\
\vspace*{2mm}
{\sl ${}^f$
    Tif lab, Dipartimento di Fisica, Universit\`a di Milano and
    INFN, Sezione di Milano,\\
    Via Celoria 16, I--20133 Milano, Italy}
\vspace*{2mm}\\
{\sl ${}^g$
    Physik-Institut, Universit\"at Z\"urich, CH-8057 Z\"urich, Switzerland}
\end{center}
\symbolfootnote[0]{{\tt e-mail addresses:}} 
\symbolfootnote[0]{{\tt emanuele.bagnaschi@desy.de,
    robert.harlander@uni-wuppertal.de,}}
\symbolfootnote[0]{{\tt hendrik.mantler@cern.ch,
    alessandro.vicini@mi.infn.it, mariusw@physik.uzh.ch}}

\vskip 0cm

\begin{abstract}
We study the prediction for the Higgs transverse momentum distribution
in gluon fusion and focus on the problem of matching fixed- and
all-order perturbative results.  The main sources of matching
ambiguities on this distribution are investigated by means of a twofold
comparison.  On the one hand, we present a detailed qualitative and
quantitative comparison of two recently introduced algorithms for
determining the matching
scale \cite{Harlander:2014uea,Bagnaschi:2015qta}. On the other hand, we
apply the results of both methods to three widely used approaches for
the resummation of logarithmically enhanced contributions at small
transverse momenta: the \mcnlo{} and \powheg{} Monte Carlo approaches,
and analytic resummation.  While the three sets of results are largely
compatible in the low-$\pt$ region, they exhibit sizable differences at
large $\pt$. We show that these differences can be significantly reduced
by suitable modifications of formally subleading terms in the Monte
Carlo implementations.  We apply our study to the Standard Model Higgs
boson and to the neutral Higgs bosons of the Two-Higgs-Doublet Model 
for representative
scenarios of the parameter space, where the top- and bottom-quark
diagrams enter the cross section at different strength.

\end{abstract}
\vfill
\end{titlepage}    
\setcounter{footnote}{0}




\section{Introduction}

After the discovery of a Higgs boson at the \lhc{} in
2012\,\cite{Aad:2012tfa,Chatrchyan:2012xdj}, many detailed studies have
analyzed its properties in order to assess its compatibility with the
Standard Model (\sm) (see,
e.g., \citeres{Aad:2015mxa,Khachatryan:2014jba}). These analyses rely on
accurate theoretical predictions of Higgs cross sections and decay
rates.  Recently, the first experimental results for the Higgs
transverse-momentum ($\pt$) distribution have been published
in \citere{Aad:2015lha,Khachatryan:2015rxa}. Such measurements, in
particular with the increased statistics expected from
\lhc{} Run II, open up many new interesting possibilities to test the
nature of the Higgs couplings to the \sm{} fields, and thus to probe for
signs of physics beyond the \sm{} (\bsm). Indeed, the gluon-gluon-Higgs
vertex could be mediated by loops of non-\sm{} particles which can
affect the shape of the $\pth$
distribution\,\cite{Langenegger:2006wu,Brein:2007da,Bagnaschi:2011tu,
Harlander:2013oja,Grojean:2013nya,
Azatov:2013xha,Banfi:2013yoa,Dawson:2014ora,Langenegger:2015lra}. 
Similarly, a modification of the Higgs Yukawa couplings can have
a significant effect on the shape of the \pt{} spectrum.  The search for
additional Higgs bosons, as predicted in extensions of the \sm{},
therefore requires the development of an accurate description of
the Higgs $\pth$ distribution. For non-\sm{} Higgs bosons, it could
exhibit sizable differences to the \sm{} prediction, even if their mass
and cross section was the same as that of the observed
particle\,\cite{Bagnaschi:2011tu,Harlander:2014uea}.

The theoretical prediction of the Higgs transverse-momentum distribution
belongs to the classic chapters of perturbative calculations.  While the
leading order\footnote{At non-zero $\pt$, ``\lo'' denotes terms of
  order $\as^3$, etc.} (\lo{}) \qcd{} result of
\citeres{Ellis:1987xu,Baur:1989cm} as well as the effects induced by
electro-weak gauge bosons\,\cite{Keung:2009bs,Brein:2010xj} cover the
full quark-mass dependence in the internal loops, the \nlo{} \qcd{}
corrections \cite{deFlorian:1999zd,Ravindran:2002dc,Glosser:2002gm} have
originally been evaluated only in the limit of an infinitely heavy top
quark. The impact of a finite top-quark mass at \nlo{} \qcd{} has been
studied subsequently in \citeres{Harlander:2012hf,Neumann:2014nha}.  The
first results towards the determination of the Higgs production at large
transverse momentum with next-to-\nlo{} (\nnlo{}) \qcd{} accuracy have
been presented in
\citeres{Boughezal:2013uia,Chen:2014gva,Boughezal:2015dra}, again in the
heavy-top limit.

It is well known that these fixed-order predictions are logarithmically
divergent as $\pth\to 0$. Only after the resummation of terms enhanced
by powers of $\log(\pth/\mphi)$ to all orders in $\als$, where $\mphi$
is the mass of the Higgs boson,\footnote{We employ the symbol $\phi$ to
  generically denote an electrically neutral Higgs boson.} does the
distribution exhibit a regular behavior towards small $\pth$. This
resummation is based on universal properties of \qcd{} radiation in the
soft and collinear limits\,\cite{Dokshitzer:1978hw,Parisi:1979se,
  Curci:1979bg,Collins:1981uk,Collins:1981va,Kodaira:1981nh,
  Kodaira:1982az,Altarelli:1984pt,Collins:1984kg,Catani:2000vq}, and can
be achieved either analytically, or numerically through the so-called
Parton Shower (\PS) in Monte Carlo (\MC) event generators.

Since the resummation of the logarithms is strictly valid only as
$\pth\to 0$ (implying $\pt\ll \mphi$), a physical prediction for the
transverse-momentum distribution which extends to $\pt\sim\mphi$
requires a matching of the resummed to the fixed-order result, while
avoiding any kind of double counting. Various matching approaches have
been proposed, both for analytic resummation
\cite{Collins:1984kg,Bozzi:2005wk,Mantry:2009qz,Becher:2012yn},\footnote{%
  Detailed studies of the approaches based on Soft-Collinear Effective
  Theory (\scet{})\,\cite{Mantry:2009qz,Becher:2012yn} concerning scale
  choices and the resulting Higgs phenomenology have been performed in
  \citeres{Ligeti:2008ac,Abbate:2010xh,Berger:2010xi}.  In this paper we
  will limit ourselves to the standard \qcd{} framework of
  \citeres{Collins:1984kg,Bozzi:2005wk}.} as well as in the framework of
Monte Carlo event generators \cite{Frixione:2002ik,Nason:2004rx}.
Common to all of these approaches is the introduction of an auxiliary,
possibly effective, momentum scale (from now on generically referred to
as ``matching scale''), which indicates the transverse-momentum region
of the transition from the resummed to the fixed-order result.  The
dependence of the distribution on this matching scale is of higher
logarithmic order. However, inadequate choices of the matching scale may
spoil the accuracy of the result, which is why its central value
requires a careful choice.  Moderate variations (typically by a factor
of two) around this central value may then be used to estimate the
residual uncertainty of the resummation/matching procedure.

The choice of the matching scale becomes particularly problematic once
the process depends on more than one mass scale. One example here is
Higgs production in gluon fusion. In the \sm{} and many extended
theories, it is predominantly described by Feynman diagrams involving
top- and bottom-quark loops. In the \sm{} it is $\mh\sim \mt$ ($\mh$ and $\mt$ 
denote the Higgs and the top-quark mass, respectively), so it
seems obvious that for the dominant top-loop induced contribution the
matching scale should be chosen to be of the order of $\mh$.  However,
despite its rather small contribution to the total cross section, the
bottom-loop plays a non-negligible role at small values of the
Higgs \pt{}.  Since it involves two very different scales ($\mh$ and the 
bottom-quark mass $\mb$), the choice of the matching scale is not at all 
obvious in this
case.  In fact, it was observed that the theoretical prediction depends
quite sensitively on the resummation method as well as on the nature of
the respective matching
scale\,\cite{Mantler:2012bj,Bagnaschi:2011tu,Grazzini:2013mca}.

Recently, two algorithmic strategies for determining an adequate central
value for the matching scale have been
proposed \cite{Harlander:2014uea,Bagnaschi:2015qta}.  It is important to
note that, even though the matching scales of the various approaches
have different origins and meanings, i.e., they deal with the unknown
higher-order terms in different ways, the proposed strategies may be
applied to all of the resummation and matching approaches mentioned
above, as will be discussed in more detail below.

From this discussion, it is clear that the resummed
transverse-momentum distribution suffers from a number of ambiguities
which are formally of higher logarithmic order. It is the goal of this
paper to study their numerical impact on the Higgs $\pt$ distribution in
and beyond the \sm{}.  Since the explicit analytic form of these
ambiguities is not fully accessible, our analysis will rely on the
numerical comparison of the results obtained in the various approaches,
assuming specific phenomenological parameters.

We consider the following three representative theoretical
approaches:\footnote{Note that all approaches feature \nlo{} accuracy
  (up to $\alpha_s^3$) on the total Higgs production cross section,
  which implies, however, a formally \lo{} accurate prediction at large
  \pt{}.}
\begin{itemize}
\item analytic resummation (\AR) as formulated
in \citeres{Collins:1984kg,Bozzi:2005wk};
\item the \powheg{} method, described
in \citeres{Nason:2004rx,Frixione:2007vw};
\item the \mcnlo{} method of \citere{Frixione:2002ik}.
\end{itemize}
The implementations of these approaches on which we base our study are
all publicly available:
\begin{itemize}
\item For the \AR{} approach, we use \moresushi{}, which includes the
description of the resummed $\pth$ distribution at next-to-leading
logarithmic (\nll{}) accuracy consistently matched to the fixed-order
cross section at \nlo{} \qcd{}
\cite{Mantler:2012bj,Harlander:2014uea,moresushiHP};
\item the \nlops{} accurate \powheg{} implementations 
of the gluon-fusion process are contained in the directories {\tt
gg\_H\_quark-mass-effects} and {\tt
gg\_H\_2HDM} \cite{Bagnaschi:2011tu} of the {\tt
POWHEG-BOX} \cite{Alioli:2010xd,powheg-box};
\item the corresponding \mcnlo{} implementation at \nlops{} is available 
in {\tt aMCSusHi} \cite{Mantler:2015vba,amcsushiHP} which
combines the {\tt MadGraph5\_aMC@NLO} package \cite{Alwall:2014hca} with
the {\tt SusHi} amplitudes \cite{Harlander:2012pb}.
\end{itemize}

Leaving aside the specific values of the associated matching scales, we
will refer to these three approaches and their implementations as
``resummation codes'' or simply ``codes''. All codes work
at \nlo{} \qcd{} accuracy in the prediction of the Higgs production
total cross section, i.e., $\order{\als^3}$.  The differences in the
$\pth$ distribution are formally subleading,\footnote{Note that the
meaning of ``subleading terms'' is somewhat different for \AR\ and
the \MC\ generators. \AR\ consistently resums \nll{} terms to all
orders, while the \PS\ in the Monte Carlo approaches strictly includes
only the leading logarithms, but resums also some logarithms beyond the
leading ones.}
but can be numerically sizable, as we will see later on.
In order to assess the impact of these differences, we compare their
numerical results using the same values of the matching scales for all
of them. On the other hand, we compare the results of a single code for
the two different strategies of setting the matching scale proposed
in \citeres{Harlander:2014uea,Bagnaschi:2015qta}.  As we will see, both
the intrinsic difference in the formulation of the codes as well as the
dependence on their matching scales are a source of sizable ambiguities
in the theoretical prediction of the Higgs $\pth$ distribution, in
particular at intermediate and large $\pt$. The sources of these
differences will be investigated in detail in the course of this paper.

In the \sm{}, the matching of fixed-order and resummed results has been
achieved with \nnlo{} \qcd{} accuracy on the top-quark induced component
of the total cross section, i.e., \nnlo{}+next-to-\nll{} (\nnll{}) in
\AR{} \cite{Bozzi:2003jy,Bozzi:2005wk,deFlorian:2011xf,Grazzini:2013mca} and {\abbrev N}\nlops{} accurate Monte Carlo \cite{Hamilton:2013fea,Hoche:2014dla,Hamilton:2015nsa,Alioli:2013hqa,Alioli:2015toa}.
Since our main focus is on the $\pth$ distribution in \bsm{} scenarios
with large bottom-quark effects and the possibility of new additional
heavy states, we refrain from including such effects in our discussion.

This paper is organized as follows: In \sct{sec:codedesc} we introduce
the gluon-fusion process at \nlo{} \qcd{} and summarize the underlying
resummation procedures of the three codes under consideration. Our focus
is on the respective matching prescriptions and corresponding matching
scales.  In \sct{sec:approaches} we recall the two strategies
of \citeres{Harlander:2014uea,Bagnaschi:2015qta} to determine proper
values for the matching scales. The two approaches are then subject to a
qualitative and quantitative comparison, where we quote values for the
matching scales in a large range of Higgs masses. Our main study of the
matching ambiguities is presented in \sct{sec:results} for the
\sm{} and for various scenarios in a generic type-II 
Two-Higgs-Doublet Model (\thdm), designed in order to emphasize specific
contributions to the gluon-fusion Higgs cross section. 
\sct{sec:conclusions} contains our conclusions.


\section{Resummation procedures}\label{sec:codedesc}

In this section, we briefly review the ingredients that are required for
the theoretical prediction of the Higgs production cross section
via gluon fusion with \nlo{} accuracy. We then summarize the main
features of three procedures that allow to resum logarithmic terms in
the Higgs transverse-momentum distribution at small $\pth$,
namely \AR, \mcnlo, \powheg. The main focus will be put on their
matching prescription and the corresponding matching scales.

Concerning the total Higgs production cross section, the three codes
considered in this paper work at the same perturbative accuracy,
\nlo\ \qcd, using the same matrix elements for the description of the
virtual corrections and the real-emission effects. We introduce a common
notation to identify these contributions in the three different
approaches for the matching of resummed and fixed-order results.

\begin{figure}[t]
  \begin{center}\hspace{-0.5cm}
    \begin{tabular}{ccccccc}
      \mbox{\includegraphics[height=.12\textheight]{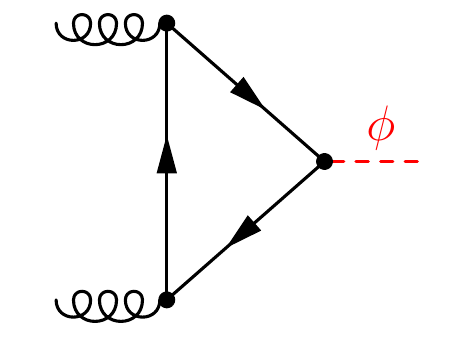}} & &                   \mbox{\includegraphics[height=.12\textheight]{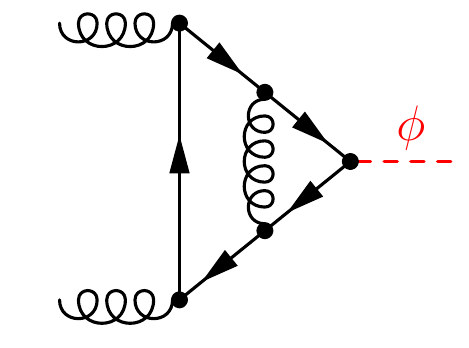}}   & &
      \mbox{\includegraphics[height=.085\textheight]{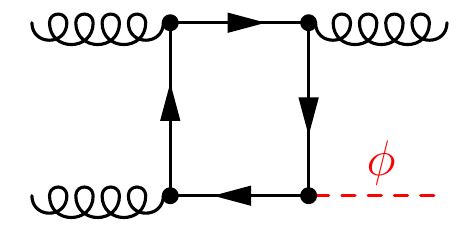}} & &
      \mbox{\includegraphics[height=.11\textheight]{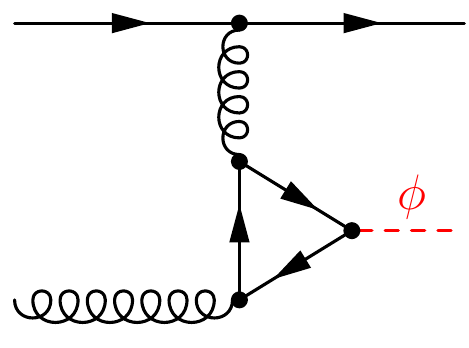}}
            \\
      (a) & & (b) & & (c) & & (d)
    \end{tabular}
    \parbox{.9\textwidth}{%
      \caption[]{\label{fig:diag}A sample of Feynman
        diagrams for
        $gg\rightarrow \phi$ contributing to the \nlo{} cross section;
        (a) \lo{}, (b) virtual and (c-d) real corrections. The
        graphical notation for the lines is: solid straight
        $\widehat{=}$ quark; curly $\widehat{=}$ gluon; dashed
        $\widehat{=}$ Higgs.} }
  \end{center}
\end{figure}

The Born-level squared matrix elements \born{} determine the \lo{} cross
section and multiply all the soft and collinear counter terms.  The
corresponding loop-induced Feynman diagrams are shown in
\fig{fig:diag}\,(a), with both the top and the bottom quark running in
the fermionic loop.\footnote{Diagrams courtesy of S.\,Liebler.}  The
interference of the \uv{}- and \ir{}-regularized virtual corrections
with the Born amplitude will be denoted by \virtuals{} in the
following.\footnote{We use a hat to indicate \ir{}-regularized
  quantities in this paper.} Its evaluation requires the computation of
the two-loop virtual diagrams
\cite{Spira:1995rr,Harlander:2005rq,Aglietti:2006tp,Anastasiou:2006hc},
e.g, in \fig{fig:diag}\,(b).  Note that there are some differences in
the definition of \virtuals{}, due to the \ir{}-regularization adopted
in each matching procedure.  We simply assume \virtuals{} to be properly
subtracted in the respective approach under consideration.  At the same
perturbative order, real-emission subprocesses need to be taken into
account \cite{Spira:1995rr,Bonciani:2007ex}. They involve an additional
final-state parton with respect to the Born-level process; their squared
matrix elements are collectively called \reals{}. Some examples of the
corresponding diagrams are shown in \fig{fig:diag}\,(c-d).

For convenience, we use the following symbolic notation
for the convolution over the parton-density functions (\pdf{}s):
\begin{align}
\left(M\otimes\pdfs\right)(q^2,\muF^2)\equiv 
\sum\limits_{i,j}
\int \dd z_1\int\dd z_2
M_{ij}(q^2,\muF^2,z_1,z_2)\pdfs{}_{ij}(\muF^2,z_1,z_2)\,,
\end{align}
where $M_{ij}(q^2,\muF^2,z_1,z_2)$ is the squared matrix element for the
scattering of partons $i$ and $j$ carrying proton momentum fractions
$z_1$ and $z_2$, respectively, $q^2$ is the momentum transfer of the
scattering process, $\pdfs{}_{ij}(\muF^2,z_1,z_2)=f_i(\muF^2,z_1)
f_j(\muF^2,z_2)$ is the product of two \pdf{}s, and $\muF$ is the
factorization scale.

\subsection{Analytic \pt{} resummation (\AR)}

The first resummation procedure that we consider is the analytic
resummation of soft and collinear logarithms in the inclusive
transverse-momentum spectrum, see
\citeres{Bozzi:2005wk,Collins:1984kg}.

For the matching of the low- and high-\pt{} Higgs cross section at
\nlo{}\plus\nll{},\footnote{Note that 
the accuracy (including terms of order $\alpha_s^3$) at large \pt{} is formally only \lo{}.} the additive procedure
of \citere{Bozzi:2005wk} is adopted, and the hadronic differential cross
section is obtained as
\begin{align}
\label{eq:ar}
\frac{\dd\sigma}{\dd\pt^2}=
\int \frac{\dd\Phi_B}{\dd\pt^2}\,(\born+\virtuals)\,{\cal F}_\text{\nll}(\qres)
+\int \frac{\dd\Phi}{\dd\pt^2}\,R\otimes\pdfs
-\int\frac{\dd\Phi_B}{\dd\pt^2}\,\born\,{\cal F}_\text{\nlo}(\qres)\,,
\end{align}
where $\dd\Phi_B$ and $\dd\Phi$ represent the phase space of the Born
process and of the process that includes one additional real parton,
respectively. By $\int{\dd X}/{\dd\pt^2}$ we denote integration over all
variables of $X$ except $\pt^2$.  The function ${\cal F}$ is defined as
follows:
\begin{equation}
\begin{split}
\label{eq:res}
{\cal F}_\text{\nll}(\qres,\pt)&= 
\frac{\mphi^2}{S} \int_0^{\infty} \dd b\, \frac{b}{2} \,J_{0}(b\,\pt)\,
{\cal S}(\als,\tilde L)\\
\times 
\sum_{i,j}\int \dd z_1&\,\dd z_2\,
\left[\delta_{z_1}\,\delta_{z_2}+\frac{\als(b_0/b)}\pi\,C^{(1)}_{gi}(z_1)\,\delta_{z_2}+\frac{\als(b_0/b)}\pi\,\delta_{z_1}\,C^{(1)}_{gj}(z_2)
\right] \pdfs{}_{ij}(b_0/b,z_1,z_2)\,,\\
{\rm with} \quad {\cal S}(\als,\tilde L) &=
 \exp\left\{\tilde L\,g^{(1)}(\als\,\tilde L)+
g^{(2)}(\als\,\tilde L)\right\}\,,
\end{split}
\end{equation}
where $\delta_{z_i}\equiv \delta(1-z_i)$ for $i\in\{1,2\}$ is a
short-hand notation, $b_0=2\exp(-\gamma_E)=1.12292\ldots$ is a numerical
constant,\footnote{$\gamma_E=-\Gamma'(1)$ is the Euler constant.}
$J_0(x)$ is the Bessel function of the first kind with $J_0(0)=1$, $S$
is the hadronic center of mass energy, and the sum over $i,j$ runs over
all kinds of partons.  In the notation of the present paper, we apply
the ``hard scheme'' as defined in \citere{Catani:2013tia} for the
collinear coefficient functions $C_{ab}(z)$; their first order
expressions for gluon-induced processes can be also found in that
reference.

The Sudakov form factor ${\cal S}(\als,\tilde L)$ accounts for the
resummation of logarithms of the form $\tilde L =
\ln(b^2\qres^2/b_0^2+1)$ at \nll{} accuracy, with $\als\,\tilde L$ being
considered of order unity.  The functions $g^{(1)}$ and $g^{(2)}$,
relevant for leading logarithmic (\llog{}) and \nll{} accuracy,
respectively, are given in \citere{Bozzi:2005wk}.  The scale $\qres$ is
conventionally called resummation scale and controls up to which values
of \pt{} the resummation is effective.  Since it thus parameterizes the
arbitrariness in the separation of the ``soft'' from the ``hard''
region, it plays the role of the matching scale in the \AR{} framework
and is usually chosen at the order of characteristic scale of the hard
scattering process.  Note that the first and the third term on the
r.h.s.\ of \eqn{eq:ar} are explicitly $\qres{}$ dependent. However, this
dependence is canceled through the matching formula order by order in
the logarithmic expansion, which renders the \pt{} distribution
independent of $\qres$ when computed to infinite logarithmic accuracy.
In \eqn{eq:ar}, ${\cal F}_\text{\nlo}$ is the \nlo{} truncation of
${\cal F}_\text{\nll}$; i.e., for gluon fusion, it includes terms up to
$\als^3$.  The last term in \eqn{eq:ar} subtracts the singular behavior
of the real corrections from the fixed-order expression, given by the
second term on the r.h.s.; it therefore avoids double counting of
logarithmic terms which are already contained in the first term.

Furthermore, the matching procedure of \eqn{eq:ar} induces unitarity on
the matched cross section, meaning that integration of \eqn{eq:ar} over
$\pt^2$ yields the total \nlo{} cross section:
\begin{align}
\int \dd\pt^2 \frac{\dd\sigma}{\dd\pt^2} = \sigma_\text{tot}^\text{\nlo}\,.
\label{eq:unitarity}
\end{align}

The \AR{} approach has been implemented, using parts of {\tt
  HqT}\,\cite{Bozzi:2005wk,Bozzi:2003jy,deFlorian:2011xf}, in the code
\moresushi{}\,\cite{Mantler:2012bj,Harlander:2014uea,moresushiHP} that
will be used in all the numerical simulations based on \AR{} in this
paper.

\subsection{\nlops{} Monte Carlo}\label{sec:mc}

An alternative to the analytic resummation is offered by \PS{} Monte
Carlo event generators, where \PS{} algorithms allow the numerical
simulation of multiple parton emissions.  A consistent matching of the
fixed-order \nlo{} \qcd{} predictions\footnote{Note also in this case
  that the description is formally only \lo{} accurate as far as large
  \pt{} are concerned.} with the \PS{} has been discussed in
\citere{Frixione:2002ik} and in
\citeres{Nason:2004rx,Frixione:2007vw,Alioli:2010xd}, and implemented in
the {\tt MC@NLO} Monte Carlo event generator \cite{Frixione:2002bd} and in
the {\tt POWHEG-BOX} \cite{Alioli:2008tz}, respectively.  We
refer the reader to the above publications for a detailed discussion of
the two implementations and summarize here the main differences of these
two approaches, from the point of view of the matching of resummed and
fixed-order results.

We introduce the ``shower operator'' ${\cal I}_n(t_1)$ to represent in a
compact form the action of the \PS{} to describe $n$ parton emissions;
the latter are ordered with respect to a parameter $t$, which for
simplicity we assume to be the transverse momentum of the emission in
this paper (other options are the virtuality, or the angle, for
example), starting from a maximum value $t=t_1$. For the $i^\text{th}$
emission of a parton, the \PS{} associates a Sudakov form factor times
an approximate emission probability, both evaluated at the same value
$t_i$ of the specific ordering parameter.  The ordering of the emissions
is a requirement for the \PS{} algorithm to reach the \llog{} accuracy.

In both the \mcnlo{} and \powheg{} approaches, the hardest parton
emission is treated retaining the accuracy of the exact matrix elements,
whereas the others are generated according to the \PS{} algorithm.

Given a hard scattering process, we describe the evaluation of the cross
section $\dd\sigma_{n}$ for the radiation of $n$ additional partons in
terms of two steps. The first step results in the cross section
$\dd\sigma_1$, which includes only the hardest emission; the remaining
$n-1$ emissions are taken into account in the second step.  This
splitting is represented symbolically as applying ${\cal I}_{n-1}(t_1)$
to $\dd\sigma_{1}$, and multiplying by the relevant phase space of the
additional $n-1$ particles: \be \dd\sigma_{n} = {\cal I}_{n-1}(t_1)\,
\dd\sigma_{1} \dd\Phi_{n-1}\,.
\label{eq:codes:nm1n}
\ee
The formula that describes the hard scattering with the emission of 0 or
1 additional parton can be written, in a sufficiently general way, as\footnote{While in the \powheg{} approach the first emission is explicitly implemented as 
given in \eqn{eq:codes:matching}, it is vital for the subsequent discussion to stress 
that, in the case of \mcnlo, the terms in the curly brackets in the
first term of \eqn{eq:codes:matching} are implicitly generated when
attaching the shower
to the Born-level configurations.}
\begin{align}
  \dd\sigma_{1} = \bar{B}^s \otimes\Gamma\, \dd\Phi_B \left\{
  \Delta^s_{t_\text{min}} + \Delta^s_t\frac{R^s}{B}\,
  \dd\Phi_r \right\} + R^f\otimes \Gamma\, \dd\Phi +
  R_{\text{reg}}\otimes \Gamma\, \dd\Phi\,,
  \label{eq:codes:matching}
\end{align}
where $t_\text{min}$ is the value of the ordering variable $t$ below
which no emissions are allowed,\footnote{The physics below the scale
  $t_{\rm min}$ is not properly accounted for by a perturbative
  description.  Its numerical value is typically set to the
  hadronization scale of the \PS, coherent with the fact that the first
  term in the curly bracket describes the no-emission probability.  }
and where we used the factorization of the real phase space into the
Born one times the one of a single additional real parton, $\dd\Phi =
\dd\Phi_B \,\dd\Phi_r$.

As defined before, $R$ contains collectively all the squared matrix
elements of the real-emission subprocesses, and we further split them
into two groups: $R_{\text{div}}$ is the sum of squared matrix elements
which are divergent in the limit of soft or collinear emissions (in our case
$gg\to g\phi$ and $qg\to q\phi$); the regular ones are denoted by $R_{\text{reg}}$ instead (in
our case $q\bar q\to g\phi$).  The squared matrix elements of the
divergent subprocesses can be further split in two parts:
\begin{align}
  R_{\text{div}} = R^s + R^f.
\label{eq:realsplit}
\end{align}
The term $R^s$ contains the soft- and collinear-singular part of
$R_{\text{div}}$, while $R^f$ is the finite remainder.\footnote{The
splitting into $R_\text{div}$ and $R_\text{reg}$ seems redundant at this
point, but will play a role in the \powheg{} approach, see
\sct{sec:powhegcode}.} Obviously, this splitting is not unique, since
finite parts can be shifted between $R^s$ and $R^f$.

The generalized Sudakov form factor\,\cite{Sjostrand:1985xi} is denoted
by the symbol $\Delta_t^s$ in \eqn{eq:codes:matching}, with $t$ the
shower ordering variable; it depends on $R^s$ and expresses the
probability of not emitting any parton with a value for the ordering
variable larger than its own argument $t$:
\begin{align}
  \Delta^s_{t} = \exp\left\{-\int \frac{\dd t'}{t'}
  \frac{R^s\otimes\Gamma}{B\otimes\Gamma} \dd\Phi_r
  \theta(t'-t)\right\}\, .
  \label{eq:sudakov}
\end{align}
The factor $\bar{B}$ in \eqn{eq:codes:matching} is defined by
\begin{align}
  \bar{B}^s = B + \hat{V}_{\text{fin}}+ \int \hat{R}^s \, \dd\Phi_r\, 
\label{eq:codes:nlonorm}
\end{align}
and includes the contributions of the Born squared matrix elements, the
corresponding virtual corrections, and the integral over the radiation
phase space of $R^s$.  The finiteness of the $\bar{B}$ factor is
guaranteed by the fact that all the divergent terms are properly
subtracted; this is possible thanks to the renormalization of the UV
divergences, to the cancellation of the IR soft singularities between
virtual and soft real contributions, and to the cancellation of the
collinear singularities, reabsorbed in the definition of the physical
proton \pdf{}s.

The curly bracket in \eqn{eq:codes:matching} describes the probability
of zero or one parton emission in those subprocesses that are
divergent in the soft/collinear limit, where $R^s$ is the singular part
of the squared matrix elements. The precise definition of $R^s$ (or,
equivalently, $R^f$) therefore directly affects the expression of the
Sudakov form factor. In the following we comment on the two choices
adopted in the
\powheg{} and in the \mcnlo{} implementations.
We stress that these two alternatives differ by terms that are formally of higher order in the perturbative expansion,
but that can nevertheless be numerically sizable.
The arbitrariness in the definition of $R^s$
can be exploited to parameterize the matching procedure uncertainties.

The last two terms in \eqn{eq:codes:matching} depend on the process ($R_{\text{reg}}$)
and on the definition adopted to split $R^{s,f}$; both yield a regular contribution in the soft and the collinear limit.

The evaluation of the exact real and virtual matrix elements guarantees,
for any observable inclusive over radiation, the \nlo{} \qcd{} accuracy.
The latter is preserved by the unitarity of the \PS{} algorithm in the
generation of each additional real parton; this feature holds also for
the first emission, described by the curly bracket in
\eqn{eq:codes:matching}.

\subsubsection{\mcnlo}
\label{sec:mcnlo}

In the \mcnlo{} formulation, the \PS{} algorithm is used to generate all
the additional parton emissions starting from the Born-level and real-emission 
configurations. The exact ${\cal O}(\alpha_s)$ corrections are applied 
in order to recover the exact matrix element description
of the first hard emission and the correct normalization, including the
effect of the virtual corrections to the underlying Born.

The Sudakov form factor implemented in the \PS{} generators uses a
universal, process-independent expression to describe parton radiation
in the soft and collinear limit, based on the Altarelli-Parisi splitting
functions $P$.  Using the notation of \eqn{eq:realsplit}, in \mcnlo{}
the singular part of the squared real-emission matrix element is
$R^s_\text{\mcnlo} \propto \alpha_s P B$, and the generalized Sudakov
form factor that appears in \eqn{eq:codes:matching} is actually not 
explicitly implemented, but generated by the first \PS{} emission 
on top of the Born-level configuration.

Given the assignment for $R^s_\text{\mcnlo}$, the definition of
$R^f_\text{\mcnlo}=R_\text{div}-R^s_\text{\mcnlo}$ follows by
construction, as the difference between the exact real matrix element
correction and its \PS\ approximation, sometimes called Monte Carlo
subtraction term; the latter is needed to avoid a double counting with
the emission described by the second term in the curly brackets of
\eqn{eq:codes:matching} that is generated by the first emission of the 
shower.

In the \mcnlo{} approach, the differential \nlops{} cross section with
respect to a variable $O$ is:
\begin{align}
\notag
\left(\frac{\dd\sigma}{\dd O}\right)_{\text{MC@NLO}}=&
\;\sum_{n\geq 0}\,\int \left[B\otimes \Gamma\,+\hat{V}_{\text{fin}}\otimes \Gamma\,+
\int \hat R^s_\text{\mcnlo}\otimes \Gamma\,\dd\Phi_r^\text{\MC}
\right] \frac{\dd\Phi_B\,\dd\Phi_n^\text{\MC}}{\dd O}\,{\cal I}_n(t_1\equiv Q_{\rm sh}^s)\\
+&\;\sum_{n\geq 1}\,\int \left[R \otimes \Gamma\, \frac{\dd\Phi\,\dd\Phi_{n-1}^\text{\MC}}{\dd O}-R^s_\text{\mcnlo}\otimes \Gamma\,\frac{\dd\Phi^\text{\MC}\,\dd\Phi_{n-1}^\text{\MC}}{\dd O}\right] {\cal I}_{n-1}(t_1\equiv Q_{\rm sh}^h)\,,
\label{eq:mcatnlo}
\end{align}
where the sum runs over all possible $n$-parton configuration after the
shower in the final state. We shall stress at this point that the shower
spectrum ${\cal I}_n$ in the first and ${\cal I}_{n-1}$ in the second
line start from different configurations, i.e., from Born-level and real
configurations, respectively, and that the observable $O$ is defined on
a different phase space in the two lines.  The superscript \MC\ is
attached when the phase-space is computed in the \PS{}
approximation. The real Monte Carlo phase space $\dd\Phi^\text{\MC}$
tends to $\dd\Phi$ in the {\abbrev IR} limits, and
$\dd\Phi_r^\text{\MC}=\dd\Phi^\text{\MC}/\dd\Phi_B$ is therefore the
\PS{} approximation of the one-particle phase space. By $\int \frac{\dd
  X}{\dd O}$ we denote integration over all variables of $X$ except $O$.

In the first line of \eqn{eq:mcatnlo} all emissions are described via
the \PS{}, denoted as ``soft'' events, while in the second line we have
the so called ``hard'' events: the exact matrix elements with one
additional real emission with respect to the Born are sampled over the
real phase space; to avoid a double counting with the first line, there
is a \MC{} subtraction term, which is evaluated over the approximated
one-particle phase space and renders the expression in the squared
brackets finite. Expanding \eqn{eq:mcatnlo} up to the first emission
(order $\alpha_s$ with respect to the Born), we indeed recover the
structure of $\dd\sigma_1$ in \eqn{eq:codes:matching}.

In the \mcnlo{} approach, the shower emissions are bounded from above by
identifying $t_1\equiv \qsh{}$. The so-called ``shower scale'' $\qsh{}$
therefore determines the hardest scale accessible to the shower
emissions and is usually set to the characteristic scale of the hard
scattering process. $\qsh{}$ plays the role of the matching scale,
since it separates the soft/collinear from the hard region in the
additive \mcnlo{} matching approach---indeed very similar to the role of
the scale $\qres$ in \AR{}. More precisely, one may choose different
shower scales for the soft ($Q_{\rm sh}^s$) and the hard ($Q_{\rm
  sh}^h$) events, as indicated in \eqn{eq:mcatnlo}. A general feature of
the default \mcnlo{} formulation is that the strict bound enforced by
$\qsh{}$ on the shower emissions suppresses the \PS{} contribution at
values of $\pt\gg \qsh{}$, resulting in the recovery of the fixed-order
distribution at sufficiently large $\pth$, which is regarded to provide
a proper prediction in that region.

\begin{figure}
\begin{center}
\includegraphics[width=0.48\textwidth]{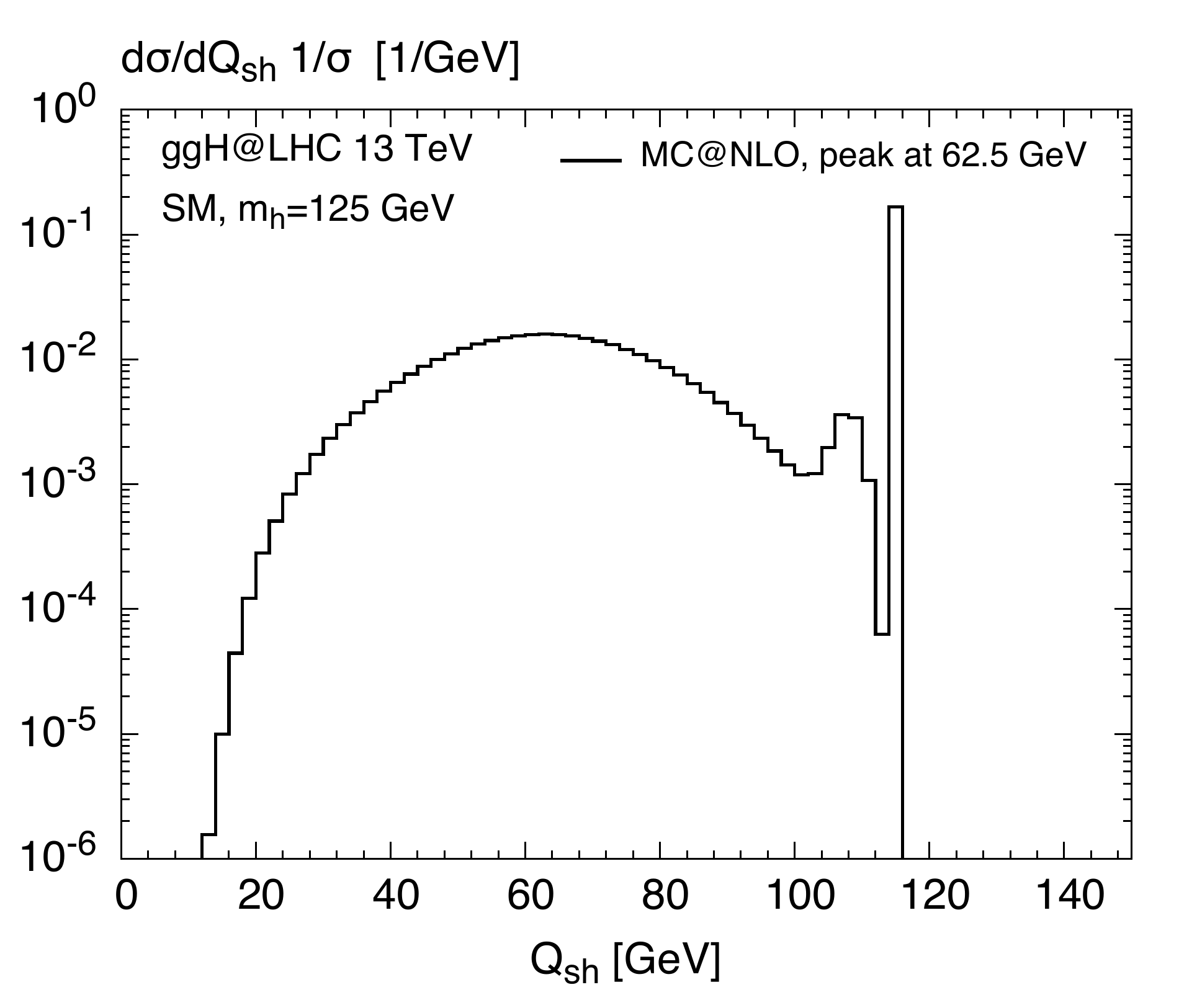}
\caption[]{\label{fig:showerscale}Sample of a shower scale distribution in 
{\tt MadGraph5\_aMC@NLO} for a $125$\,GeV Higgs boson produced in gluon 
fusion. The distribution is normalized such that it integrates to one.}
\end{center}
\end{figure}

The original choice for $\qsh{}$ in the \mcnlo{} code
\cite{Frixione:2002ik,Frixione:2002bd} was to set its value to the
shower scale of the \PS{} Monte Carlo, which corresponds to a narrow
distribution around the Higgs boson mass. 
In {\tt MadGraph5\_aMC@NLO}, on the other hand, the shower scale for the
soft events, $Q_\text{sh}^s$, is statistically extracted from a
probability distribution with a range that can be defined by the
user.\footnote{Further details can be found in \citere{Alwall:2014hca};
  this technique has been used sparingly also in {\tt MC@NLO\,v3.3}
  \cite{Frixione:2006gn} and higher.}  The shower scale for the hard
events, $Q_{\rm sh}^h$, is fixed to a specific value (i.e., using a
$\delta$-distribution) that by default is chosen to be the upper bound
of the probability distribution applied to the soft events.  For
reference, \fig{fig:showerscale} shows an example of the shower scale
distribution of all events for a \sm{} Higgs boson ($\mh=125$\,GeV) with
the restriction $\qsh{}/{\rm GeV}\in[11.425,114.25]$. The spike at the
upper end of the distribution is due to the hard events.  The peak at
the center of the distribution can be considered as ``effective'' shower
scale; we will often refer to it simply as ``shower scale'' in what
follows.  E.g., \fig{fig:showerscale} has an (effective) shower scale of
$\qsh{}=\mh/2=62.5$\,GeV.  In this paper, we always use a distribution
with a shape as shown in \fig{fig:showerscale}, centered around the
respective matching scale, and with ratio of the endpoints equal to
ten\,\cite{Mantler:2015vba}.  The \mcnlo{} results in this paper have
been obtained with the code {\tt
  aMCSusHi}\,\cite{Mantler:2015vba,amcsushiHP} which combines the {\tt
  MadGraph5\_aMC@NLO} framework with the matrix elements of the code
{\tt SusHi}.

\subsubsection{\powheg}
\label{sec:powhegcode}
The fully differential cross section of an event with the emission of
additional partons is obtained by inserting \eqn{eq:codes:matching} in
\eqn{eq:codes:nm1n}.  For a generic observable $O$ we have
\begin{align}
\notag
\left(\frac{\dd\sigma}{\dd O}\right)_{\text{\powheg}}= \;\sum_{n\geq 1} \int \bigg[& \bar{B}^s\,\dd\Phi_B \left\{
  \Delta^s_{t_\text{min}} + \Delta^s_t \frac{R^s_{\text{\powheg}}}{B}
  \dd\Phi_r \right\} \\
+&  R^{f}_{\text{\powheg}} \otimes \Gamma\, \dd\Phi\ + R_{\text{reg}} \otimes \Gamma\, \dd\Phi\ \bigg] \frac{\dd\Phi^{\text{MC}}_{n-1}}{\dd O} {\cal I}_{n-1}(t_1\equiv p^{\text{rad}}_{\bot}) \,,
\label{eq:powheg}
\end{align}
where the shower scale $t_1$ is set equal to the transverse momentum
$p^{\text{rad}}_{\bot}$ of the parton radiated using the \powheg{}
approach.\footnote{For the regular events the shower scale is defined by
  the user on a process-by-process basis. In our case we follow the same
  prescription used for all the other event classes, i.e. we set it
  equal to the transverse momentum of the radiation.} In the \powheg{}
formulation, the splitting of \eqn{eq:realsplit} is obtained through a
dynamical, i.e. $\pth$-dependent, damping factor $D_h$
\begin{equation}
  \begin{split}
    \label{eq:hfact}
    D_h &\equiv \frac{h^2}{h^2+(\pth)^2}\,,\\ R^s_\text{\powheg} &\equiv D_h
    R_{\text{div}}\,,\quad\quad
    R^f_\text{\powheg} \equiv \left(1-D_h \right) R_{\text{div}}\,.
  \end{split}
\end{equation}
In this case, the role of the matching scale is assumed by $h$.
For Higgs transverse momenta larger than the scale $h$, the damping
factor suppresses $R^s$, while the Sudakov form factor quickly
approaches 1 and the spectrum is described by the finite remnant
$R^f_\text{\powheg}$.  Instead, when
$\pth\to 0$, $R^s_\text{\powheg}$ tends to $R_{\text{div}}$.  In this
limit, $R_{\text{div}}$ factorizes into the product of the underlying
Born multiplied by the universal Altarelli-Parisi splitting functions,
and the \powheg{} Sudakov form factor yields a suppression of the
transverse-momentum distribution.  The role of the scale $h$ can be
understood from two points of view: $i)$ it is the maximum value of
Higgs transverse momenta for which the curly bracket in
\eqn{eq:codes:matching} is appreciably different
from zero; the normalization factor $\bar B$ multiplies the curly
brackets and rescales them in this $\pth$ interval; $ii)$ considering
that the (\powheg) Sudakov form factor is a function that varies between
zero and one, the scale $h$ controls the region of $\pth$ where the
suppression is active.

A general feature of the \powheg{} approach is that it generates a tail
of the \pt{} distribution that is higher than the fixed-order prediction
in this region. The description of the enhanced $\pth$ tail and, in
particular, of the weight assigned to each high-$\pth$ event, deserves
some discussion.  The emission probability given by the squared
real-emission matrix elements $R$ in \eqn{eq:codes:matching} is
proportional to $\als$ and is multiplied by the overall factor $\bar B$,
which starts at \lo{} but includes also ${\cal O}(\als)$ corrections.
The latter are related to the total $K$-factor and enhance the first
emission weight.\footnote{In the gluon-fusion process, this enhancement
  is sizable, because of the large \nlo{} $K$-factor.} The damping
factor $D_h$ allows to reduce the $\pth$ interval where this reweighting
is active.  A second relevant element to understand the enhancement of
the large-$\pth$ tail is given by the impact of multiple emissions
beyond the first one.  In \powheg{} the phase space available for the
second emission is limited only by the $\pth$ value of the first one and
changes on an event-by-event basis. In some cases, the second and
following emissions can still be very hard.  This explains why the
\powheg{} high-$\pth$ tail tends to be larger than the one obtained in
the other two approaches and at fixed order.  This formulation differs
in particular from the one adopted in \mcnlo{}, where instead all the
\PS\ emissions are limited by the shower scale, so that the \pt{}
distribution merges into the fixed-order one at sufficiently large
transverse momenta. We will return to this discussion in
\sct{sec:largeb}.

The \powheg{} results in this paper have been obtained with the {\tt
  gg\_H\_quark-mass-effects} and {\tt gg\_H\_2HDM} generators
\cite{Bagnaschi:2011tu}. They are both part of the {\tt POWHEG-BOX}
framework\,\cite{powheg-box}.

\clearpage

\section{Determination of the matching scale}
\label{sec:approaches}

In this section we describe and compare two recently proposed algorithms
to determine the matching scales, defined in \citere{Bagnaschi:2015qta}
and \cite{Harlander:2014uea} and referred to as \bv\ and \hmw,
respectively, in what follows.  In both approaches, the matching scale
$\mu_i$ ($i=t,b,\text{int}$) is determined separately for the component
of the cross section involving only the top- or the bottom-quark loop
($\mu_t$, $\mu_b$), and for the top-bottom interference contribution
($\mu_{\mathrm{int}}$).  The resummed results for each of these terms
are then added in order to yield the best prediction for the $\pth$
distribution:
\begin{align}
\frac{\dd\sigma}{\dd\pth} = \frac{\dd \sigma_{t}}{\dd \pth}\bigg|_{\mu_t} + \frac{\dd \sigma_{b}}{\dd \pth}\bigg|_{\mu_b} + \frac{\dd \sigma_{\mathrm{int}}}{\dd \pth}\bigg|_{\mu_{\mathrm{int}}}\,.
\label{eq:tbsplit}
\end{align}
It is worth mentioning that the integral of this equation over $\pth$
reduces it to an identity, i.e.\ all matching scales drop out from the
equation and the correct normalization to the cross section with both
the top and the bottom loop is maintained.  The interference term, at
variance with the first two, is not positive definite; in particular, it
may vanish for a specific value of the Higgs mass. This will become
relevant for the discussion in \sct{sec:qualdiff} and \ref{sec:quandiff}.

Note that due to the fact that the scales are determined separately for each
component, it is possible to use them in any model with arbitrary
relative strength of the couplings of the Higgs boson to the top and
bottom quarks.  On the other hand, the presence of any other colored
particle running inside the loop would require a separate
consideration. This case could be treated in the very same fashion using
the methods described in this paper though.


\subsection{Matching scale determination \`a la \hmw{}}
\label{sec:HMW}

The idea behind the \hmw{} approach is the fact that (a) for $\pt\gtrsim
\mphi$, the $\pt$-spectrum should be well described by fixed-order
perturbation theory, and (b) one would like to have an all-order result
for an as large range of $\pt$ as possible. Let us first discuss
condition (b). In all the approaches described above, the matching scale
($\qres$ for \AR{}, $\qsh{}$ for \mcnlo{}, and $h$ for \powheg) can be
seen as a measure up to which value of $\pt$ the all-order resummation
is effective in the matched $\pt$-distribution. Formally, one would thus
like to choose the matching scale as large as possible.  However, the
resummation is strictly valid only in the limit $\pt\to 0$, so one
cannot expect to obtain a sensible result once the matching scale gets
too large. \hmw{} therefore uses condition (a) to determine a maximum
value for the matching scale. The basic idea of the \hmw{} prescription
applies in principle to any resummation/matching approach. Nevertheless,
let us focus on its application to \AR{} in what follows.

Even though the matched expression of \eqn{eq:ar} will eventually
converge to the fixed-order result for $\pt\to\infty$,\footnote{This
  follows from the fact that $\tilde L\to 0$ as $b\to 0$, see
  \eqn{eq:res}.} this transition may happen only at very large
$\pt$. Typically, for large values of the matching scale, the integral
over $\pt$ from zero to $\mphi$ becomes rather large, and may even
overshoot the total \nlo{} cross section. Due to the unitarity
constraint of \eqn{eq:unitarity}, it follows that $\dd\sigma/\dd\pt$ can
deviate significantly from the fixed-order result for $\pt\sim\mphi$ and
may even turn negative in order to compensate for the excess at small
$\pt$. This spoils the whole idea behind matching the resummed with the
fixed-order result and thus defines an upper limit on the matching
scale.

There is certainly quite an amount of arbitrariness in this procedure,
in particular: in what range should the matched result agree with the
fixed-order result, and to what degree? This arbitrariness has to be
taken into account in the estimate of the theoretical uncertainty. In
this paper, we define the so-called \hmw{} approach by following
\citere{Harlander:2014uea}, where a particular set of these parameters
was defined, from which the maximal matching scales were determined.

To be precise, $\qres^\text{max}$ is defined as the maximum value of
$\qres$ for which the resummed $\pt$-distribution stays within the
interval \mbox{$\left[0,\!2\right]\!\cdot\!
  [\dd\sigma\!/\dd\pt^2]_\text{f.o.}$} for $\pt\ge\mphi$. The default
matching scale $Q$ is then defined to be half of that maximum value.  As
it turns out, the choice of the central matching scale as defined above
indeed leads to a behavior of the matched result in the large $\pt$
region which is very close to the fixed-order result.

As pointed out above, this procedure is applied separately to the top- and
the bottom-quark induced contribution to the cross section, and to the
top-bottom interference term, resulting in three different \hmw\ scales
$\qrest$, $\qresb$, and $\qresint$, respectively.


\subsection{Matching scale determination \`a la \bv{}}

The validity of the resummation formalism relies on the soft and
collinear factorization of the squared matrix elements describing real
parton emissions.  The factorization in the soft limit can be
demonstrated in a straightforward way thanks to the fact that, for
increasing radiation wavelengths, the details of the hard scattering
process are not resolved, independently of all the other kinematic
details of the emitted parton.  The discussion of the collinear
factorization is more complex.  In the \bv{} approach, the accuracy of
the collinear approximation in the gluon-fusion process is discussed, at
partonic level, in the presence of an exact description of the top and
bottom quarks running in the virtual loop.  The procedure to determine
the scales is described by the following steps.  The exact squared
matrix elements of the subprocesses $gg\to gH$ and $qg\to qH$ are
compared with their collinear approximation.  A deviation by more than
10\% from the exact result signals that the collinear approximation
breaks down.\footnote{A detailed discussion of the dependence of the
  results on the threshold value can be found in
  \citere{Bagnaschi:2015qta}, with a direct proportionality between the
  threshold parameter and the resulting value of the matching scale.  }
The upper limit $w$ of the range of Higgs transverse momenta where the
collinear approximation is accurate is chosen as the value for the
matching scale in any hadron level calculation, either with analytic
resummation or in a \PS\ Monte Carlo.  The two partonic subprocesses
initiated by $gg$ and by $qg$ have a different collinear behavior, which
leads to two different scales $w^{gg}$ and $w^{qg}$; the final scale $w$
is computed as the average of the two previous values, weighted
differentially by their relative importance to the transverse momentum
distribution of the Higgs, in the $\pt$ range between $w^{gg}$ and
$w^{qg}$.

The three \bv{} scales associated with the top, the bottom and the
interference term will be denoted by $w_t, w_b$ and $w_\text{int}$,
respectively.


\subsection{Qualitative comparison of the two approaches}
\label{sec:qualdiff}

Since the matching scale (or specifically $\qres$, $\qsh{}$, $h$) is
unphysical, its choice is formally arbitrary, and any prescription for
its determination is necessarily heuristic.  The \bv{} and the \hmw{}
approach are complementary in at least two aspects.  While \bv{} works
at the {\it partonic} level and considers the {\it low-$\pth$} region,
the \hmw{} approach uses the {\it large-$\pth$} region of the {\it
  hadronic} distribution in order to choose a value for the matching
scale.  Furthermore, \bv{} does not make any reference to the specific
method of resumming the logarithmic terms at small $\pth$. The resulting
scales may be interpreted as \powheg's $h$, \mcnlo's $\qsh{}$, the
resummation scale $\qres$ of \AR, or any other matching scale
characteristic for the separation of the soft-collinear from the hard
region.  The scales determined through the \hmw{} approach, on the other
hand, do in principle depend on the underlying resummation technique.
We will discuss this issue in the light of the three resummation codes
considered here in the next section.

On the other hand, since both approaches separately treat the top-,
bottom-, and the interference term, the resulting scales obtained here
and in \citeres{Harlander:2014uea,Bagnaschi:2015qta} are independent of
the respective Yukawa-couplings and can be applied to the \sm, the
\thdm, and other models where the gluon-Higgs coupling is predominantly
mediated by the third generation of quarks.  In fact, the scales only
depend on the \cp{} parity and on the mass of the Higgs boson.

Considering the differences between the two approaches, it is not
surprising that the numerical values of the resulting scales are
different.  
Since the constraints that are adopted by the two groups act
on different parts of the $\pth$ spectrum (low $\pth$ in the \bv{} case,
large $\pth$ in the \hmw{} case),
the spread of the results is likely to cover in a quite conservative way
the ambiguities of this scale determination.
The hierarchy of the Higgs and of the quark masses determines a moderate (top) or a very good (bottom) agreement between the two groups.
There is one exception where the scales of \bv{} and \hmw{} may differ by many
orders of magnitude, namely when the \lo{} term is much smaller than the
\nlo{} term.  This only happens for the interference contribution which
is not required to be positive definite.  Since the resummed
contribution is always proportional to the \lo{} term (apart from
corrections due to the virtual contributions which are small compared to the total cross section),
it will also be small in these cases, and the distribution
will be given almost completely by the hard emission from the \nlo{}
term.  It follows that the \bv{}-scale will vanish as the \lo{} term tends
to zero, because the collinear approximation fails for any value of
$\pth > 0$.  On the contrary, since the resummed curve is almost
identical to the fixed-order one in this case, and since the \hmw{}
algorithm looks for the largest scale that fulfills the \hmw{} criteria,
the resulting matching scale will tend to be very large.


\subsection{Quantitative comparison of the two approaches}
\label{sec:quandiff}

\begin{figure}
\begin{center}
\includegraphics[width=0.9\textwidth]{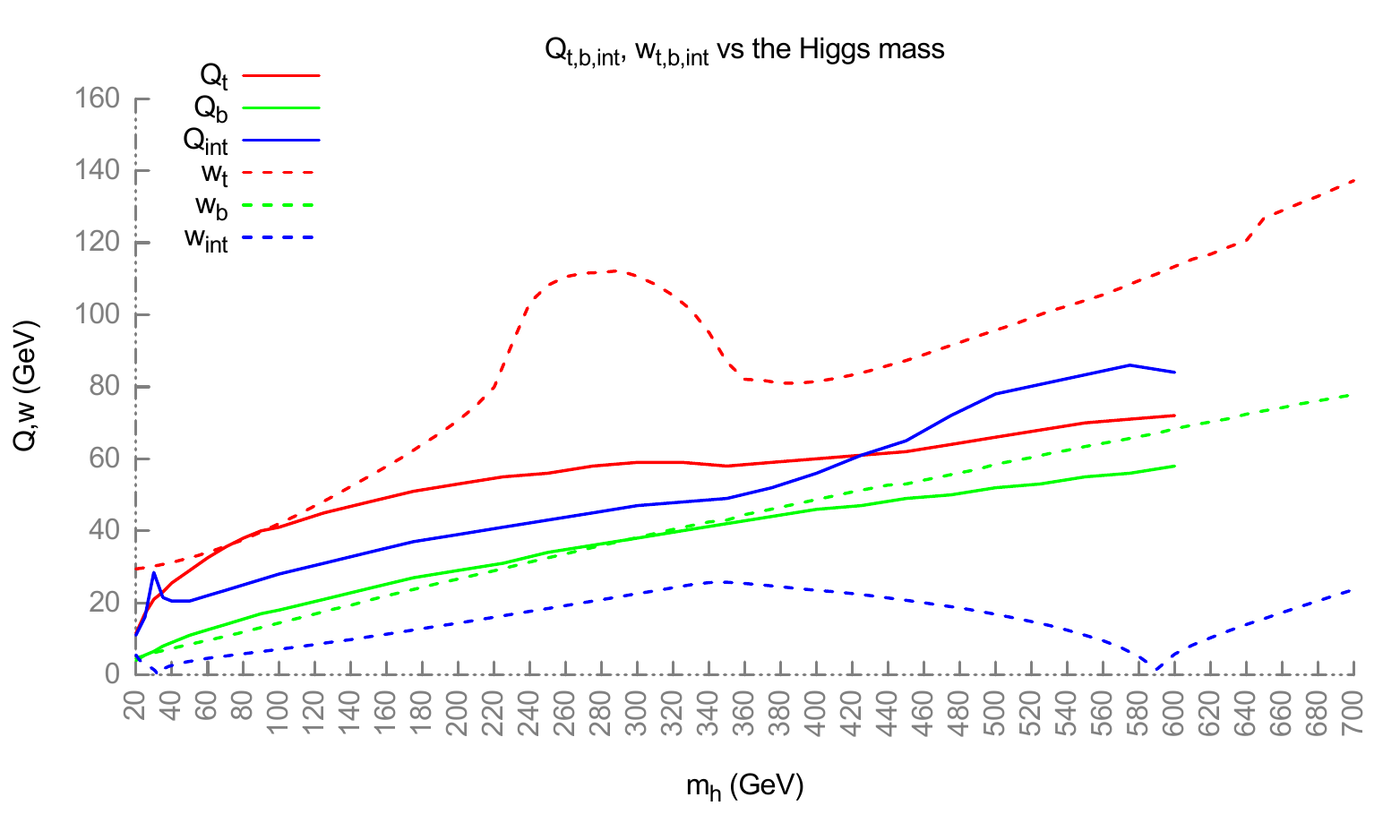}\\
\includegraphics[width=0.9\textwidth]{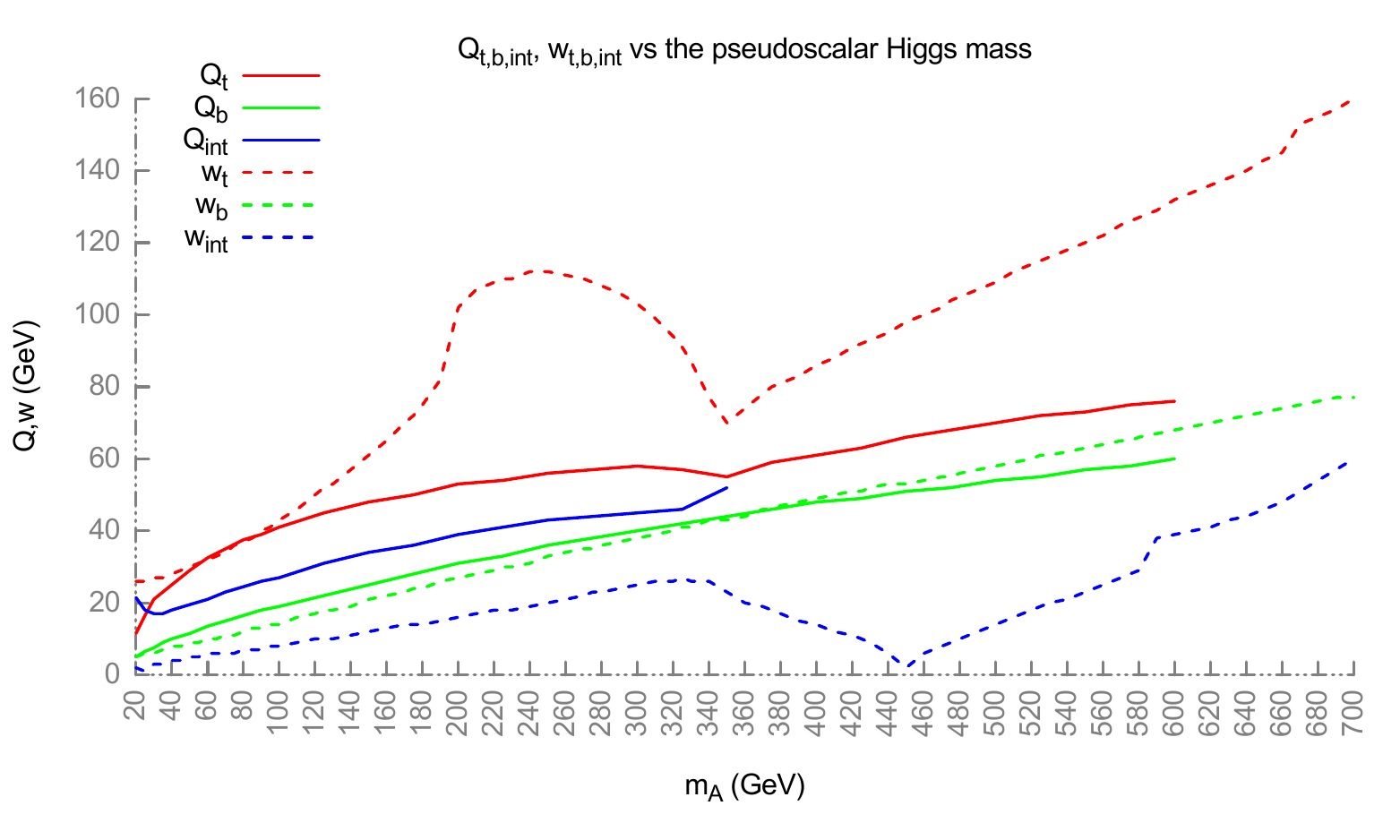}
\caption{\label{fig:scales-compare}On the top (bottom) comparison of the matching scales
  in the \bv{} and the \hmw{} approach for the scalar (pseudo-scalar). Solid (dashed) curves correspond to the \hmw{} (\bv{}) scales. The scale corresponding to the top (bottom) quark squared matrix element is shown in red (green), while
the values to be used for the interference term are in blue.}
\end{center}
\end{figure}

After clarifying the conceptual differences between the two approaches,
we can now study the actual numerical values of the matching scales.
The upper plot in \fig{fig:scales-compare} shows the three scales for
the $t$, $b$, and interference contributions for scalar Higgs production
in the two approaches, with $\mphi\le 600$ GeV in the \hmw{} case and
$\mphi\le 700$ GeV in the \bv{} case. As outlined above, while the \bv{}
scales are independent of the resummation procedure, this is not the
case for \hmw{}. The numbers shown in \fig{fig:scales-compare} are based
on \AR{}.  In principle, separate \hmw{} matching scales should be
determined for \mcnlo{} and \powheg.  For \mcnlo{}, it turns out though
that the \hmw{} scales of \fig{fig:scales-compare} lead to numerical
results close to the ones obtained in the \AR{} approach, as the
matching procedure is indeed rather similar in the two cases.  On the
contrary, \powheg{} consistently exceeds the fixed-order $\as^3$
distribution (referred to as \fnlo{}\,\cite{Alwall:2014hca} in the
following) in the high-$\pth$ tail even for very small matching scales
$h$. Any \hmw{} criterion which requires a transition to \fnlo{} at
large $\pt$ becomes questionable in this case. We therefore do not
attempt to determine separate \hmw{} scales for \powheg{}, but simply
use the ones of \fig{fig:scales-compare}.  Let us add that the situation
changes significantly when applying the so-called m\powheg{}
modification, to be defined in \sct{sec:largeb}. Similar to \mcnlo{},
the \hmw{} scales of \fig{fig:scales-compare} lead to reasonable results
in this case.

Concerning the top-induced contribution, $\wrest$ exhibits a non-trivial
structure in a broad $\mphi$ range around the top-quark threshold,
between about $220$ and $380$\,GeV.  The corresponding \hmw{} scale
$\qrest$, on the other hand, increases quite steadily with the Higgs
mass, and the top-quark threshold only has a very mild effect, if any.
In addition, the overall growth of $\wrest$ with $\mphi$ is stronger
than for $\qrest$.  Nevertheless, over the whole $\mphi\le 600$ GeV
region considered here, the two scales never differ by more than a
factor of two.  Since the resummation uncertainty in the two approaches
will be estimated by varying the matching scales by a factor of two, we
can expect consistent results in cases where the top contribution is
dominant.

On the other hand, the \bv{} and \hmw{} scales for the bottom-induced
contribution are in much better agreement.  The \bv{} scale $\wresb$
exhibits a slightly steeper dependence on $\mphi$ than $\qresb$, but the
difference between the two remains quite small for all $\mphi\lesssim
600$\,GeV. Except for very small Higgs masses, $\wresb$ and $\qresb$ are
considerably larger than $m_b$. In the \bv\ approach one observes that
this result is driven mainly by the behavior of the partonic $gg\to \phi
g$ channel\,\cite{Bagnaschi:2015qta} (see also \citere{Banfi:2013eda}),
while the $qg\to \phi q$ channel would suggest a value for the bottom
matching closer to $m_b$ \cite{Grazzini:2013mca}.

For the scales of the interference contribution, both approaches lead to
a very similar slope in their respective matching scales as $\mphi$
increases from about 40 to 320\,GeV, even though their absolute values
differ significantly. While $\wresint$ remains below about 25\,GeV,
$\qresint$ is always {\it larger} than 20\,GeV as long as $\mphi\gtrsim
30$\,GeV. For larger values of $\mphi$, the two interference scales show
a very different behavior: The \bv{} scale $\wresint$ slowly decreases
until it reaches the value $\wresint=0$ at about 590\,GeV, which happens
precisely where the interference term of the total cross section at
\lo{} vanishes. In contrast to that, the slope of the \hmw{} scale {\it
  increases} beyond $\mphi\approx 340$\,GeV, and $\qresint$ assumes its
maximal value of about 80\,GeV at $\mphi\approx 570$\,GeV. A similar
feature is observed around $\mphi=30$\,GeV. The reasons for the
qualitative differences in this case have already been discussed in
\sct{sec:qualdiff}; they are the clearest manifestation of the different
ideas behind the \bv{} and the \hmw{} method.

The corresponding plot for a \cp-odd Higgs boson is shown in
\fig{fig:scales-compare}; the general behavior for all the three
contributions is quite similar to the scalar case and requires no
separate discussion.

\clearpage

\section{Numerical results} \label{sec:results}


In this section, we present a quantitative comparison of the impact of
the \bv{} and of the \hmw{} scale determinations, using three different codes:
{\tt MoRe-SusHi}, which implements the analytic resummation of the
$\pth$ spectrum, {\tt gg\_H\_2HDM} and {\tt aMCSusHi}, which apply the
\powheg\ and the \mcnlo\ method, respectively.  Although the theoretical
basis underlying these three codes is the same, namely resummation based
on soft and collinear factorization, the specific algorithms involved
are quite different, see \sct{sec:codedesc}. Therefore, we will try to
disentangle effects due to these different implementations from those
arising from the different matching scales.

The uncertainty band due to a variation of the matching scale is
obtained by the following procedure: Given a set of reference values
$(\mu_{t},\mu_{b},\mu_\text{int})$ for the three matching scales
according to \eqn{eq:tbsplit} (adopting either the \bv{} or the \hmw{}
approach) we consider all the possible combinations which can be
generated by taking half and twice these values, or the reference values
themselves; for each setting, we compute the transverse momentum
distribution; collecting all these results, we take the envelope for
each $\pth$ value, i.e.\ the minimum and the maximum values among all
the simulations.\footnote{Recall that the accuracy of \AR{} is \nll{},
while the \nlops{} \MC{}s only resum consistently all {\abbrev LL} terms
and partially the \nll{} ones. Therefore, the higher order terms probed
by the scale variation procedure are slightly different in the two
cases.}

For the analytic resummation, which consistently matches the fixed-order
results at large transverse momenta, we
follow \citere{Harlander:2014uea} and apply an additional factor
\begin{equation}
\begin{split}
d(\pt) =
\left\{1+\exp\left[\alpha\left(\pt-\mphi\right)\right]\right\}^{-1}\,,\qquad
\alpha = 0.1\,\mathrm{GeV}^{-1}\,,
\label{eq:damping}
\end{split}
\end{equation}
to the error band which damps it towards large values of $\pt$. This
takes into account condition (a) from \sct{sec:HMW}, according to which
resummation should not have a big impact on the large-$\pt$ region, and
thus also not on its theoretical ambiguities.

As we will see, the shape of the uncertainty band as derived from the
variation of the matching scales has a feature common to all the
codes. In a region just above the peak of the $\pth$ distribution, the
band is relatively narrow. This is a consequence of the unitarity
constraint, which establishes an anticorrelation between the low- and
the high-$\pth$ tails of the distribution.  The precise position of this
region depends on the position of the peak of the resummed distribution,
on the total variation of the cross section, as well as on the central
value of the matching scale; the interplay of these factors determines
the precise shape of the uncertainty band.


\subsection{Setup and representative scenarios in the \thdm}
\label{sec:defscenarios}

This section defines the phenomenological scenarios considered in this
paper. They have been designed to highlight possible interplays between
the top-quark and the bottom-quark mediated amplitudes.  

We start with the \sm{}, where the $\pt$ spectrum of the Higgs boson is
known through \nlo{}+\nll{} including the full dependence on the quark
masses \cite{Mantler:2012bj} and through \nnlo{}+\nnll{} for the top-quark induced terms and by
assuming the limit
$\mtop\to\infty$\,\cite{Bozzi:2003jy,Bozzi:2005wk,deFlorian:2011xf}.\footnote{Recall 
that the {\abbrev (N)NLO} accuracy denoted here refers to the underlying total 
cross section (including terms up to $\alpha_s^3$ and $\alpha_s^4$, respectively) and corresponds to {\abbrev (N)LO} accurate predictions at large 
transverse momenta.}
In the present study, for uniformity with the \bsm\ codes where only
\nlo\ \qcd\ accuracy on the total cross section and 
  \nlo{}+{\abbrev(N)LL} results on the Higgs transverse momentum distribution are
available, we restrict also the \sm\ analysis to this level of accuracy.

While there is hardly any controversy that the ``characteristic scale''
for the top-quark induced contributions in the \sm{} should be of the
order of the Higgs boson mass ($m_h=125$\,GeV), already in this case the
\bv\ and the \hmw\ methods provides us with a more quantitative result
for the matching scale which turns out to be close to the often adopted
choice of $\mh/2$ in the \sm{}, but becomes significantly smaller than
$\mphi$ towards larger values of $\mphi$ (see \fig{fig:scales-compare}).
One of the main subjects of this paper, however, is the question of how
to take into account the bottom-quark induced contribution. In the
\sm{}, the effect of the bottom quark is suppressed by the Yukawa
coupling, and therefore differences in this treatment have limited
effect on the overall momentum distribution. In models with an extended
Higgs sector, however, the bottom-quark Yukawa coupling can be
significantly enhanced, at least for some of the Higgs bosons of such
theories.

One of the simplest extensions of the \sm{} in this respect is the
\thdm{}. Therefore, we focus on \thdm{} scenarios which we devise in
order to enhance specific contributions to the cross
section.\footnote{More precisely, we only consider the type-II \thdm{},
  in which one doublet generates the masses of the up-type quarks and
  the other of down-type quarks and charged leptons. In any case our
  results are directly applicable also to the other \thdm{} types.}  The
conclusions of our study, however, trivially generalize to other models
where the gluon-Higgs coupling is predominantly mediated by top and
bottom quarks (e.\,g.\ most of the experimentally viable parameter space
of the \mssm).  Since the aim of this paper is a conceptual one, we
disregard any phenomenological constraints on the \thdm{} parameter
space, in general; we do not consider the case of a light Higgs boson of
mass $\mh=125$\,GeV with enhanced bottom-quark Yukawa coupling though,
due to the obvious conflict with experimental observations.  We do,
however, respect the {\it theoretical} constraints due to unitarity and
triviality of the theory, as well as stability of the physical
vacuum. We check these constraints with the help of the program {\tt
  2HDMC}\,\cite{Eriksson:2010zzb,Eriksson:2009ws}.  In all scenarios,
except for the \lowma\ scenario to be introduced at the end of this
section, we set the mass of the two \cp-even Higgs bosons to
$\mh=125$\,GeV and $\mH=300$\,GeV, respectively, while the mass of the
\cp-odd Higgs boson is set to $\ma=270$\,GeV.

The first scenario we consider is ``scenario B'' as defined in
\citere{Harlander:2013qxa}. For consistency with the notation in the
rest of this paper, however, we will refer to it as ``\largebot{}
scenario'' in what follows. Since it induces a \sm{}-like light Higgs
boson, we only study the production of the heavy and the \cp{}-odd Higgs
boson for that scenario, both of which have strongly enhanced couplings
to the bottom quark.

As a modification of this, we use the same parameters as in the
\largebot{} scenario, except that we set $\tan\beta=1$. This will be
referred to as ``\largetop{} scenario'' in what follows. It is designed
to result in a top-quark dominated cross section for the heavy and
pseudo-scalar Higgs boson. Again, the light Higgs is very \sm{}-like and
will not be considered in this scenario.

Finally, we devise a set of rather pathological scenarios where the
\lo{} cross section receives a large top-bottom interference
contribution.  The parameters of these ``\largeint{} scenarios'' have to
be chosen differently for each of the neutral Higgs bosons.

\begin{table}
\begin{center}
\begin{tabular}{|c|cc|c|rr|rr|rr|}
\hline
\multirow{2}{*}{scenario} & 
\multirow{2}{*}{$\tan\beta$} & 
\multirow{2}{*}{$\sin(\beta-\alpha)$} &
\multirow{2}{*}{$\phi$}
&\multicolumn{2}{c|}{$\sigma_t$/pb} & 
\multicolumn{2}{c|}{$\sigma_b$/pb} & 
\multicolumn{2}{c|}{$-\sigma_\text{int}$/pb}\\
&&&& \lo{} & \nlo{} & \lo{} & \nlo{} & \lo{} & \nlo{}\\
\hline
\multirow{2}{*}{\sm} & \multirow{2}{*}{---} & \multirow{2}{*}{---} &
$H$ & 20.027 & 33.400 & 0.220 & 0.268 & 2.410 & 2.433\\
&&& $A$& 46.355 & 78.125 & 0.244 & 0.291 & 4.202 & 4.506\\
\hline
\multirow{2}{*}{\largebot{}} & \multirow{2}{*}{50} & \multirow{2}{*}{0.999} &
$H$ & 0.002 & 0.005 & 5.085 & 7.089 & $0.163$&$0.199$\\
&&& $A$& 0.005 & 0.010 & 9.984 & 13.408 & 0.334 & 0.412\\
\hline
\multirow{2}{*}{\largetop{}} & \multirow{2}{*}{1.0} & \multirow{2}{*}{0.999} &
$H$ & 3.715 &6.788& 0.002&0.003 & $-0.132$&$-0.168$\\
&&& $A$ & 12.844 & 23.832 & 0.004 &0.005&0.334&0.428\\
\hline
\multirow{2}{*}{\largeint{}} & 3.2 & $-0.6$ & $h$ &2.453&4.091&2.192&2.674&$2.665$&$2.677$\\
 & 7.1 & $-0.26$ & $A$ & 0.255&0.473 & 0.201&0.270 &
$0.334$&$0.430$\\
\hline
\lowma{} & 36.9 & $0.998$ & $A$ &0.399 &0.552 &$2.480\cdot
10^5$&$2.292\cdot 10^5$&89.70&$-693.6$\\
\hline
\end{tabular}
\end{center}
\caption{Cross sections for the three \thdm{} scenarios considered in
  our study, obtained with \sushi{} (the integration error at \nlo{} is
  of the order of 0.1\%, and negligible at \lo{}). See the text for a
  description of their characteristics.\label{tab:scenarios}}
\end{table}

The precise definition of all scenarios, together with the top-, bottom-
and interference component of the total inclusive cross sections at
\lo{} and \nlo{}, is given in Table\,\ref{tab:scenarios}.  Note that,
while for the light and the pseudo-scalar Higgs the absolute value of the
interference term in the \largeint{} scenarios amounts to more than
100\% of the total cross section, we did not manage to find a parameter
point for the heavy Higgs which has a similarly large interference term
while still respecting the theoretical constraints of unitarity,
stability, and perturbativity. 

Let us emphasize again that most of these scenarios are in vast conflict
with experimental observations; they only serve as theoretical
benchmarks for the study of resummation ambiguities in the Higgs $\pth$
distribution. For phenomenologically viable \thdm{} benchmark points we
refer the reader to \citere{Haber:2015pua}.

We further investigate one phenomenologically interesting scenario with
a very low pseudo-scalar Higgs boson mass. In this case we chose a
scenario of \citere{Bernon:2014nxa} that meets all theoretical as well
as experimental constraints. This scenario is referred to as \lowma{} in
Table\,\ref{tab:scenarios}. The masses of the three Higgs bosons are
$\mh=125.5$\,GeV, $\mH=507$\,GeV and $\ma=29.9$\,GeV. For such a low
Higgs-boson mass, the gluon-fusion process is particularly important,
since its cross section is highly enhanced with respect to Higgs
production in association with bottom quarks\footnote{For details on
  this process, see \citeres{Harlander:2014hya,Wiesemann:2014ioa} and
  references therein.} and dominant even at large
$\tan\beta$\,\cite{Bernon:2014nxa}.

All the numerical results are computed for the \lhc{}, with a
center-of-mass energy of $\sqrt{S}=13$\,TeV.  We use the {\tt
  MSTW2008nlo68cl} \pdf{} set\,\cite{Martin:2009iq} through the {\tt
  LHAPDF6} library\,\cite{Buckley:2014ana} and the corresponding value of
$\alpha_s(M_Z)=0.120179$. The renormalization and factorization scales
are both identified with $\mphi$. The pole masses of the top and the
bottom quark are fixed at $m_t=172.5$\,GeV and $m_b = 4.75$\,GeV,
respectively. We used the on-shell renormalization scheme for the Yukawa
couplings. For simulations involving a parton shower, we apply {\tt
  Pythia8}~\cite{Sjostrand:2007gs}.


\subsection{Numerical results in the different scenarios}


This section contains the results obtained with the three codes under
study, in the various scenarios defined in \sct{sec:defscenarios}.
Since our focus is on the uncertainties inherent to the matching
procedure, we compute the uncertainty band by varying only the matching
scales, as described in \sct{sec:results}.

For each scenario we performed two different studies:
\begin{enumerate}
\item
The predictions for \AR, \mcnlo~and \powheg{} are compared by using the
same numerical values for the respective matching scales; in this way we
can assess the impact of the higher-order \qcd{} terms which are
included in different ways in the three codes, due to the different
matching procedures.
\item
For each code, the predictions obtained by setting the matching scale to
the \bv{} and to the \hmw{} values are compared; this allows us to
assess the sensitivity of each code to a matching scale variation, the
impact of the \bv{} {\it vs.}\ the \hmw{} prescription, and the overlap
of the respective uncertainty bands.
\end{enumerate}


\subsubsection{Results in the Standard Model}
\begin{figure}[!h]
\centering
\includegraphics[width=0.45\textwidth]{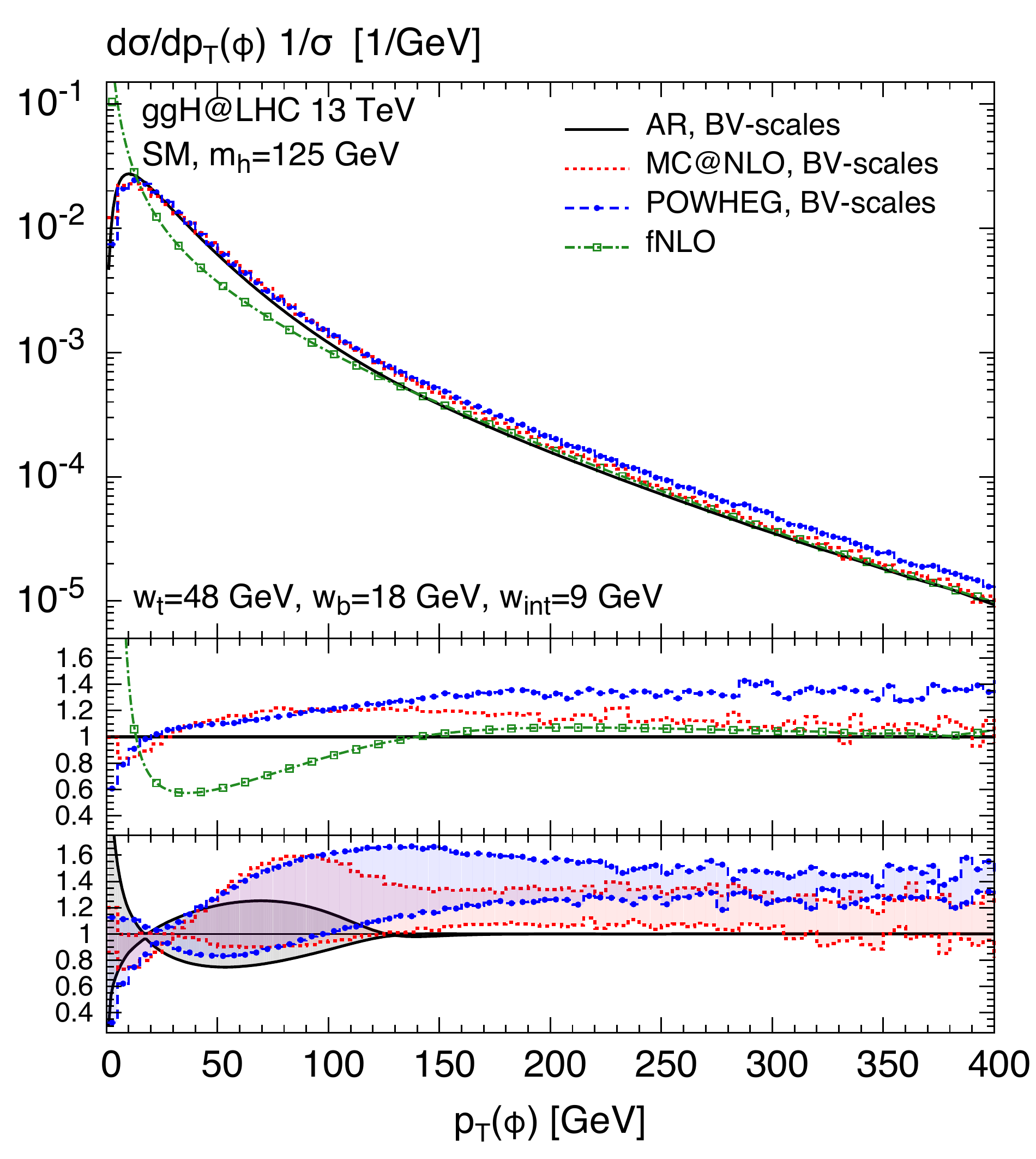}
\includegraphics[width=0.45\textwidth]{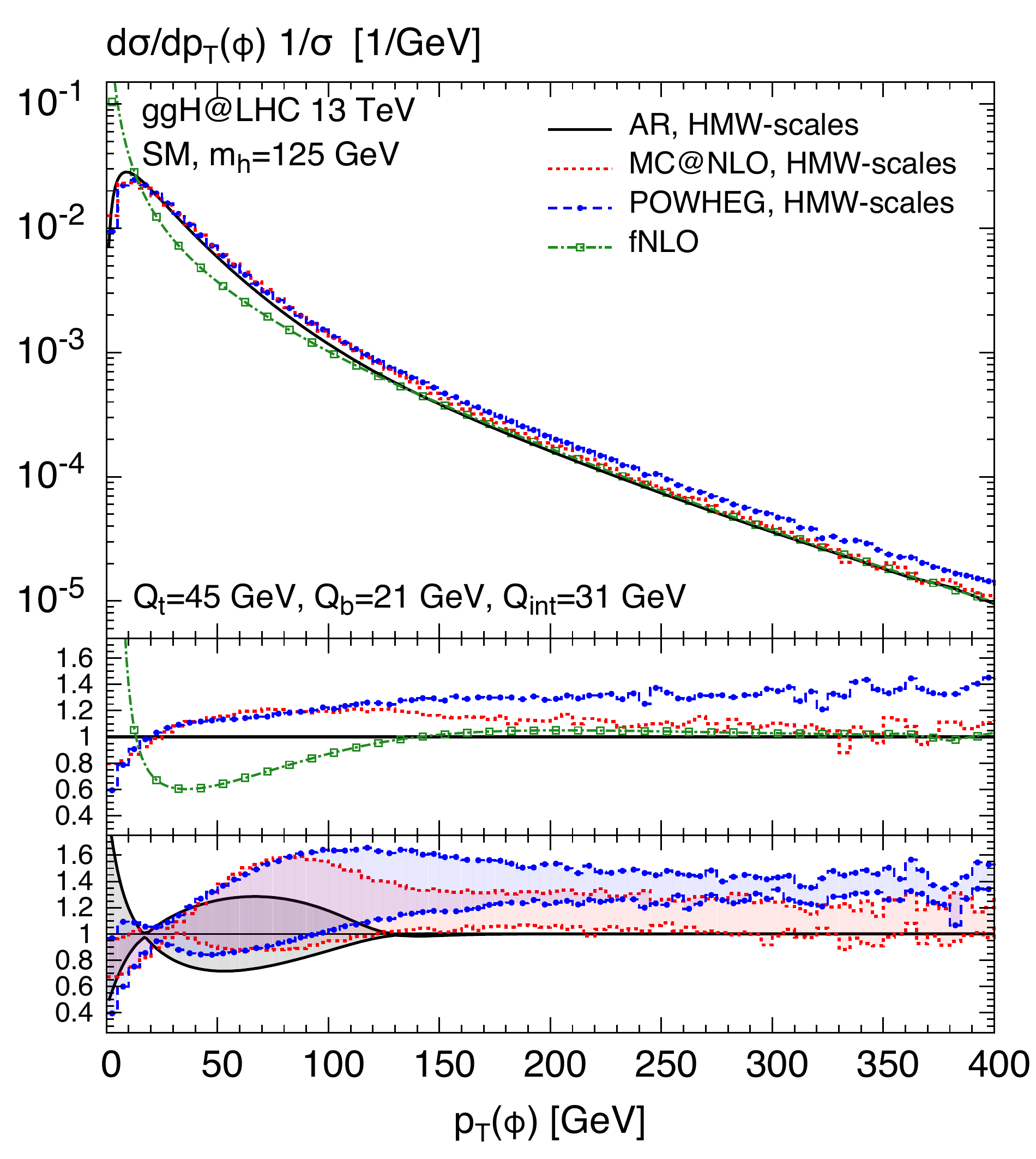}\\
\hspace{-0.25cm}
\includegraphics[width=0.371\textwidth]{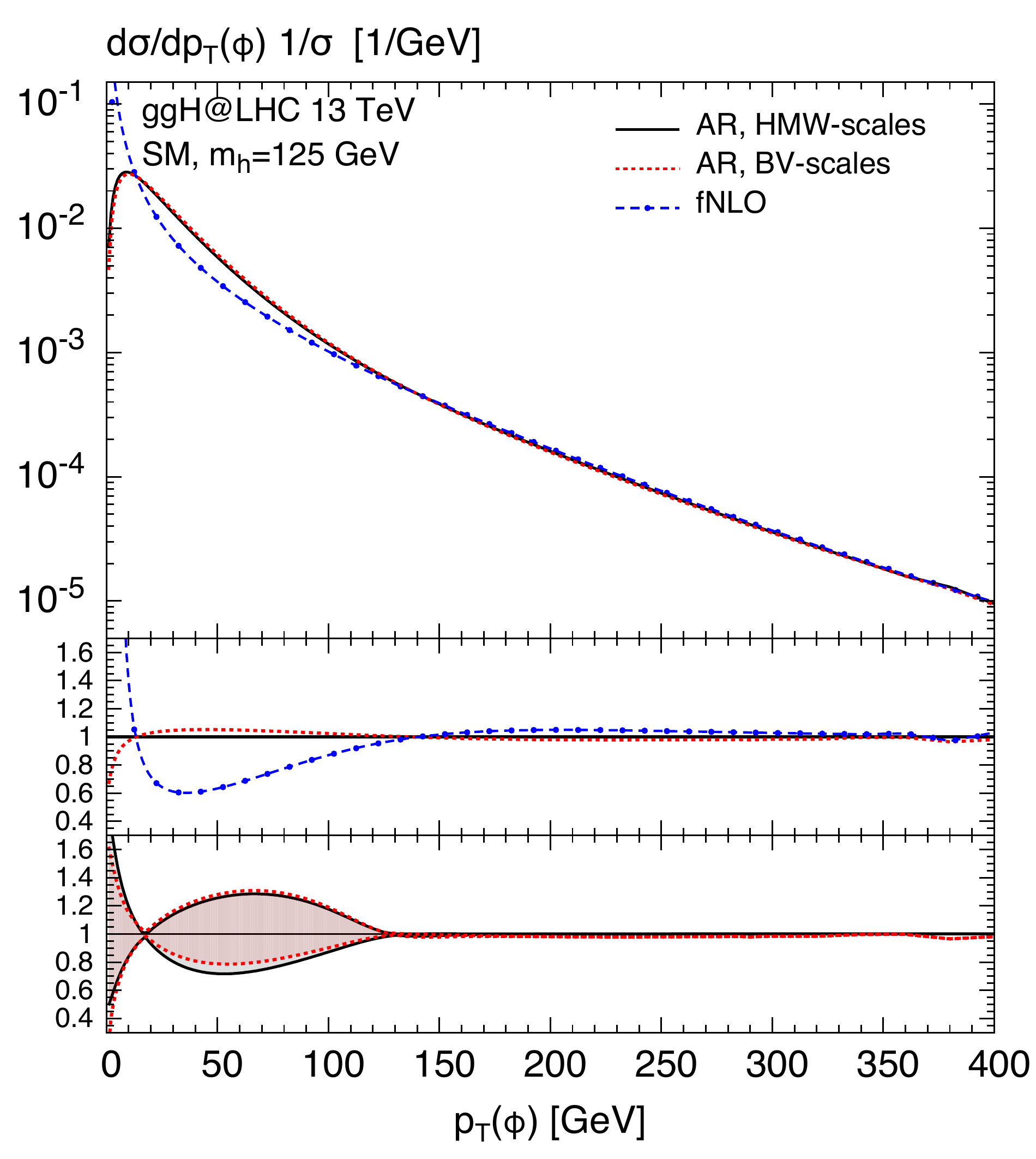}\hspace{-1.02cm}
\includegraphics[width=0.371\textwidth]{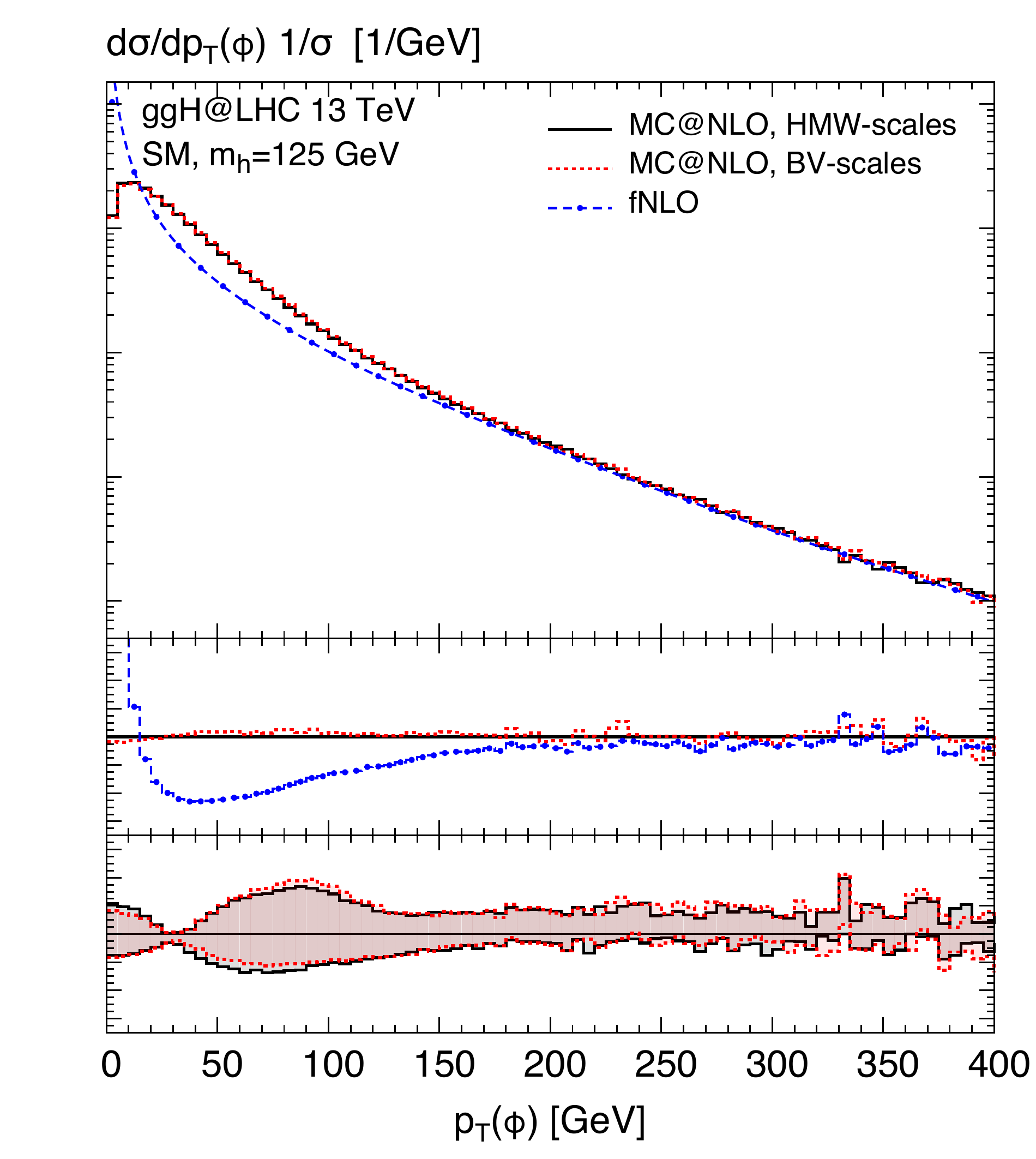}\hspace{-1.02cm}
\includegraphics[width=0.371\textwidth]{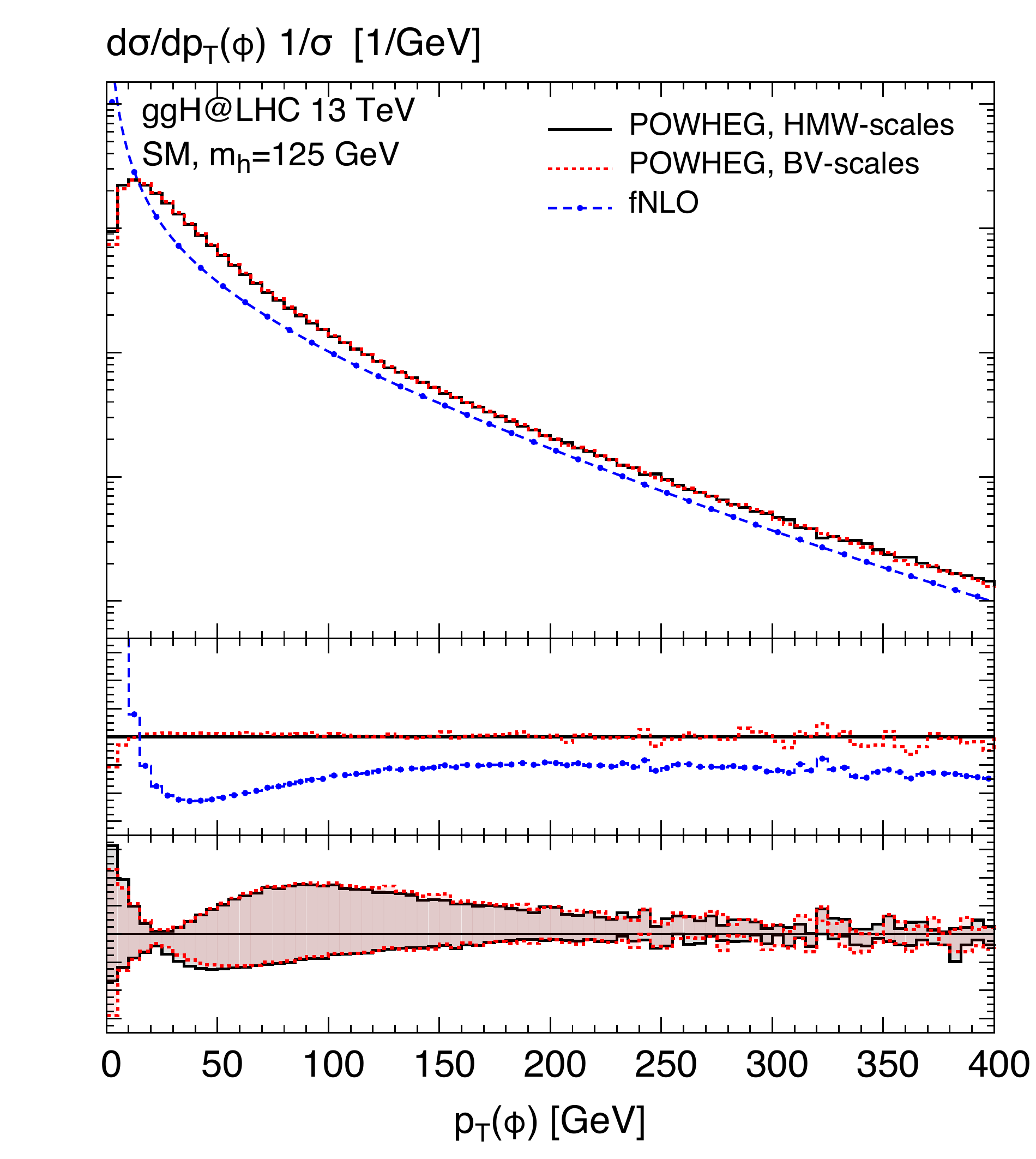}
\caption{Shapes of the transverse-momentum distributions (i.e., normalized such 
that the integral yields one) for a \sm{} Higgs
  boson with $\mh=125$\,GeV.  In the upper plots we show the
  distributions computed with \AR{} (black, solid), \mcnlo{} (red, dotted) and \powheg{} (blue, dashed overlaid by points), setting the
  matching scales to the \bv{} values (left) or the \hmw{} values (right).
  For reference, we also show the fixed-\nlo{} (\fnlo{}) prediction (green, dash-dotted with open boxes).
  The main frame shows the absolute distributions, the
  first inset the shape-ratio of the central values to the \AR{} distribution,
  and the second inset the uncertainty bands, normalized
  again to the central \AR~value.  In the lower three plots we compare the
  results within each code, using for the matching scales the \bv{}
  values (red, dotted) and the \hmw{} values (black, solid), taking the \hmw{} results as
  reference for the ratios of the insets.
\label{fig:results-sm}
}
\end{figure}

\begin{figure}[!h]
\centering
\includegraphics[width=0.45\textwidth]{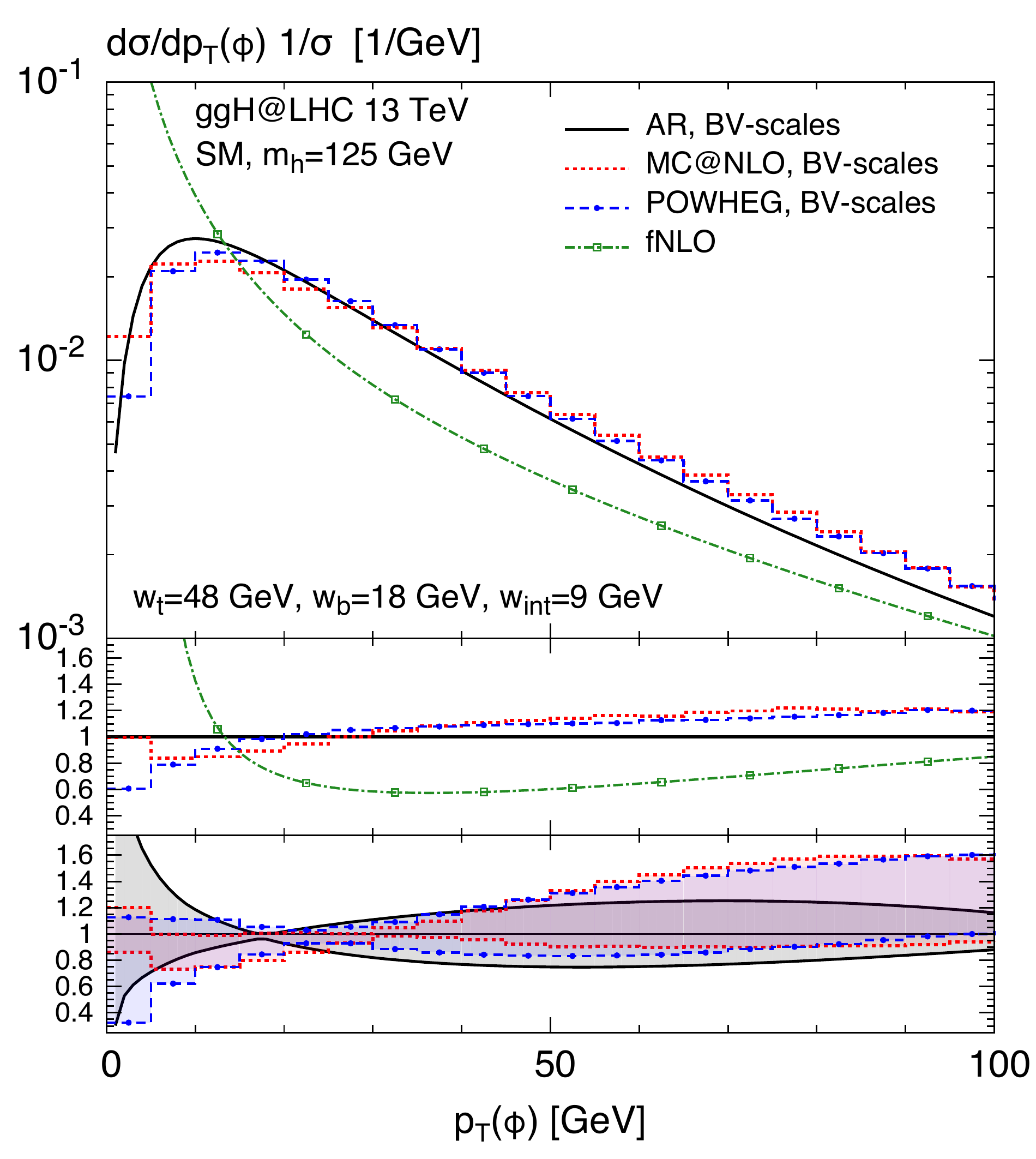}
\includegraphics[width=0.45\textwidth]{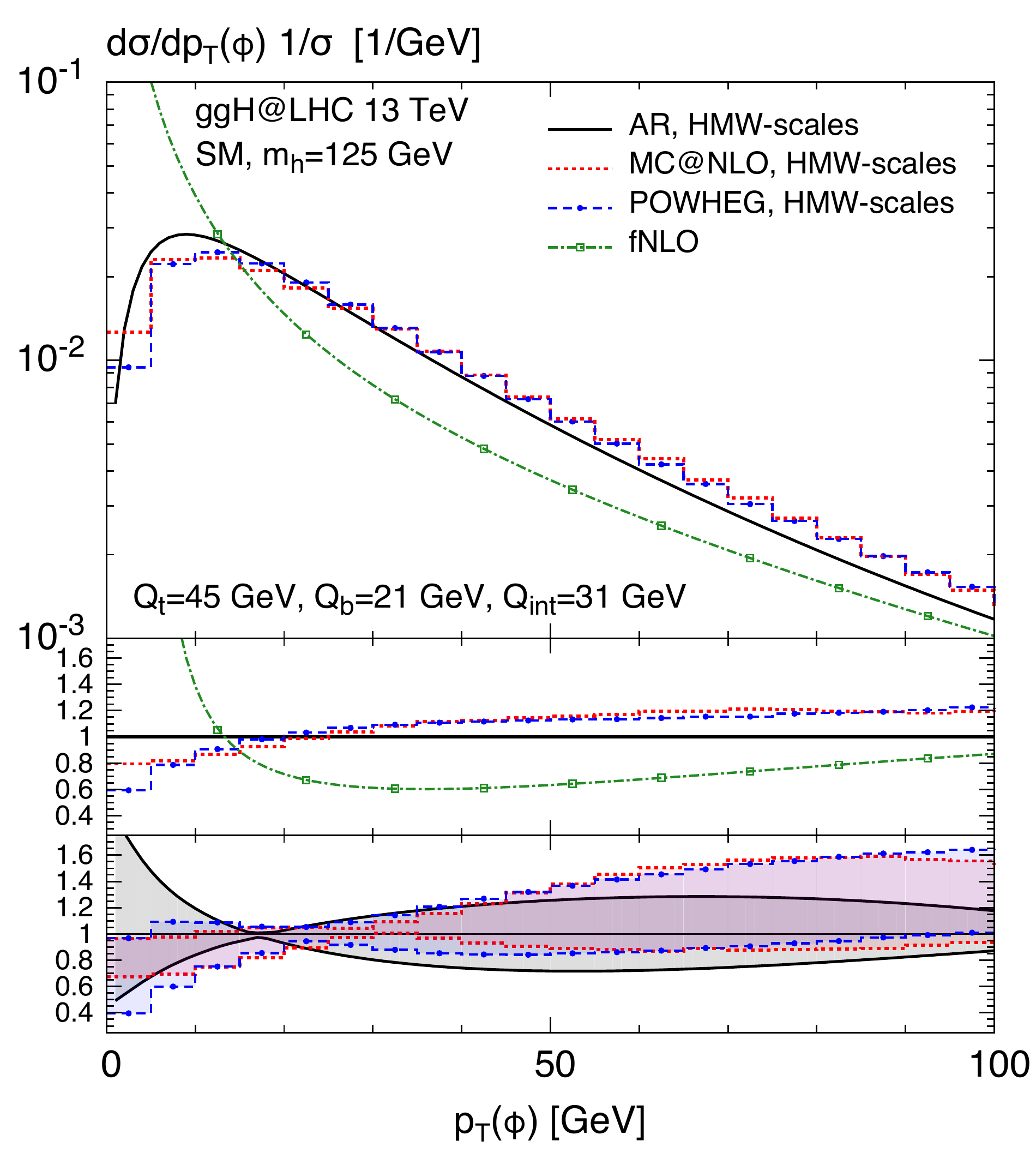}\\
\hspace{-0.25cm}
\includegraphics[width=0.371\textwidth]{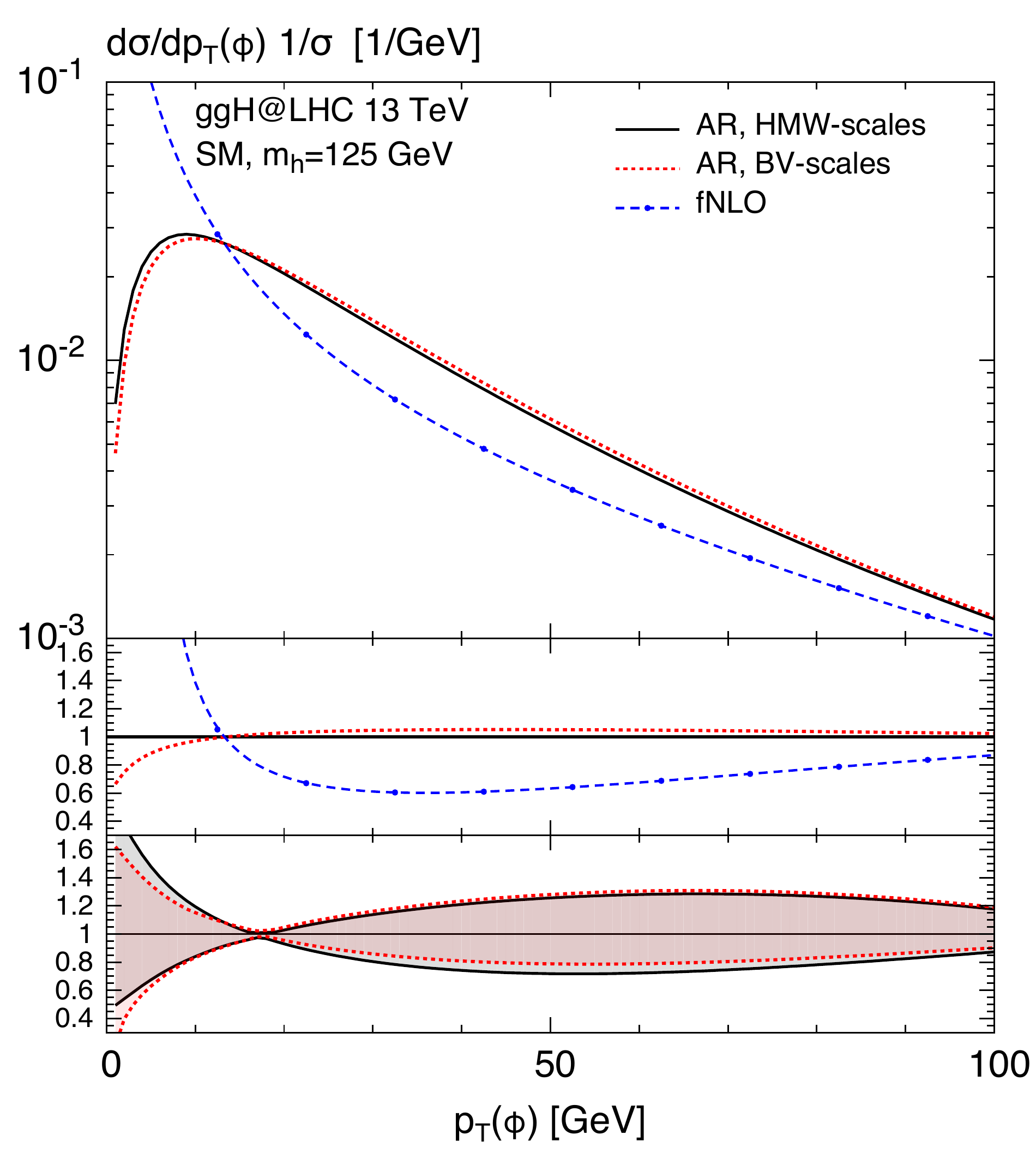}\hspace{-1.02cm}
\includegraphics[width=0.371\textwidth]{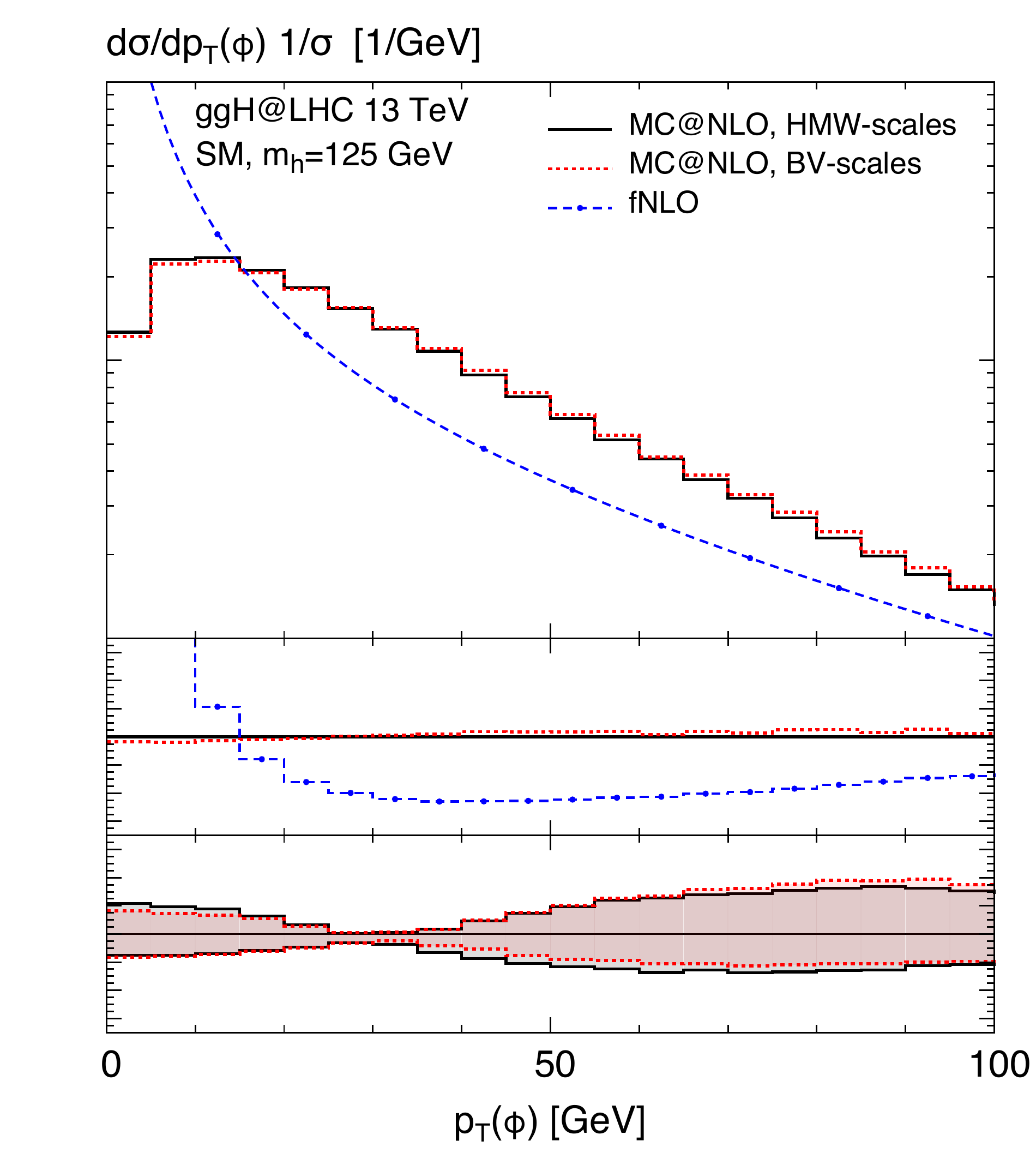}\hspace{-1.02cm}
\includegraphics[width=0.371\textwidth]{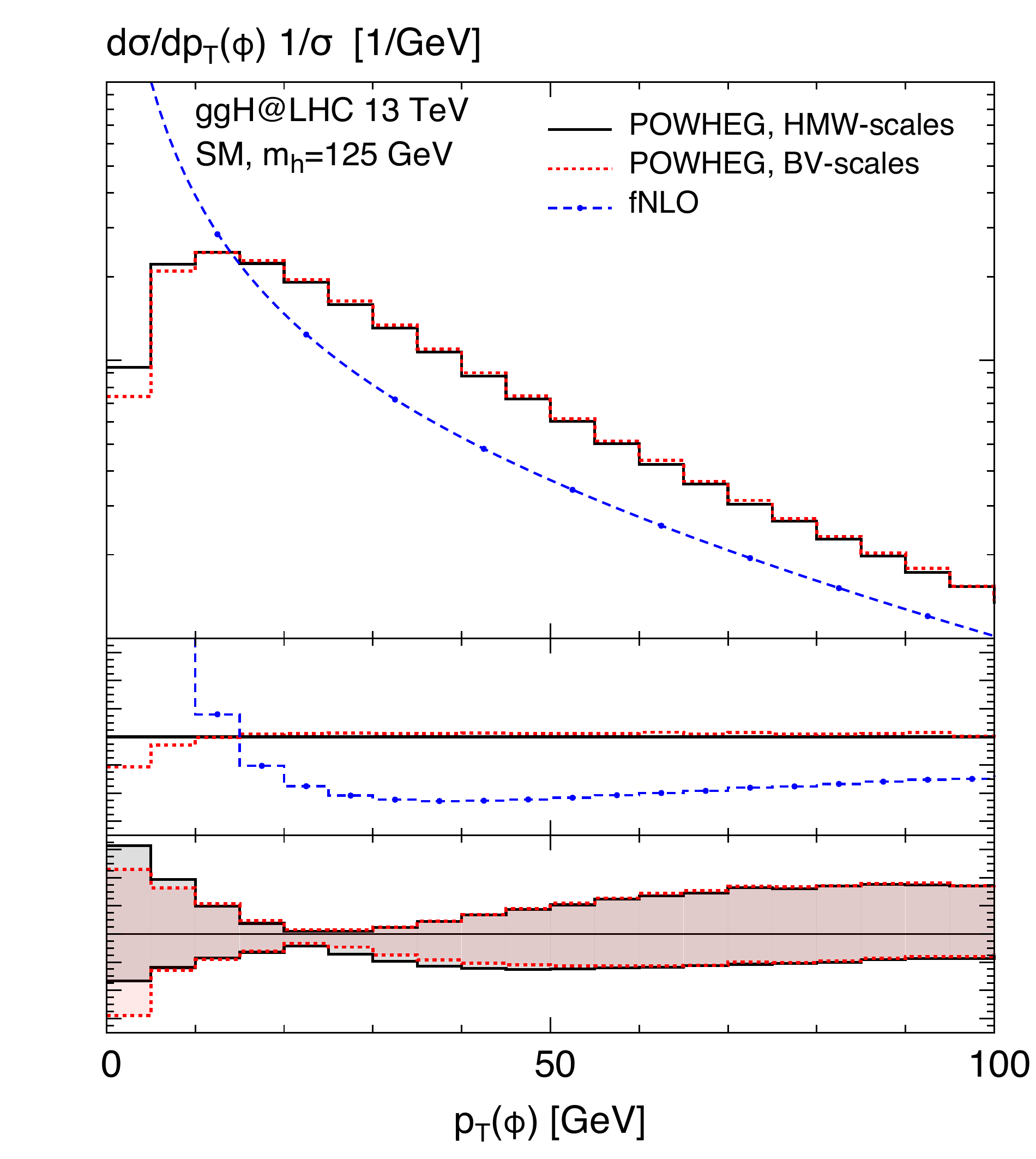}
\caption{Same as \fig{fig:results-sm}, but with enlarged low-\pt{} region.
\label{fig:results-sm2}
}
\end{figure}

Even though for the \sm{} one may expect consistency among the three
codes, it will be instructive to start our discussion with this case,
because it highlights certain generic features of the individual
approaches which will be carried forward also to some of the other
scenarios considered in this paper. \fig{fig:results-sm} shows the shape
of the transverse-momentum distribution (i.e., the integral of each
curve is normalized to one) for a \sm{} Higgs of $\mh=125$\,GeV in the
range $0\le \pt/{\rm GeV}\le 400$.  In the two upper plots, we compare
the results of the three codes, setting the matching scales to the same
numerical values: \bv{} scales in the left and \hmw{} scales in the
right panel (cf.\,item {\it i)} above). Each of the lower plots, on the
other hand, was obtained with one particular code (left: \AR{}; center:
\mcnlo{}; right: \powheg{}); the different lines correspond to different
values of the matching scales, \bv{} and \hmw{} (cf.\,item {\it ii)}
above). All plots contain the \fnlo{} result as a reference.  In order
to facilitate the discussion, we show the same plots in
\fig{fig:results-sm2}, but with enlarged low-\pt{} region ($0\le
\pt/{\rm GeV}\le 100$).

Apart from the fact that, as expected, the three codes yield compatible
results within uncertainties (at least for $\pt\le \mh$), it is worth
noting the following characteristics:
\begin{itemize}
\item The central \powheg{} and the \mcnlo{} spectra are in excellent agreement (at the few-percent
  level) between about $10<\pt/\text{GeV}<130$, while they differ by about
  20\% from the central \AR{} prediction in most of this region.
\item The peak position for both \powheg{} and \mcnlo{} is at around
  $\pt=12$\,GeV, while the peak for \AR{} is about 2\,GeV below that.
\item The central value of \AR{} approaches \fnlo{} at the level of
  $\lesssim$\,5\% above $\pt\approx 130$\,GeV; for \mcnlo{}, such a
  transition to the fixed-order curve occurs at about $\pt\approx
  180$\,GeV, while \powheg{} always remains about 20\% above \fnlo{} in
  the tail of the distribution.  The latter characteristic is a general
  feature of the default \powheg{} matching, as described in Section
  \ref{sec:powhegcode}, and will be analyzed in detail in Section
  \ref{sec:largeb}.
\item Above $\pt\approx 130$\,GeV, the uncertainty band for \AR{}
  is suppressed due to the damping factor introduced in
  \eqn{eq:damping}. Concerning the two \MC{}s, in \mcnlo{}
  the uncertainty, expressed in unit of the \AR\ central value, 
  is of the order of $\pm 10\%$ from $\pt=130$\,GeV all the way up to $\pt=400$\,GeV,
  while for \powheg{} it decreases uniformly from $\pm 20$\% to $\pm 10$\%.
\item The \mcnlo{} and \powheg{} bands nicely overlap for $\pth\leq
  100$\,GeV; above this value the overlap is only partial, because of
  the different central predictions.
\item Towards smaller values of $\pt$, the uncertainty bands for all
  three codes develop a bulgy structure with a maximum of about $\pm
  20\%$ $(\pm 35\%)$ for \AR{} (the Monte Carlos) and a minimum of a few percent slightly above the peak position.
\item Towards even smaller values of $\pt$, the \AR{} uncertainty band
  quickly grows to the 100\% level, and the \powheg{} band to about $\pm
  40\%$. Only the \mcnlo{} band increases to a moderate $\pm 15\%$.
\end{itemize}

Since, for a \sm{} Higgs of $\mh=125$\,GeV, the \bv{} and \hmw{} scales
are quite close to each other (see \fig{fig:scales-compare}), the two
upper plots in \fig{fig:results-sm} are very similar. In order to see
this more explicitly, we study the impact of the different scale choices
(\bv{} or \hmw{}) within one particular code (left: \AR{}; right:
\mcnlo) in the lower three plots of \fig{fig:results-sm}.  From the
latter we can indeed see that the results obtained in the two cases are
in very good agreement with each other, both with respect to the central
value and the uncertainty band; the only difference being at very low
transverse momenta ($\pt{}<10$\,GeV) in the central predictions of \AR{}
and \powheg{}. This can be traced back to the different matching scales
for the interference term, which constitutes the largest contribution
induced by the bottom-quark loop in the \sm{}.

\clearpage
\newpage
\subsubsection{Results in the \largetop{} scenario}

\begin{figure}[!h]
\centering
\includegraphics[width=0.45\textwidth]{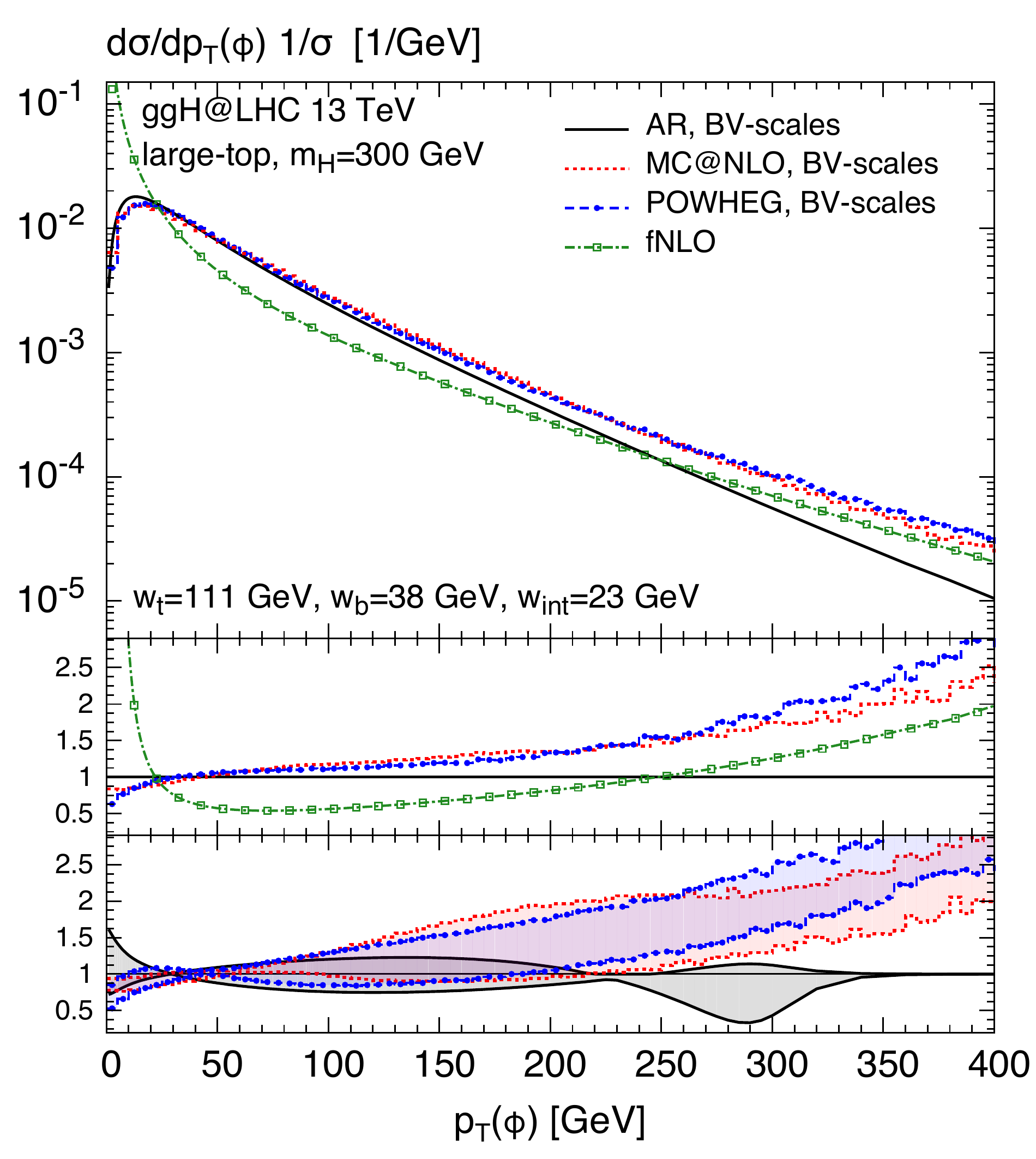}
\includegraphics[width=0.45\textwidth]{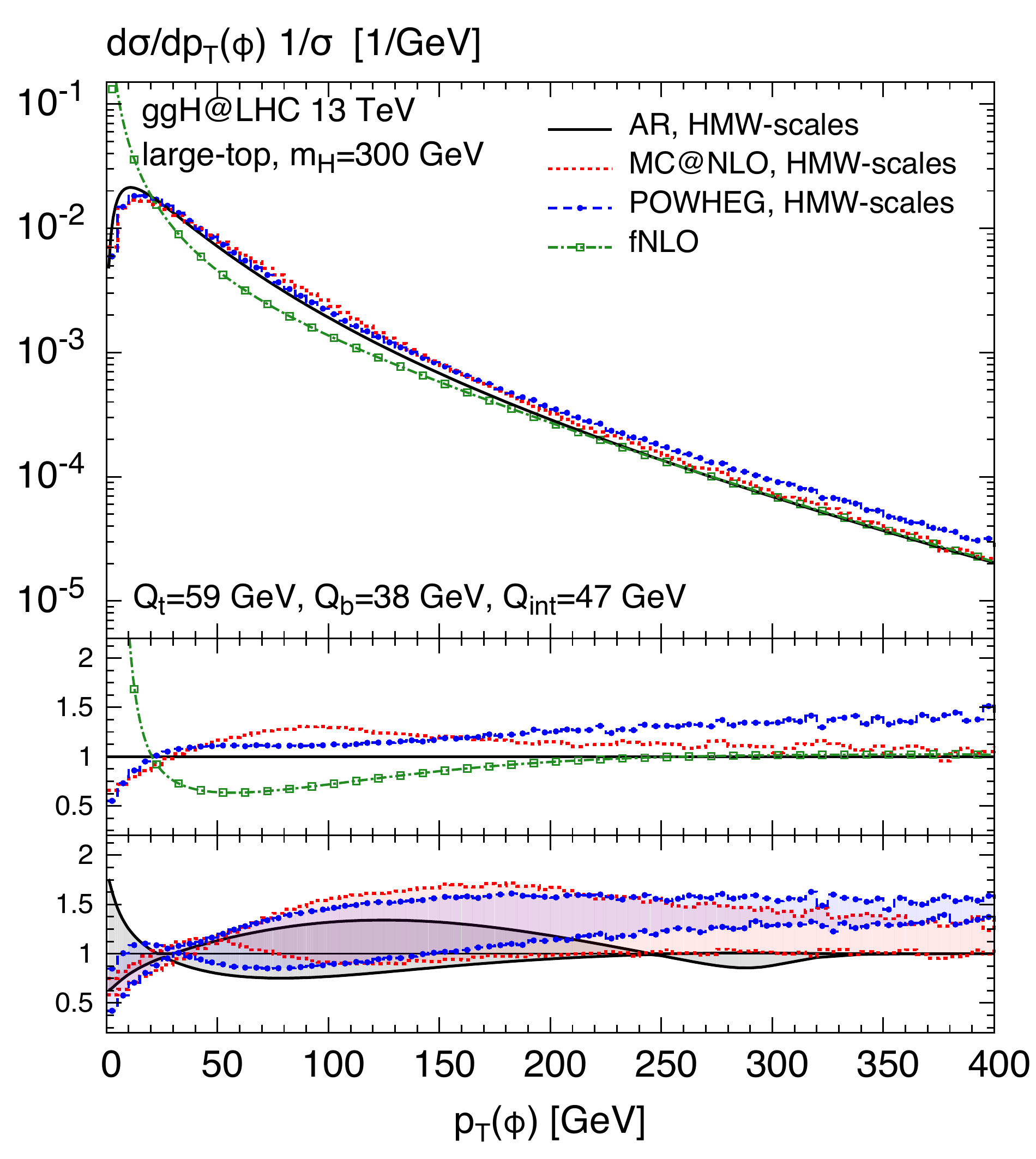}\\
\hspace{-0.25cm}
\includegraphics[width=0.371\textwidth]{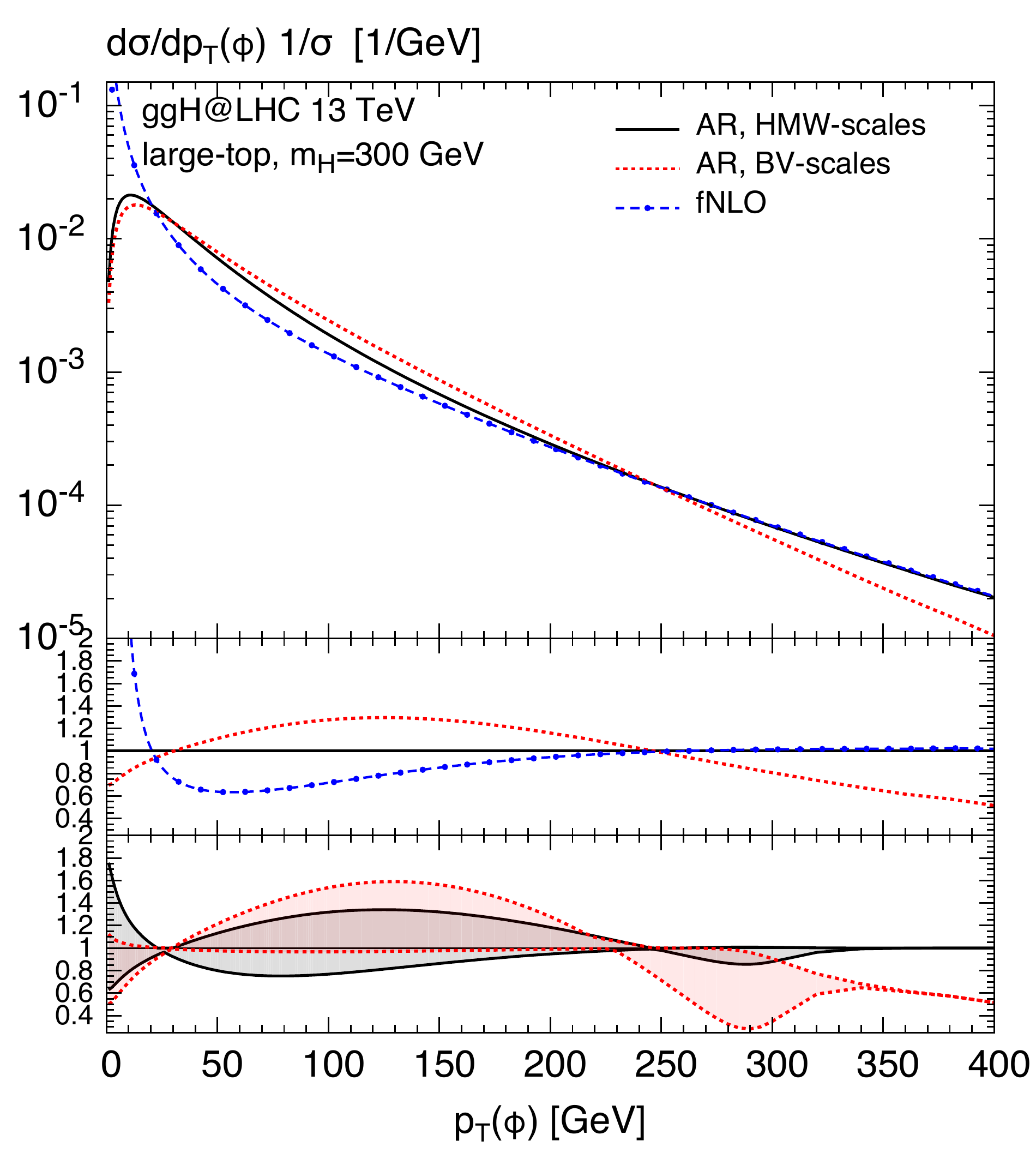}\hspace{-1.02cm}
\includegraphics[width=0.371\textwidth]{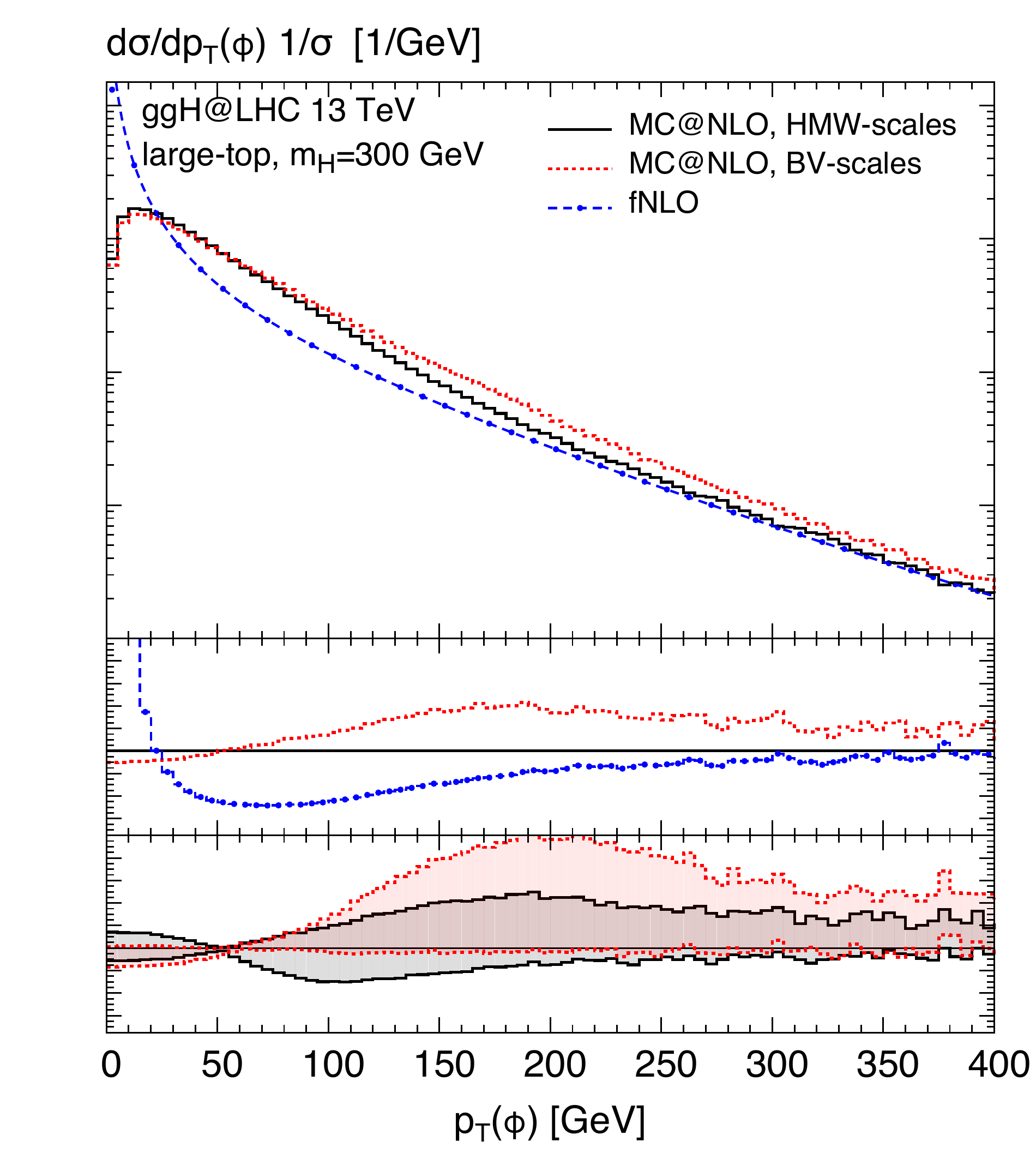}\hspace{-1.02cm}
\includegraphics[width=0.371\textwidth]{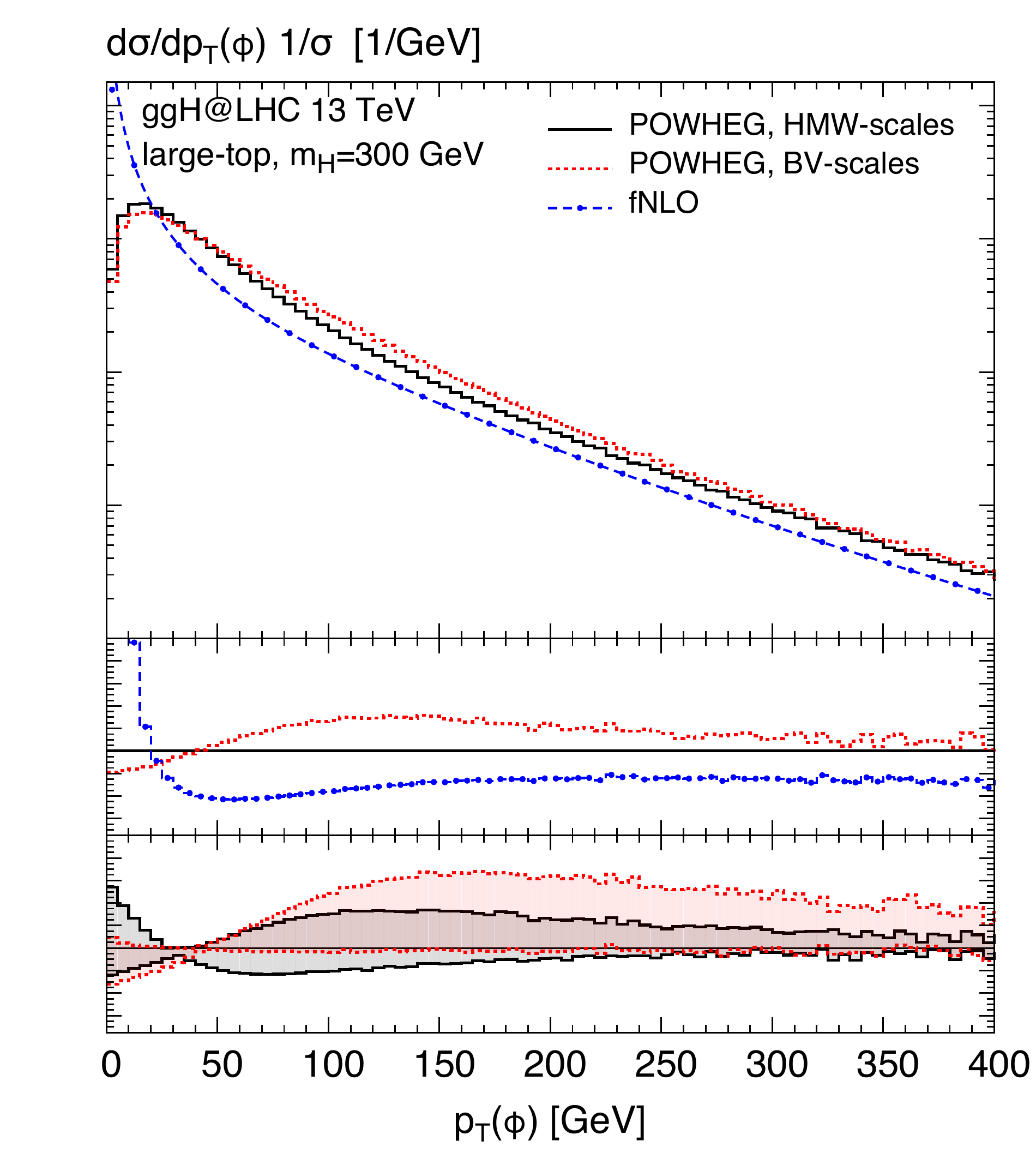}
\caption{Same as \fig{fig:results-sm}, but for a \thdm\, heavy scalar Higgs boson with $\mH=300$\,GeV in the top dominated scenario.
\label{fig:results-larget-hH}
}
\end{figure}

Let us now consider the production of a heavy Higgs with $\mH=300$\,GeV
in the \largetop{} scenario, where, similar to the \sm{}, the
bottom-loop and the interference contribution play only a minor role. In
contrast to the \sm{}, however, the matching scales for the top-loop
contribution derived using \bv{} and \hmw{} differ by almost a factor of
two. The results for the $\pth$ distribution are shown in
\fig{fig:results-larget-hH}.

Using \bv{} scales, one notices that \AR{} deviates quite significantly
($\sim 50$\%) already at $\pt{}= 400$\,GeV from the \fnlo{} result; this
deviation tends to further increase towards larger $\pt$ values. This is
not unexpected since the \hmw{} scales are designed to guarantee
similarity between the resummed and the \fnlo{} curve at large $\pt$.
Scale choices larger than the values determined by \hmw{} will therefore
necessarily lead to a deviation from the \fnlo{} in that region.  In the
case of \powheg{} and \mcnlo{}, their canonical high-$\pt$ behavior
starts to appear only for relatively large values of $\pth$, the reason
for this being again the relatively high value of the scale for the top
contribution ($w_t = 111$\,GeV).  Indeed, the agreement between the two
Monte Carlos turns out to be excellent, at least up to \pt{} values as
large as the Higgs mass.  Despite the large deviations of \AR{} in the
tail and the much softer \AR{} spectrum, all approaches are compatible
within uncertainties at small to intermediate transverse momenta
($\pt{}\lesssim 200$\,GeV).  It should be noted that this is partly due
to the fact that the uncertainty bands are significantly larger (almost
by a factor of two) than in the \sm{}.

Using \hmw{} scales, the transition to the high-$\pth$ region is more
similar to the \sm{} case (including the consistent overshooting of the
\powheg{} spectrum) because of the relatively flat dependence on the Higgs mass of the \hmw{} scales,
implying $Q_t^{\text{large-t}}/Q_t^\text{\sm} \simeq 1.3$.
The results from the three codes appear to be more compatible, in particular in the tail of the distributions, mainly due to the different \AR\ behavior.
The bulges of the uncertainty bands above the peak position extend to considerably larger values of $\pt$, and
their width is similar to what we observed for the \sm{}.

Despite the apparent differences between the left and the right upper
plots, the dedicated analysis of the impact of the scales choice in the
lower plots reveals that the results are nicely compatible within the
respective uncertainty bands. The only exception from this occurs for
\AR{} at large $\pt$. But one should bear in mind that the uncertainty
band for \AR{} is manually suppressed at large $\pt$, a procedure that
should strictly be applied only when \hmw{} scales are used.  The
observation that the width of the \bv{}-bands is larger than for \hmw{} is due
to the fact that $\wrest$ is more than twice as large than $\qrest$, and
thus the variation by a factor of two has a bigger impact on the final
result.

The results for the \cp{}-odd Higgs boson in the \largetop{}
scenario are very similar to the ones of the heavy \cp{}-even
Higgs shown here. We therefore refrain from showing them here.

\clearpage
\newpage

\subsubsection{Results in the \largebot{} scenario}
\label{sec:largeb}

\begin{figure}
\centering
\hspace{-0.2cm}\includegraphics[width=0.48\textwidth]{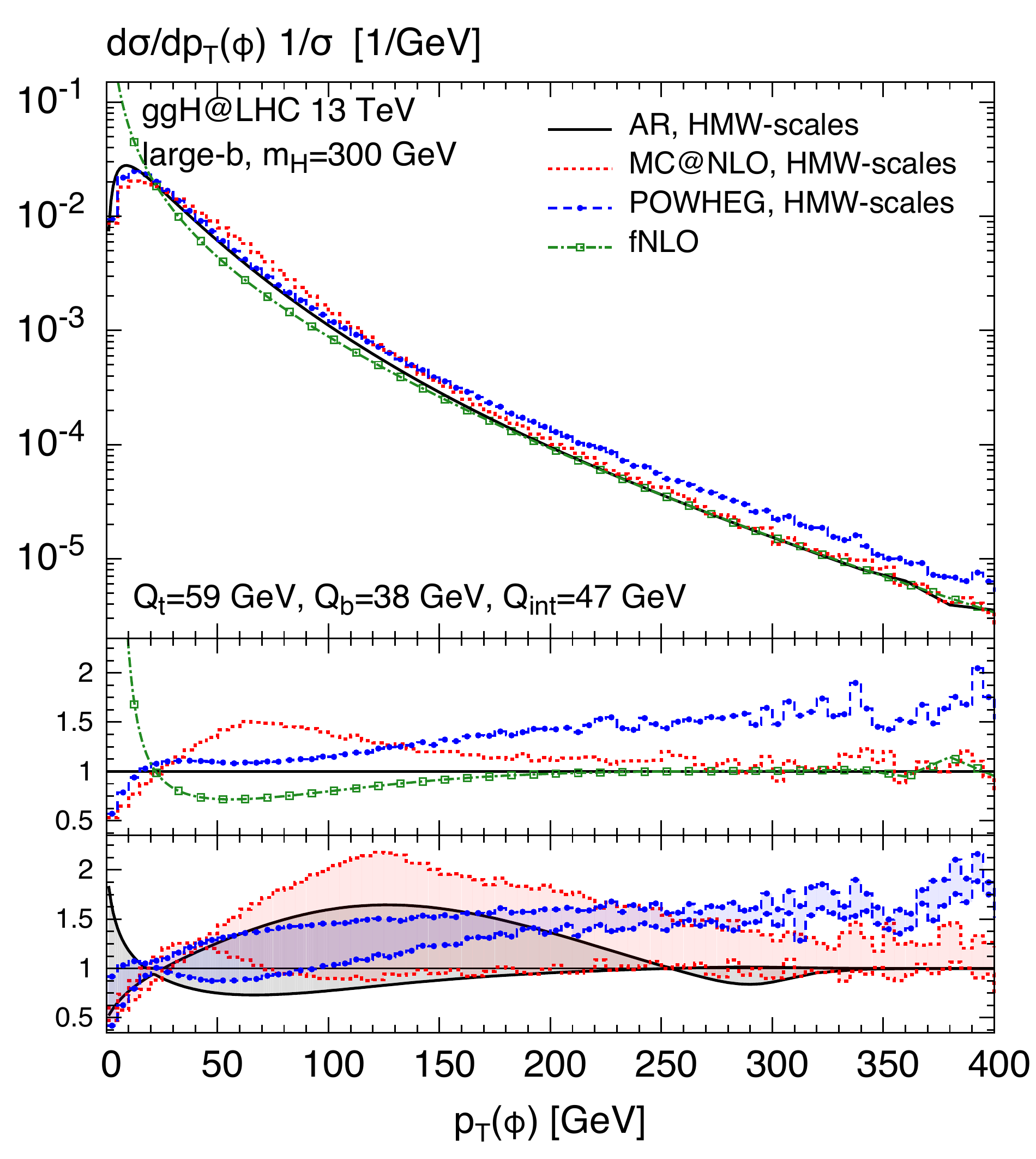}
\includegraphics[width=0.48\textwidth]{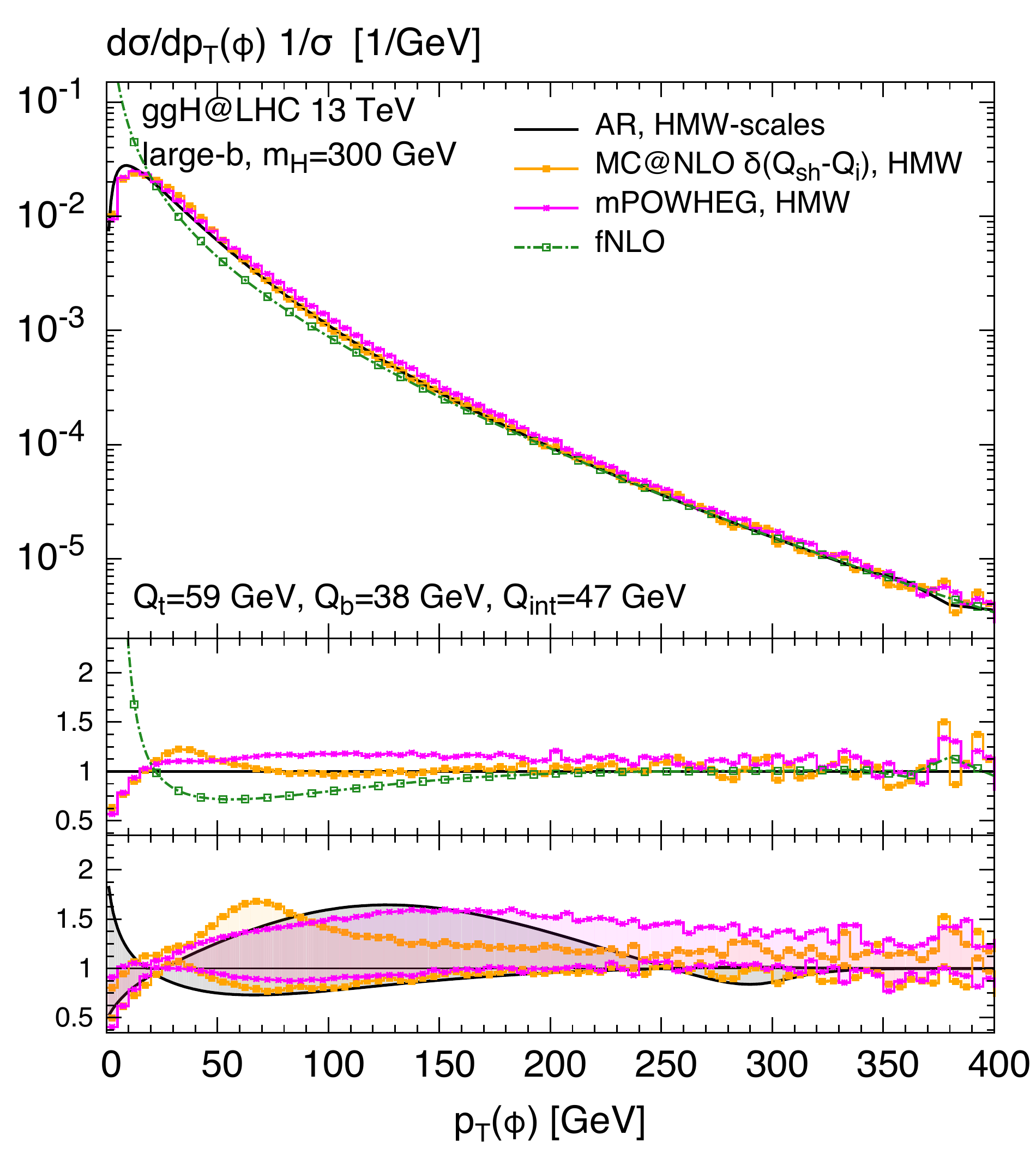}\\
\hspace{-0.2cm}\includegraphics[width=0.48\textwidth]{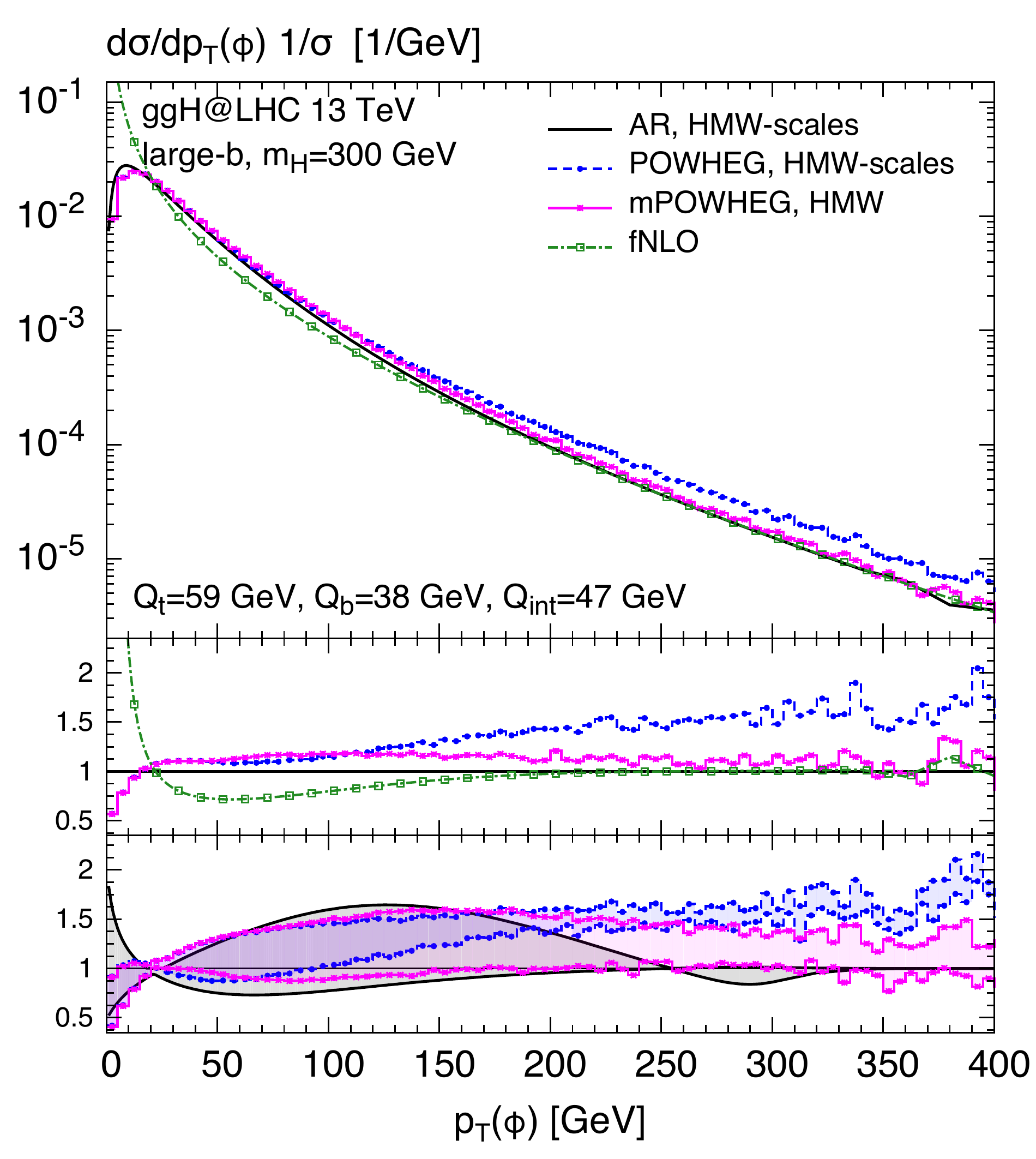}
\includegraphics[width=0.48\textwidth]{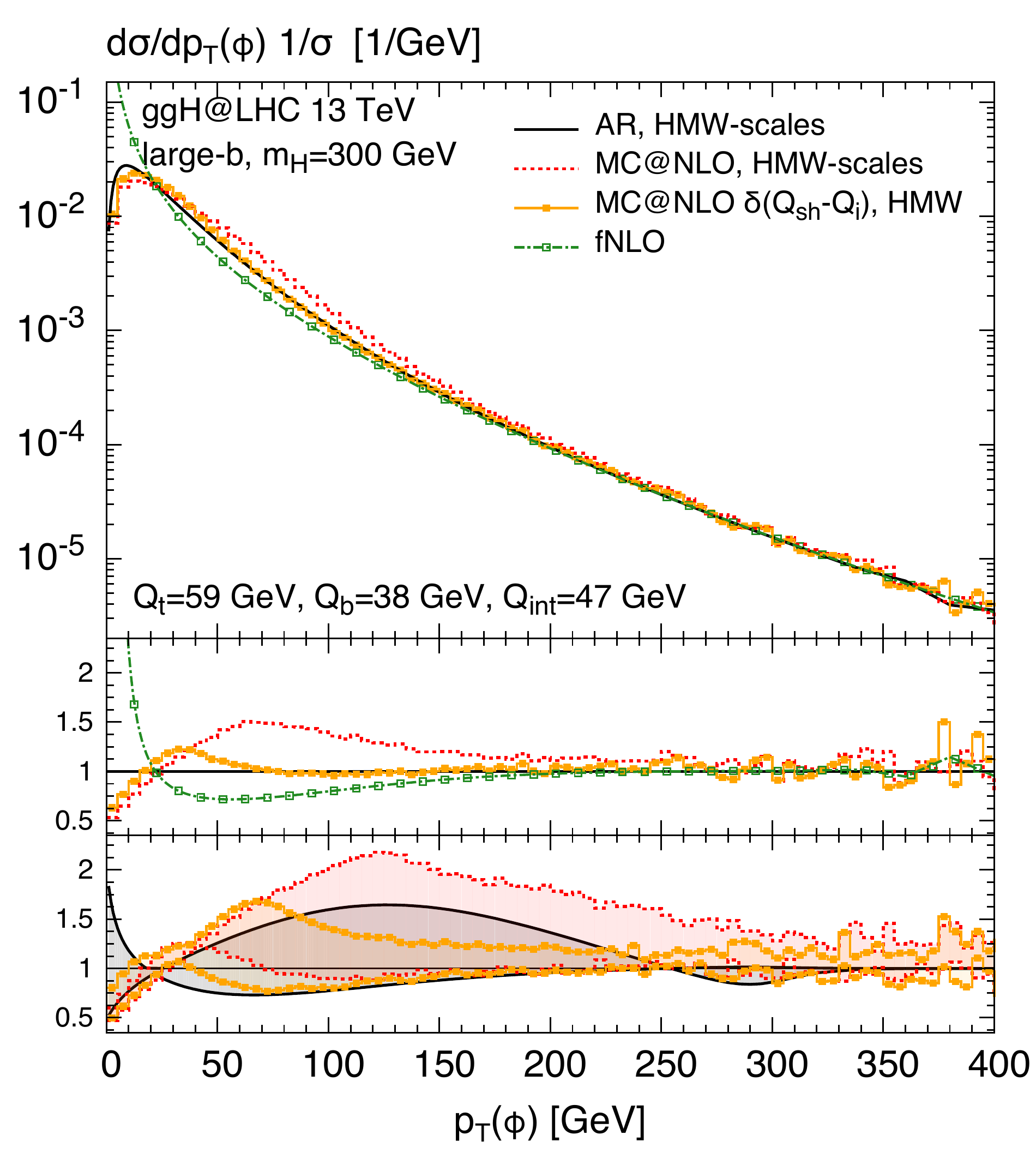}\\
\caption{The first plot is the same as the upper right plot of
  \fig{fig:results-sm}, but for a \thdm\, heavy scalar Higgs boson with
  $\mH=300$\,GeV in the bottom dominated scenario. The other plots show
  two additional curves: \mcnlo{} applying a fixed value
  ($\delta$-distribution) to the shower scale of each contribution using
  the \hmw{} values (orange, solid with full boxes); a
  modified-\powheg{} (m\powheg{}) approach requiring emissions in all
  remnant events to be bounded by the matching scales (\hmw{} in this
  case) from above (magenta, solid with stars).
\label{fig:results-scenB-hH}
}
\end{figure}

The \largebot{} scenario produces large bottom-Yukawa couplings for the
heavy and the pseudo-scalar Higgs boson which renders the bottom-loop
induced contribution by far dominant. Since the associated matching
scales $\wresb$ and $\qresb$ are very close to each other, any
difference in the $\pt$ distributions are due to the conceptional
variants of the matching in the three codes under
consideration. \fig{fig:results-scenB-hH} shows the results for a heavy
Higgs boson using \hmw{} scales (the corresponding plots for \bv{}
scales are identical for all practical purposes).

Let us first discuss the upper left plot of \fig{fig:results-scenB-hH},
where the curves are displayed in the same way as in the corresponding
plots of \fig{fig:results-sm} and \fig{fig:results-larget-hH} for the
\sm{} and the \largetop{} scenario, respectively.  While the large-$\pt$
behavior for \AR{} and \mcnlo{} is similar to the \larget{} scenario,
\powheg{} produces a spectrum that is significantly harder, exceeding
the \fnlo{} result by about 50\% for $\pt\gtrsim 200$\,GeV.  Apparently,
the specific matching procedure of \powheg{} has a significant impact on
the large \pt{} region, where the parton shower, based on the
soft/collinear approximation, is outside its region of validity.
Between $10\lesssim\pt/\text{GeV}\lesssim 130$, on the other hand,
\powheg{} and \AR{} agree within 10\%, while \mcnlo{} is significantly
higher in that region. The central predictions of the two Monte Carlo
results are in reasonable agreement only at small transverse momenta
($\pt\lesssim 30$\,GeV).  Moreover, the size of the error bands is very
different in the two Monte Carlo approaches: the \mcnlo{} band blows up
to $\order{100\%}$ around $\pt{}\sim 125$\,GeV; the \powheg{} band
remains very narrow over the whole range, indicating an uncertainty even
smaller than in the \sm{}.  The fact that all approaches lead to
compatible predictions below $\pt\approx 200$\,GeV is mainly due to the
large \mcnlo{} band.

Since the \largebot{} scenario reveals the differences between the three
codes under study in the most striking way, it will be instructive to
investigate them in more detail within this scenario.  Consider first
the \powheg{} approach. As it turns out, both observed
features---enhanced high-\pt{} tail and small uncertainty band---can be
tackled by the same modification of the matching procedure: In the
original \powheg{} approach, the scale $t_1$ (see \sct{sec:mc}) for each
event is identified with the transverse momentum of the first emission.
If the latter is very large, the shower will act up to scales which are
way beyond the validity range of the underlying approximations.  If
instead we restrict $t_1$ {\it for all remnant events} (the $R^f$-term
in \eqn{eq:codes:matching}) to remain below the matching scale
(e.g.\ \bv{} or \hmw{}), one obtains the result shown in the lower left
plot of \fig{fig:results-scenB-hH} (magenta, solid curve with
stars). Since the high-$\pth$ tail of the distribution is driven by the
remnant events, the above restriction of $t_1$ ensures a transition to
the \fnlo{} curve. We will refer to this modified approach as
``m\powheg{}'' in what follows.\footnote{Modifications of the shower
  scale setting were studied also in \citere{Nason:2013uba} in order to
  improve the theoretical prediction for dijet production.}  It happens
that also the uncertainty band is more similar to the other approaches
for the m\powheg{} result (see again the lower left plot of
\fig{fig:results-scenB-hH}).

The fact that the m\powheg{} modification applies only to the remnant
events ensures that the formal accuracy of the original \powheg{}
approach remains unaffected. In this respect, it is reassuring that at
low transverse momenta ($\pt\lesssim 100$\,GeV), the modifications have
practically no effect (compare the \powheg{} with the m\powheg{} curve
in \fig{fig:results-scenB-hH}), as this region is controlled by the soft
events (the first term in \eqn{eq:codes:matching}), which remain
unchanged.  We checked that the \fnlo{}-like high-$\pt$ behavior of
m\powheg{} is a generic feature and not specific to the underlying
phenomenological scenario or the details of the parton shower. A more
quantitative study of its numerical impact has to be deferred to a
future study though.  Stressing again that, to our understanding,
m\powheg{} is only a logarithmically subleading modification of
\powheg{} (or rather its interface with the parton shower), one may
consider it as a viable alternative whenever an \fnlo-like high-$\pt$
behavior is required.

Let us now discuss the \mcnlo{} approach in this scenario, featuring a
peculiarly large uncertainty regarding shower scale variations, which in
turn leads to very different shapes of both the central predictions and
the error band with respect to the other two approaches. Recall that in
the \mcnlo{} implementation of {\tt MadGraph5\_aMC@NLO}, the values for
the shower scale of the soft events follow a specific distribution with
a peak at the matching scale $\mu_i$ ($i=t,b,\text{int}$; e.g.\ \bv{} or
\hmw{}), see \sct{sec:mcnlo}.  We find that, in the \largebot{}
scenario, a restriction of the range of that distribution has a
significant effect on the central \mcnlo{} prediction, which is not
surprising given the large associated uncertainty. In the limit where
this distribution turns into a $\delta$-function $ \delta(\qsh{}-\mu_i)$,
also the size of the uncertainty band is strongly reduced, as can be
seen in the lower right plot of \fig{fig:results-scenB-hH} (orange,
solid curve with full boxes). We checked that, besides a slightly
earlier matching to the \fnlo{} prediction controlled by the remnant
events, the observed features are due to the restricted shower scale
range accessible to the soft events. This study simply shows that the
predictions in bottom-quark dominated scenarios depend strongly on
details of the matching procedure and thus should be attributed with
large uncertainties.

Finally, the upper right plot of \fig{fig:results-scenB-hH} shows the
comparison of the results obtained with the modifications done for
\mcnlo{} and \powheg{}.  Indeed, the modified predictions turn out to be
in much better agreement, in terms of both the central curves and the
uncertainty bands.

Similar to the \largetop{} scenario, the distributions for the
pseudo-scalar Higgs in the \largebot{} scenario largely resemble the
ones of the heavy Higgs shown in \fig{fig:results-scenB-hH}, and we do
not need to discuss them separately at this point.


\subsubsection{Results in the \largeint{} scenario}

The matching scales of the interference term in the \bv{} and the \hmw{}
approach exhibit quite a different behavior, as shown in
\fig{fig:scales-compare}. It will thus be interesting to see the
distributions in the various approaches in a scenario with a
particularly large interference term. Note, however, that the
interference term in the \largeint{} scenarios defined in
Table\,\ref{tab:scenarios} is always negative and competes with a
similarly large top and/or bottom contribution. Our \thdm{} parameter
scan did not reveal a truly interference-{\it dominated} scenario, where
both the top and the bottom contribution are small compared to the
interference term, and which passes the unitarity, stability, and
perturbativity checks.

\begin{figure}
\centering
\includegraphics[width=0.45\textwidth]{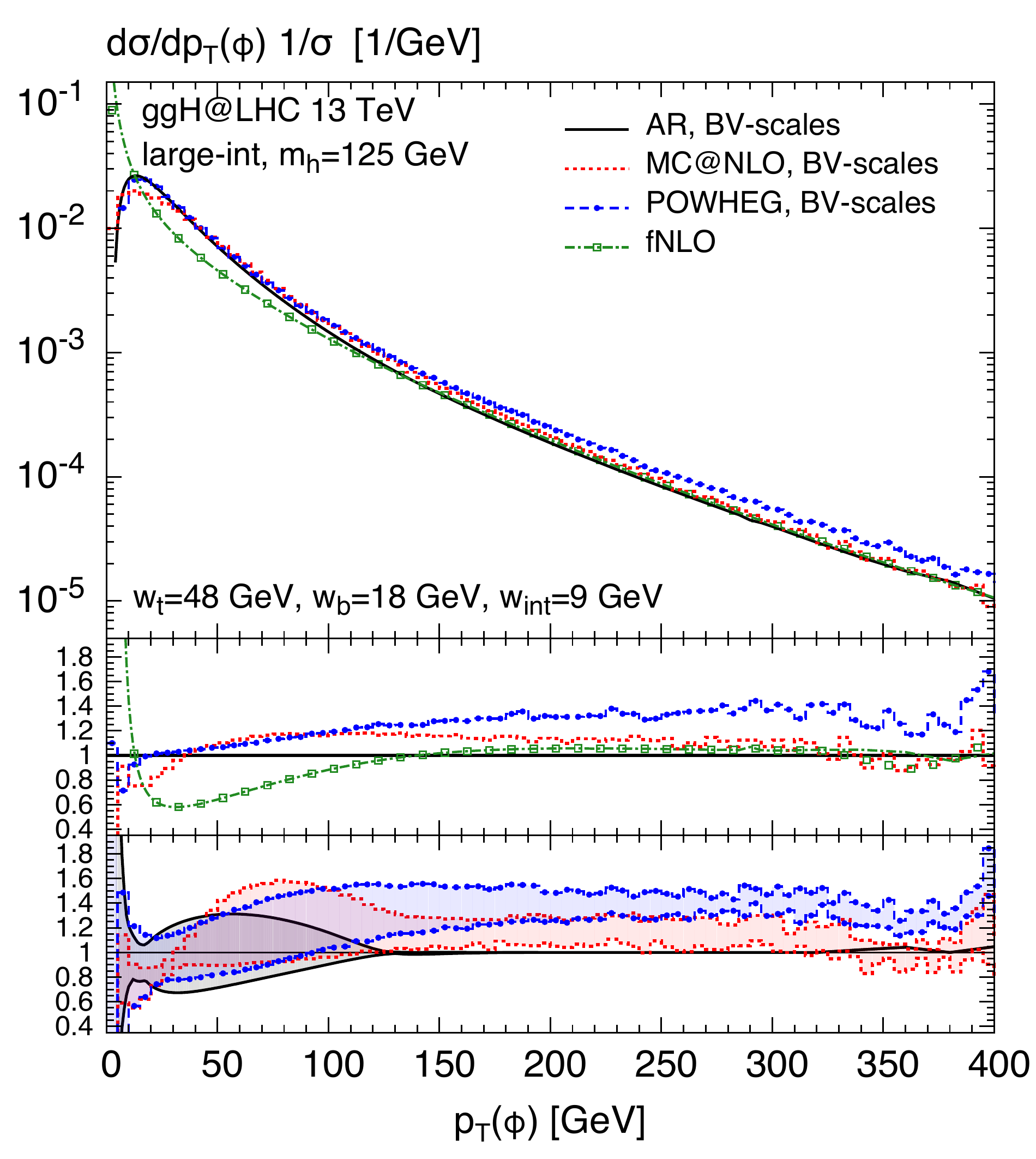}
\includegraphics[width=0.45\textwidth]{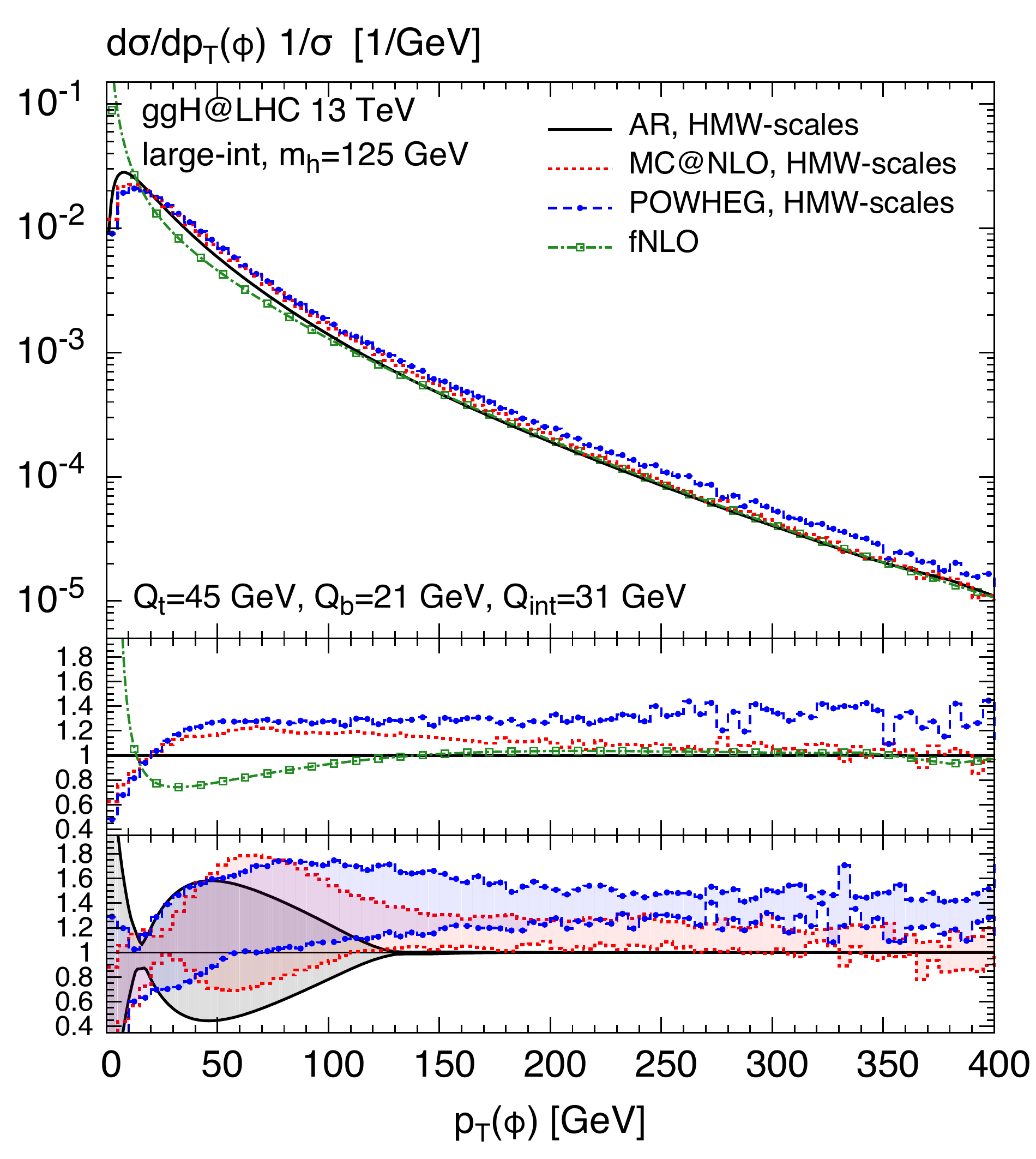}\\
\hspace{-0.25cm}
\includegraphics[width=0.371\textwidth]{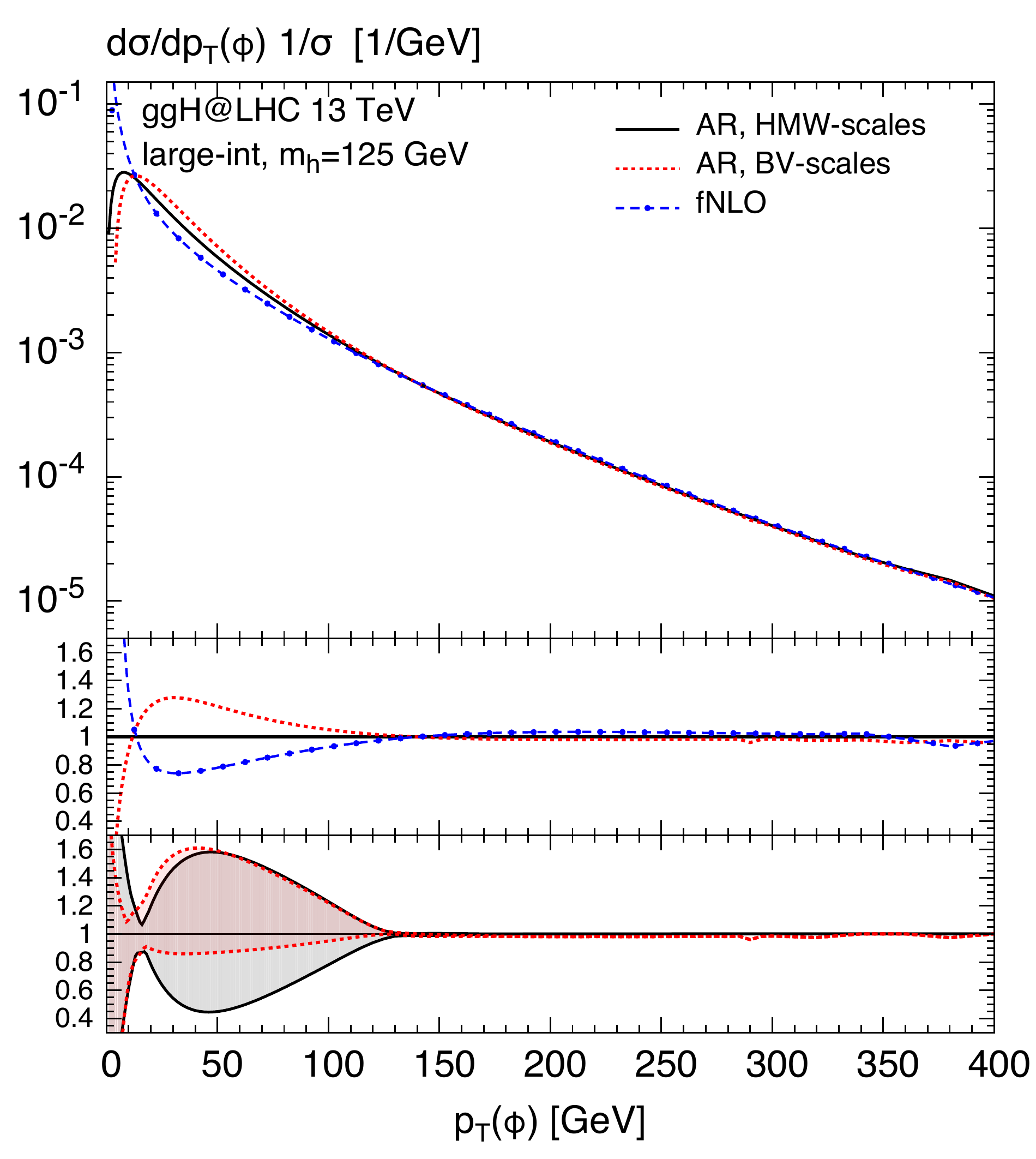}\hspace{-1.02cm}
\includegraphics[width=0.371\textwidth]{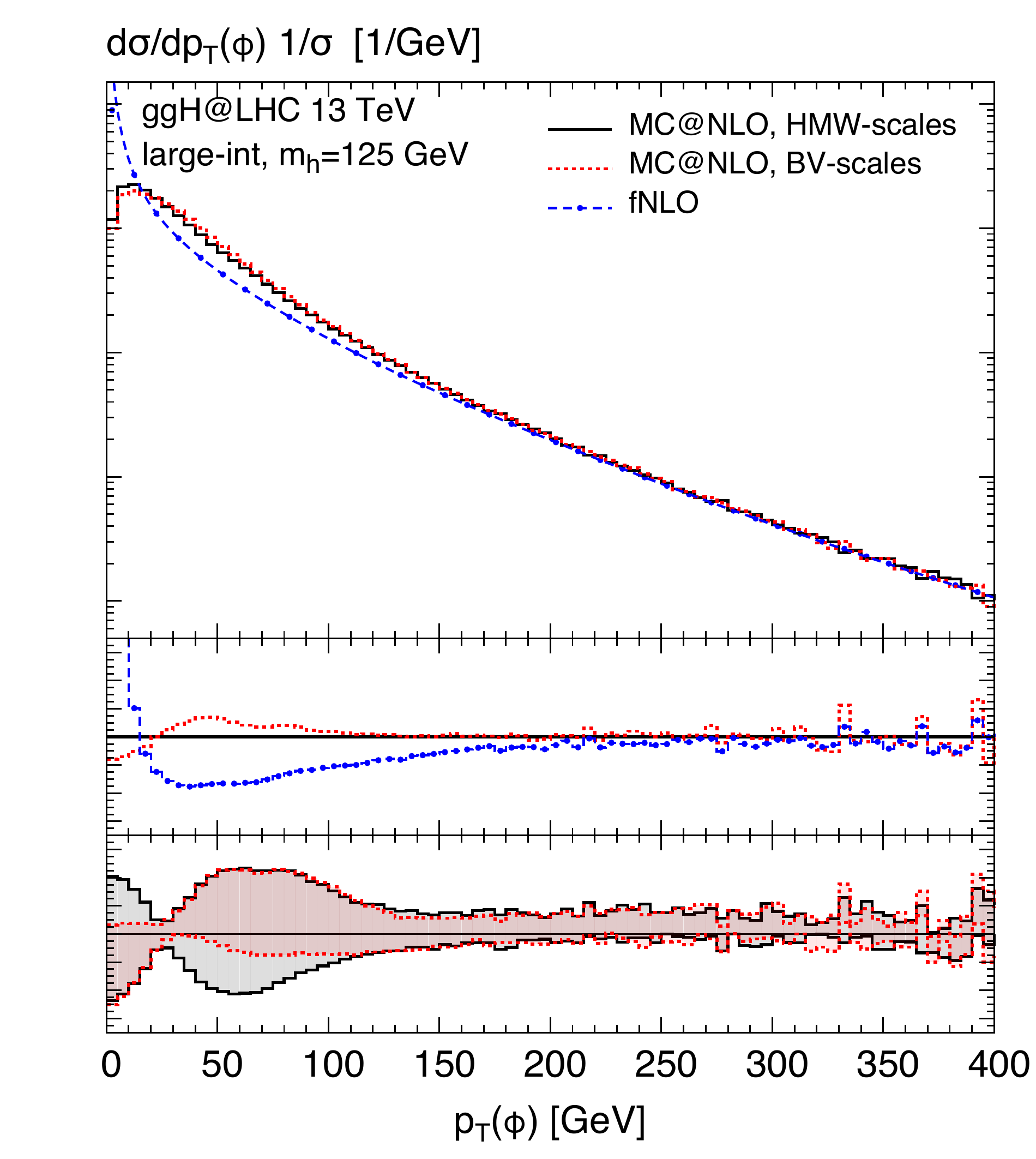}\hspace{-1.02cm}
\includegraphics[width=0.371\textwidth]{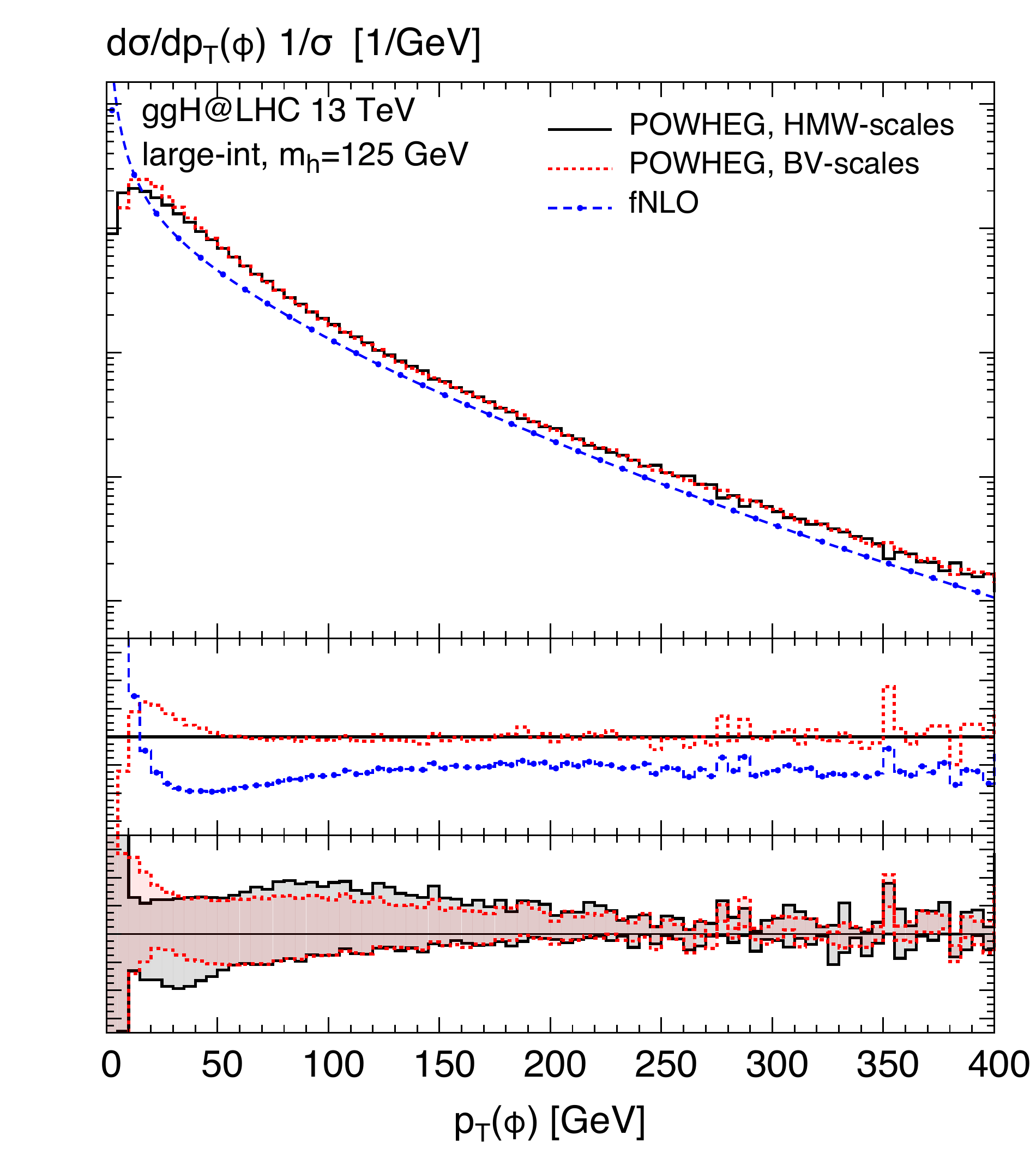}
\caption{Same as \fig{fig:results-sm}, but for a \thdm\, light scalar Higgs boson with $\mH=125$\,GeV in the large interference scenario.
\label{fig:results-largeint-lh}
}
\end{figure}

\fig{fig:results-largeint-lh} shows the results for the light Higgs
boson, where the modulus of the top, the bottom, the interference term, and
the total cross section each are roughly of the same size ($\sim 2$\,pb,
see Table\,\ref{tab:scenarios}). Note that the top and the bottom
matching scales agree well in the \bv{} and the \hmw{} approach for the light
Higgs, while the interference matching scale is about a factor of three
larger in the \hmw{} approach.

Using \hmw{} scales, the comparison of the distributions in the various
approaches leads to a picture that is quite similar to the one that we
found in the \sm{}: good agreement among the Monte Carlos, softer
spectrum of \AR{}, compatible results within uncertainties of all
approaches (at least for $\pt\le \mh$). The error bands, on the other
hand, are significantly larger than in the \sm{}. They amount up to
about\footnote{We disregard the region very close to the threshold
  ($\pt=0$), where, as stated before, the \AR{} uncertainty band blows
  up.} $\pm 60\%$ for \AR{} and \mcnlo{}, and about $\pm 40\%$ for
\powheg{}.

\begin{figure}[t]
\centering
\includegraphics[width=0.45\textwidth]{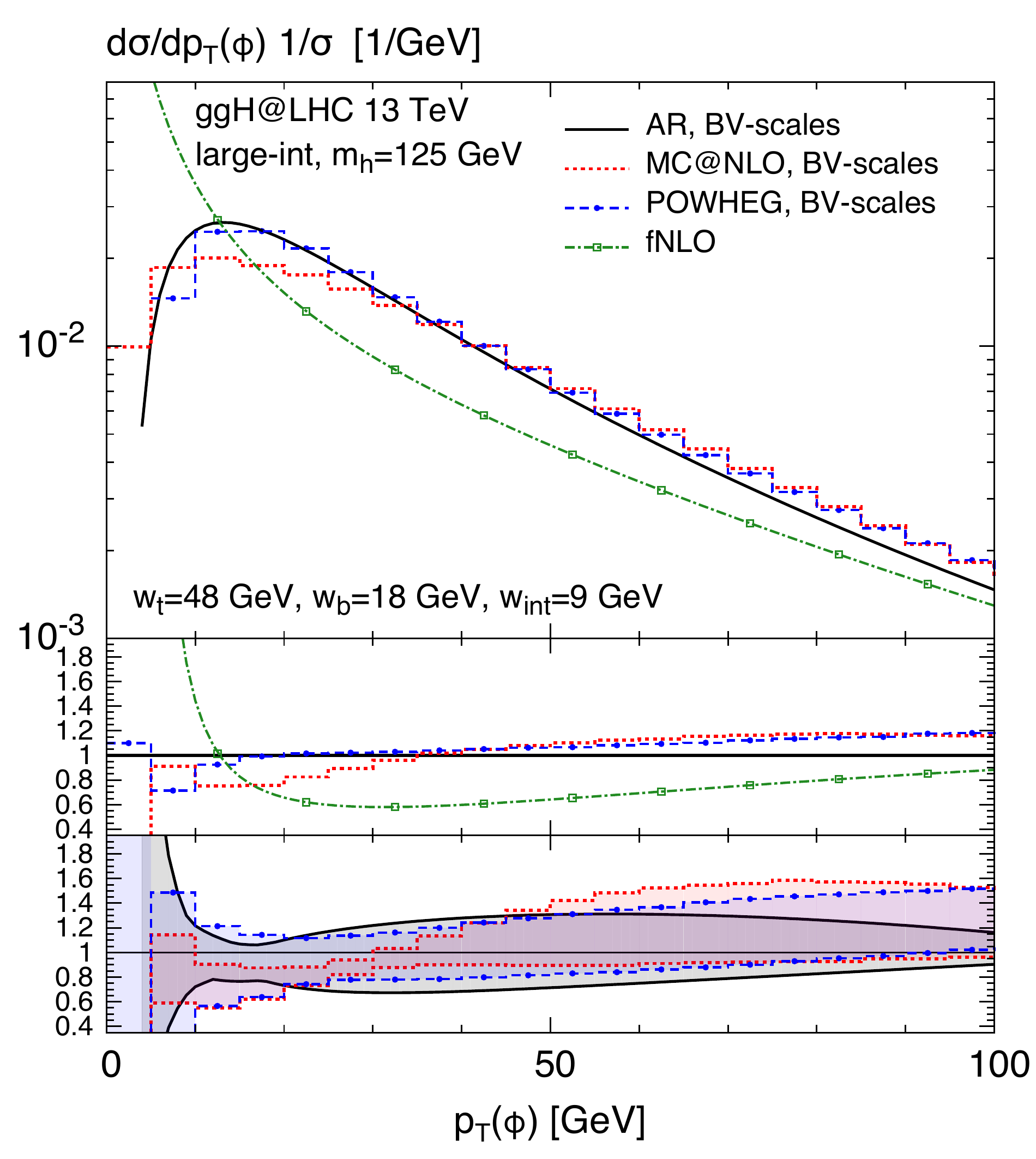}
\includegraphics[width=0.45\textwidth]{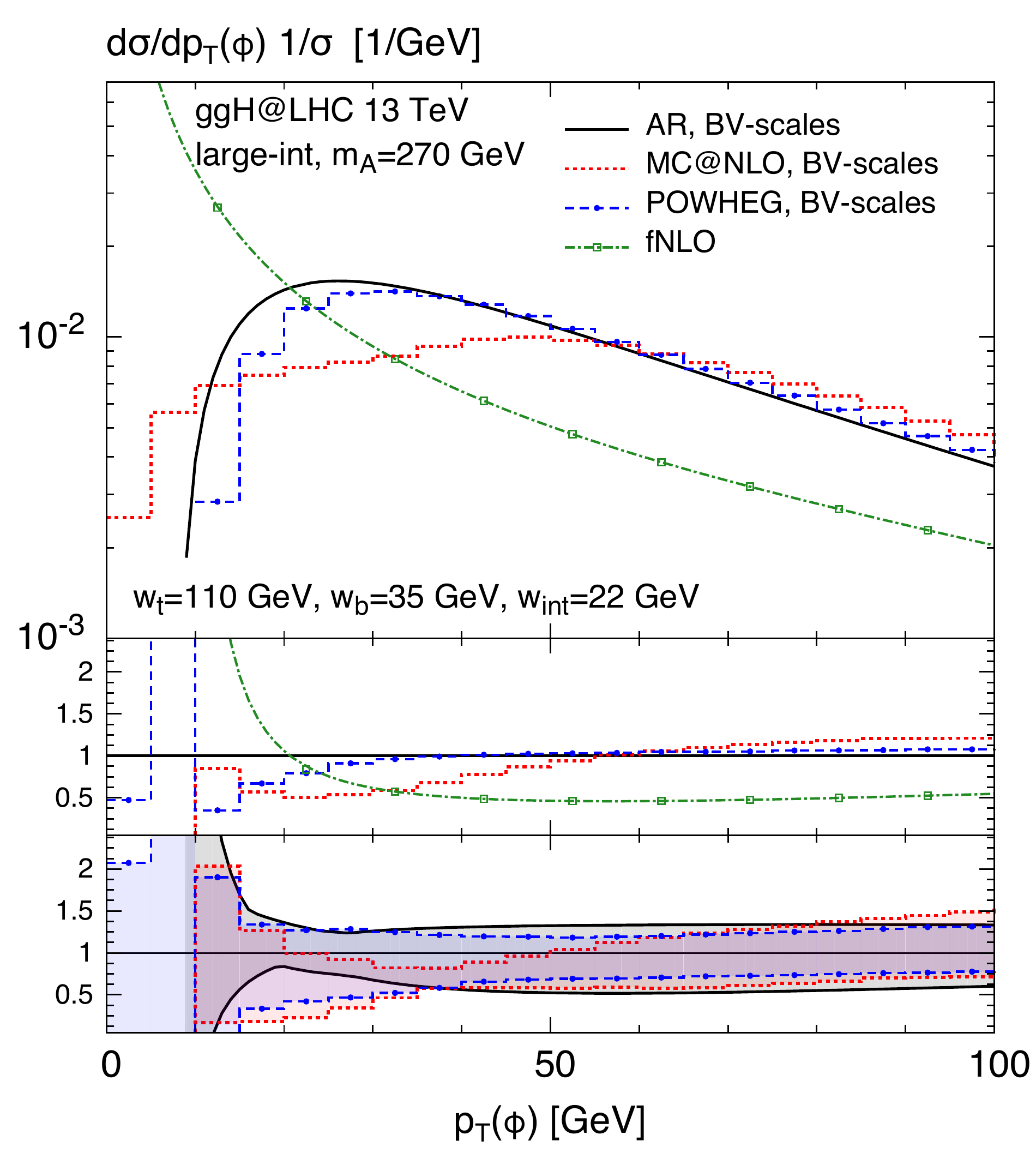}
\caption{Left: same as upper left plot of \fig{fig:results-largeint-lh},
  but with enlarged low-\pt{} region. Right: Same scenario and notation,
  but for a pseudo-scalar Higgs boson with $\ma=270$\,GeV.
\label{fig:results-largeint-bv}
}
\end{figure}


\begin{figure}
\centering
\includegraphics[width=0.45\textwidth]{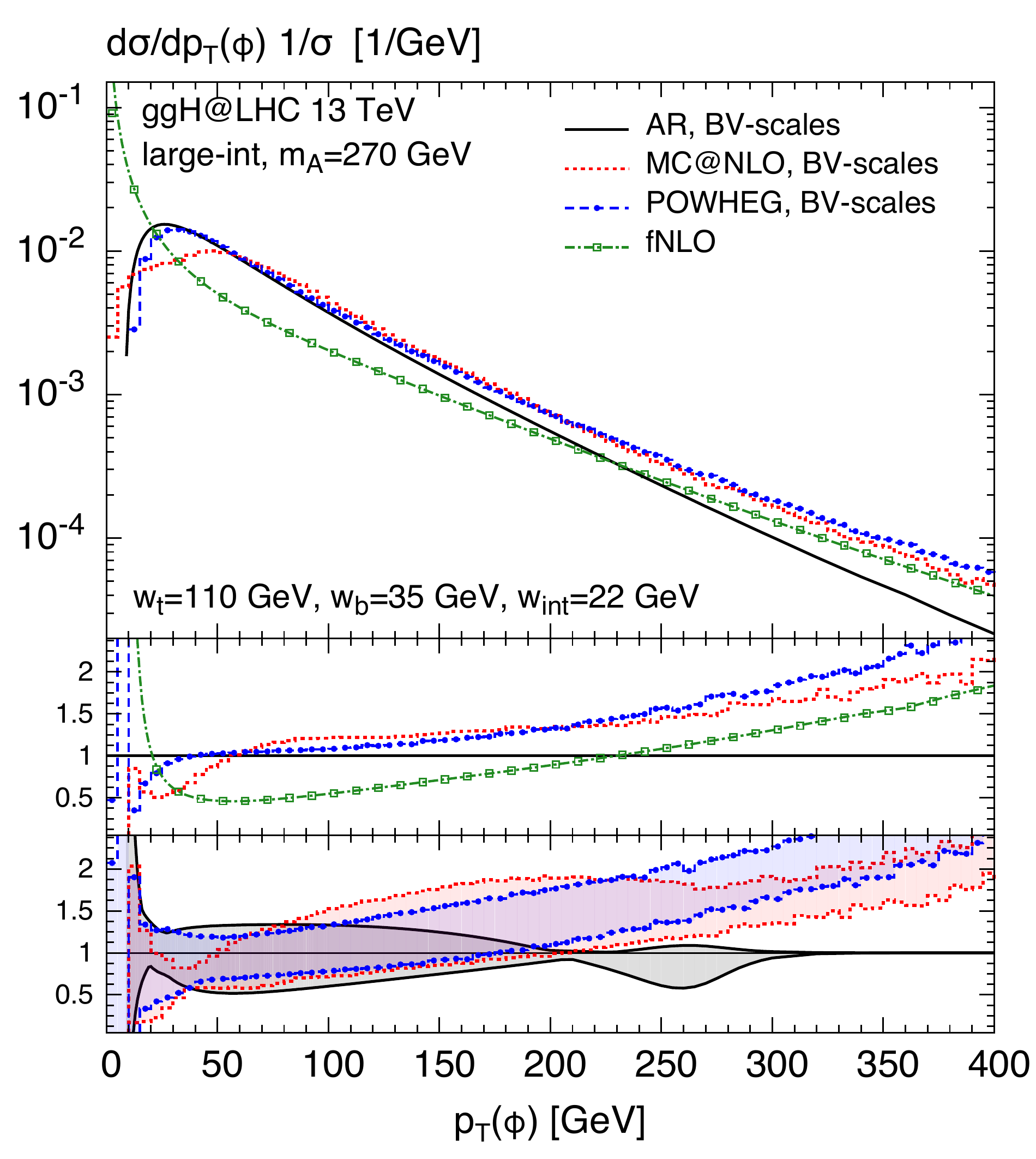}
\includegraphics[width=0.45\textwidth]{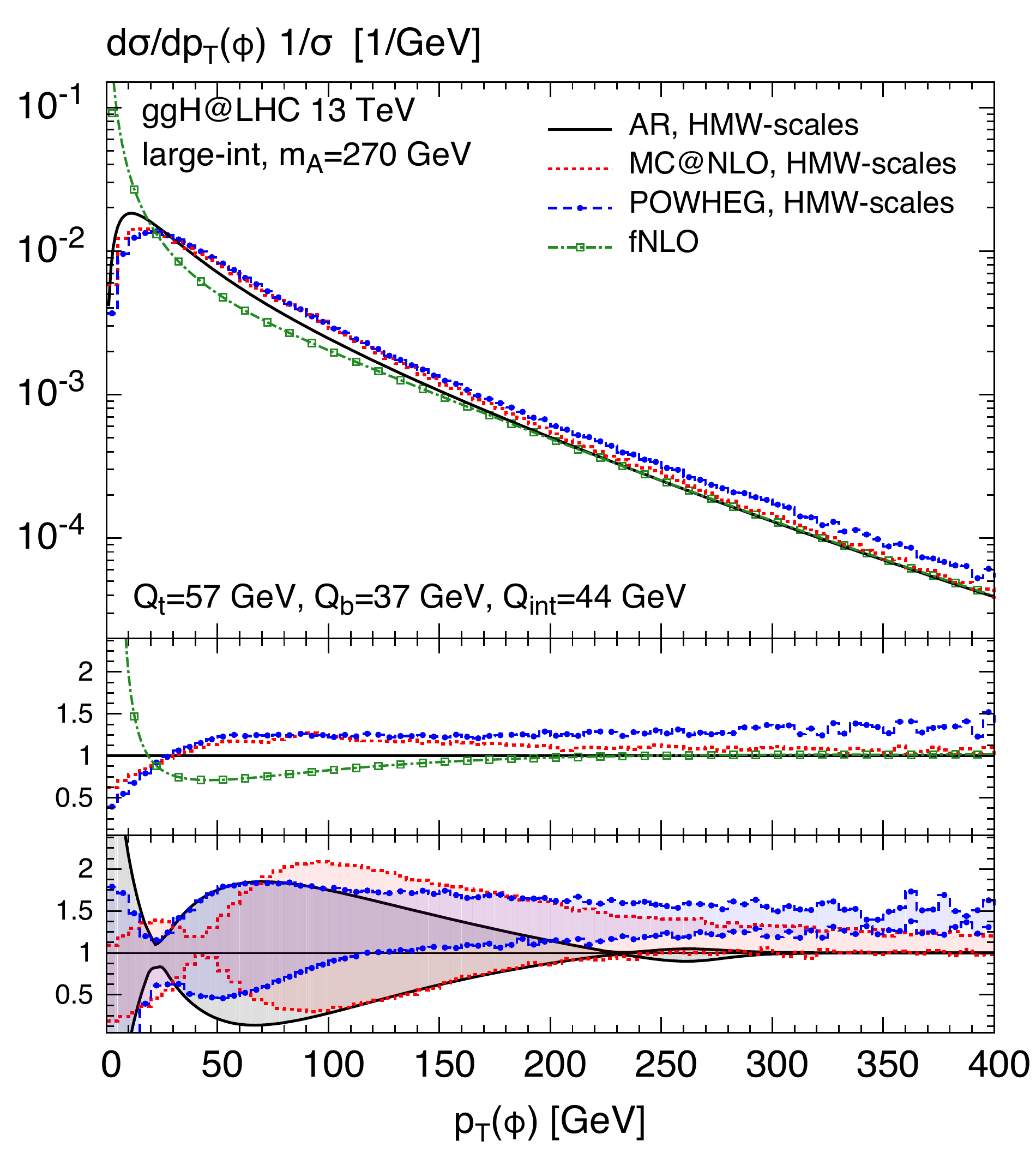}\\
\hspace{-0.25cm}
\includegraphics[width=0.371\textwidth]{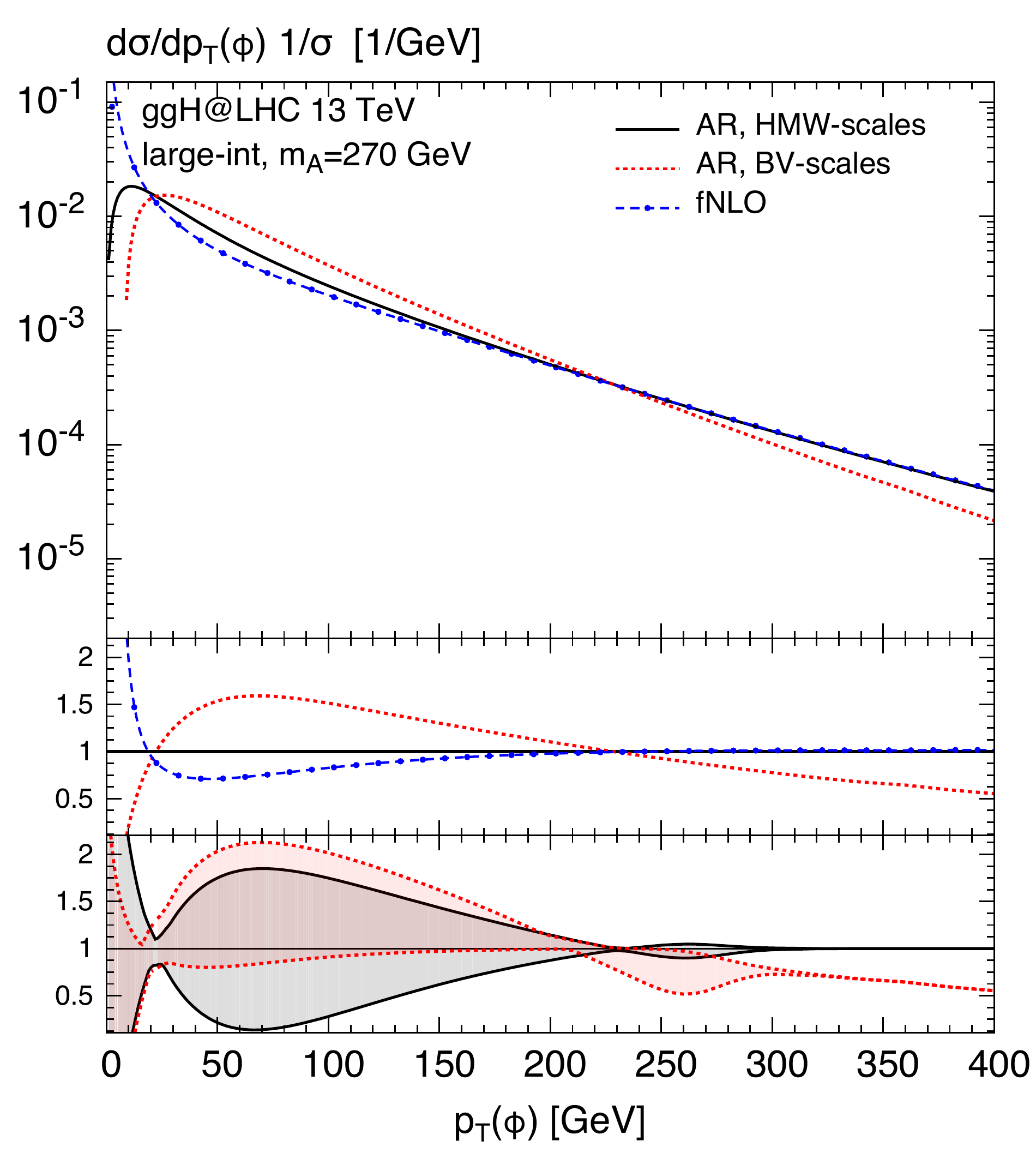}\hspace{-1.02cm}
\includegraphics[width=0.371\textwidth]{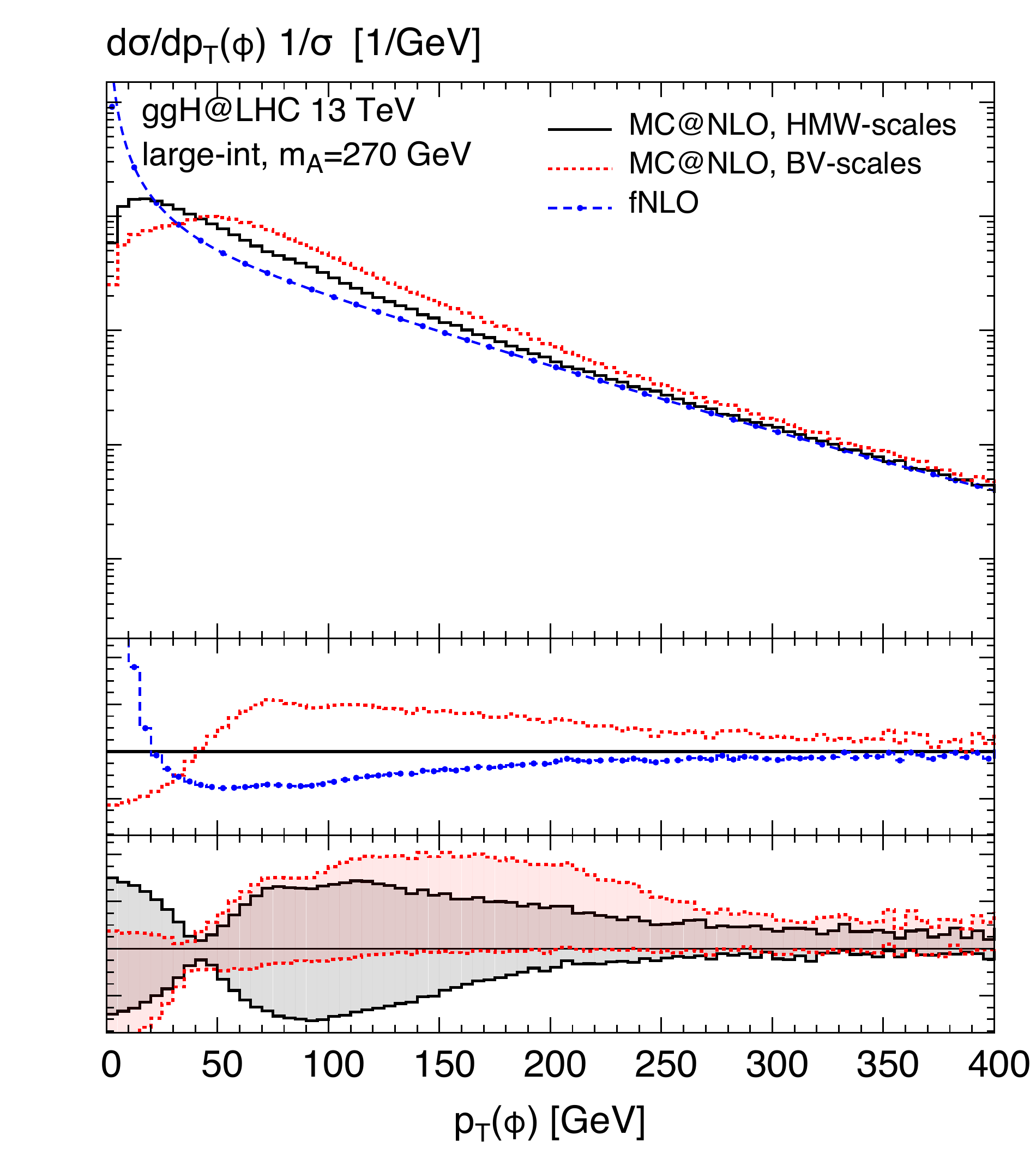}\hspace{-1.02cm}
\includegraphics[width=0.371\textwidth]{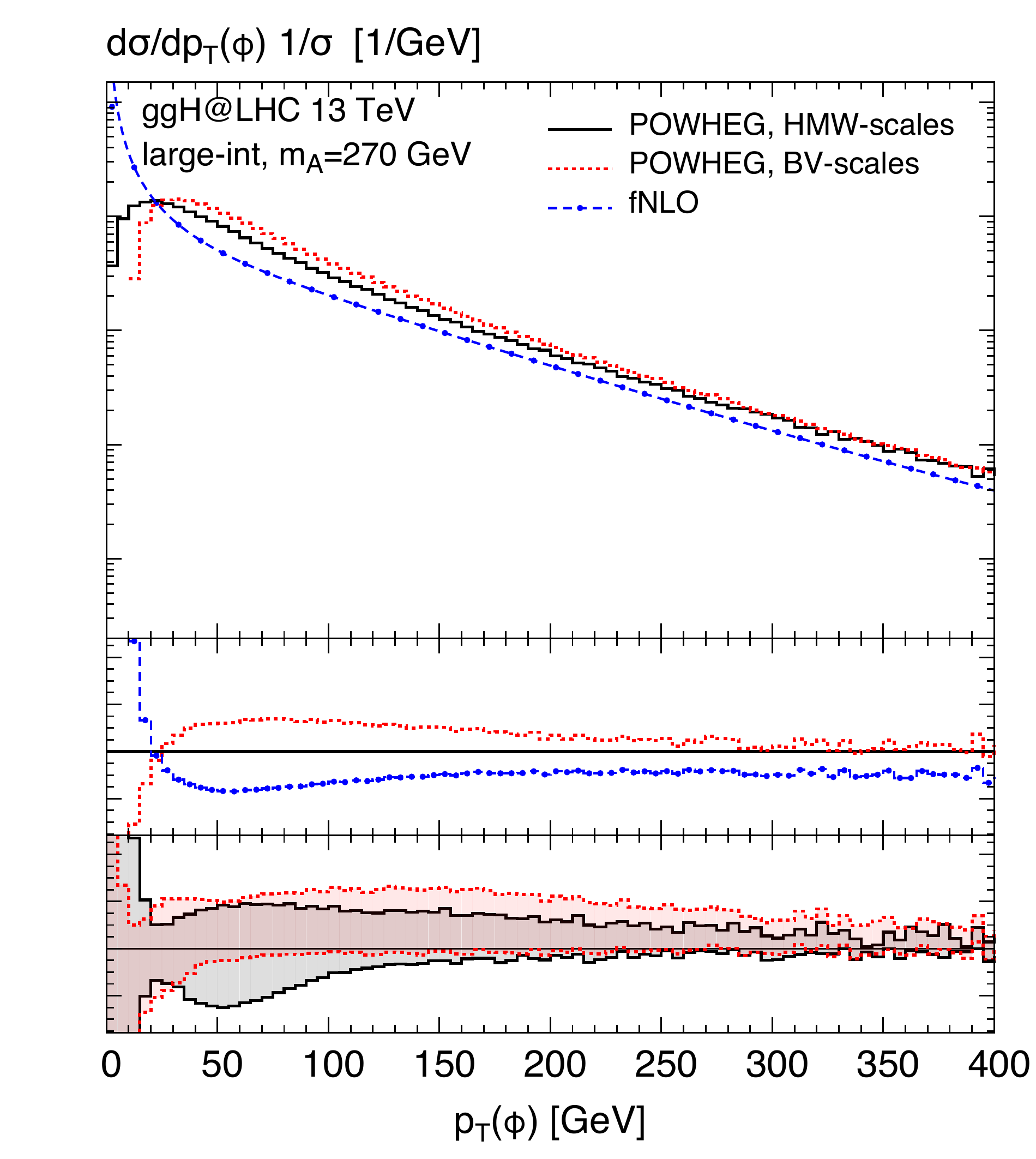}
\caption{Same as \fig{fig:results-sm}, but for a \thdm\, pseudo-scalar Higgs boson with $\ma=270$\,GeV in the large interference scenario.
\label{fig:results-largeint-A}
}
\end{figure}

Using \bv{} scales, on the other hand, the width of all uncertainty
bands is strongly reduced (up to about $\pm 20\%$ and $\pm 30\%$ for
\AR{} and the Monte Carlos, respectively), the reason being again the
fact that a large $\qresint$ (as in \hmw{}) also induces a large
interval $[\qresint/2,2\qresint]$.  While also for \bv{} the results at
small \pt{} are compatible within uncertainties and the high-\pt{}
behavior is very reminiscent of the one observed in the \sm{}, there are
some substantial differences at very small transverse momenta
($\pt{}<40$\,GeV) between the shapes predicted by the two Monte Carlos.
For better visibility, \fig{fig:results-largeint-bv} shows in the left
panel the \bv{} plot with an enlarged low-\pt{} region. Clearly, the
\mcnlo{} spectrum is harder, while the \powheg{} one actually gets
negative in the first bin.\footnote{Note also that the \AR{} curve for
  {\abbrev BV} scales turns negative at very low \pt{}.}  Since the
interference term is not positive definite, due to its definition by
subtraction, the distribution may turn negative in scenarios where its
contribution is large. Clearly, this behavior appears to be amplified at
small \pt{} if the matching scale is particularly small.  At least this
specific case, therefore, leads to a similar conclusion as pointed out
in \citere{Mantler:2015vba}, that very low scales are not well suited
for \nlo{} matched parton shower predictions.

Indeed, the behavior just observed is even enhanced in the \largeint{}
scenario of the pseudo-scalar Higgs with $\ma=270$\,GeV, see
\fig{fig:results-largeint-bv} (right panel). The \mcnlo{} distribution
displays some odd behavior between about $10<\pt{}/{\rm GeV}<60$,
developing an almost linear behavior rather than the ordinary Sudakov
shoulder. The \powheg{} prediction, on the other hand, becomes negative
in the first and the second bin of the distribution (i.e., below
$\pt=10$\,GeV). There are mainly two reasons for these features being
intensified in the \largeint{} scenario of the pseudo-scalar Higgs:
First, the ratio of the interference matching scale to the Higgs mass is
even smaller than for the light Higgs; second, the relative contribution
of the interference term is larger than in the light Higgs case.  The
latter can be inferred from Table\,\ref{tab:scenarios}: For the
pseudo-scalar Higgs, the \largeint{} scenario corresponds to an
interference term whose modulus is about 30\% larger than the top, 50\%
larger than the bottom contribution, and 270\% larger than the total
cross section.

The other results in the large-int scenario of the pseudo-scalar Higgs
are shown in \fig{fig:results-largeint-A}. The conclusions are
essentially the same as the ones for the light Higgs, which will thus
not be repeated here.  It is worth mentioning though that the relative
size of the uncertainty bands for both \bv{} and \hmw{} scales are
increased with respect to the light Higgs case, the reason being again
the larger value of the matching scale and the thereby enlarged
variation range.

\subsubsection{Results in the \lowma{} scenario}

\begin{figure}
\centering
\hspace{-0.2cm}
\includegraphics[width=0.48\textwidth]{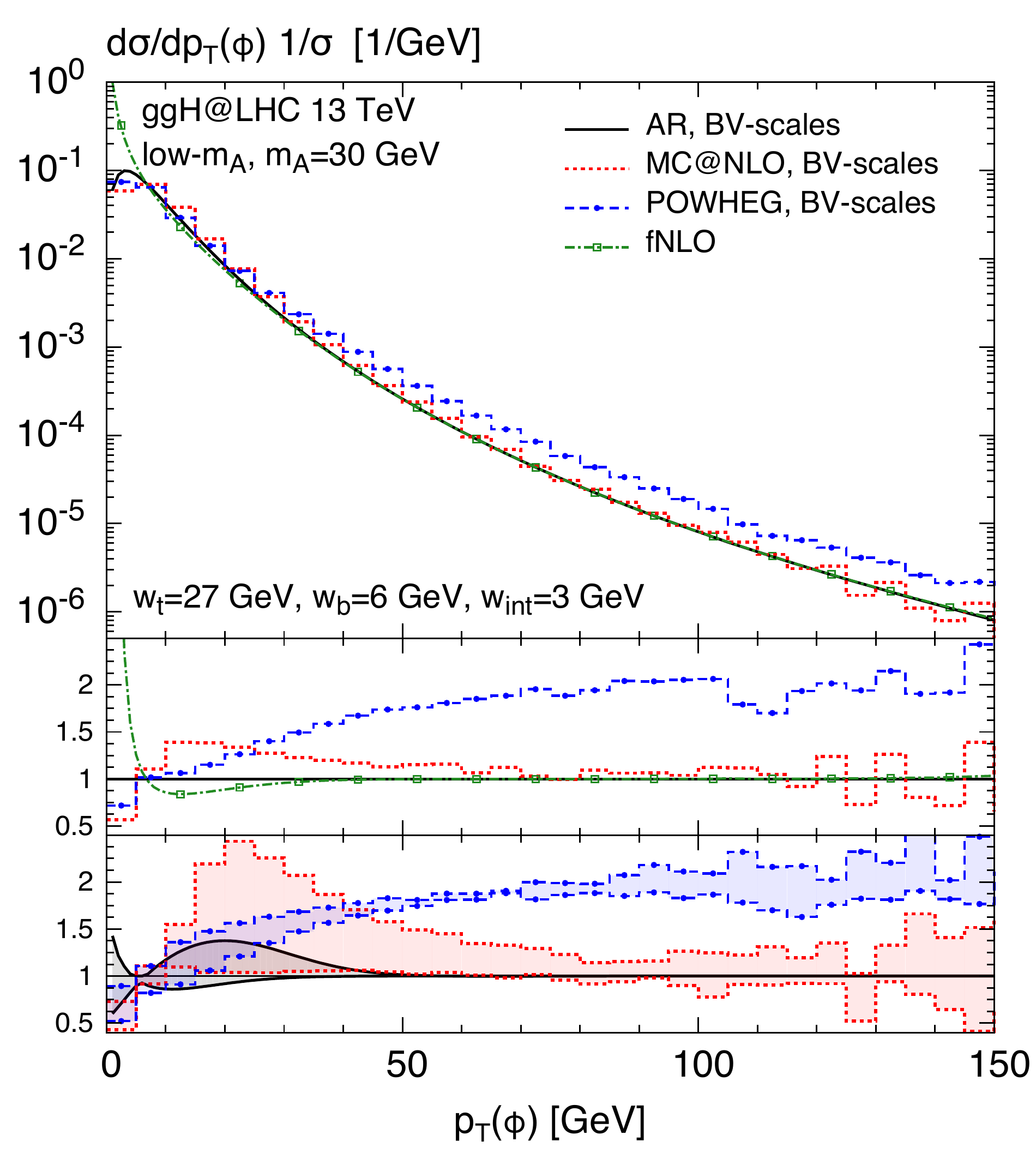}
\includegraphics[width=0.48\textwidth]{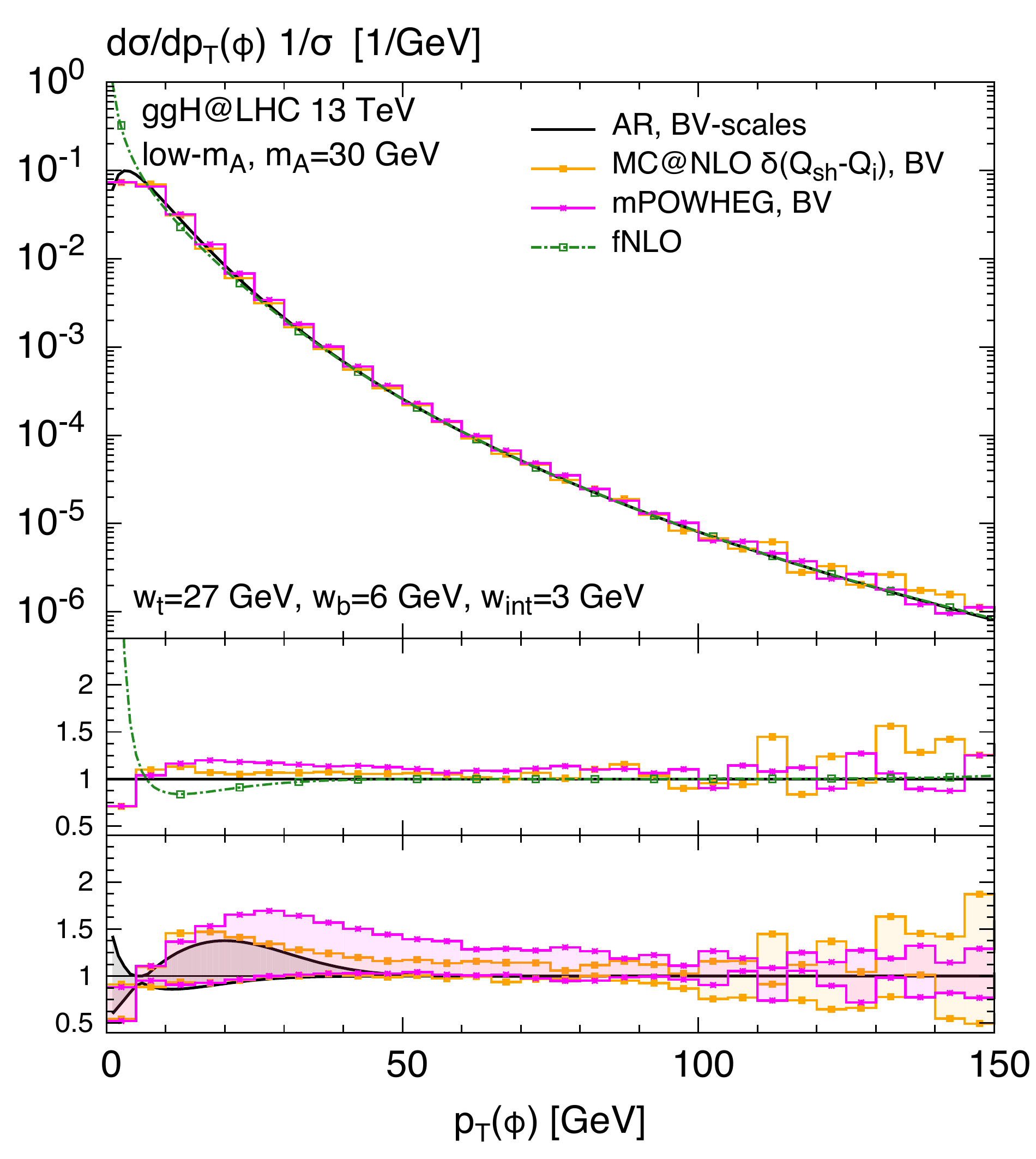}
\caption{Same as \fig{fig:results-scenB-hH}, but for \bv{} scales and a \thdm\, pseudo-scalar Higgs boson with $\ma=30$\,GeV in the \lowma{} scenario.
\label{fig:results-lowmA-A}
}
\end{figure}

Also the \lowma{} scenario produces large bottom-Yukawa couplings, but
in this case the Higgs and the bottom mass are significantly closer than
in the \largeb{} scenario. Thus, all contributions besides the
bottom-loop induced one are negligible and $\qresb$ is quite close to
$\wresb$. \fig{fig:results-lowmA-A} shows, therefore, only the
comparison among the three codes for the \bv{} scales, while the
corresponding \hmw{} results are identical for all practical purposes.

Looking first at the left plot in \fig{fig:results-lowmA-A}, we notice
that the characteristic features observed in the \largeb{} scenario
appear to be amplified in this case: the \mcnlo{} uncertainty band blows
up to $\order{150\%}$ around $\pt{}\sim 25$\,GeV; the \powheg{} curve
exceeds the \fnlo{} result by about 100\% for $\pt\gtrsim 50$\,GeV with
a very small uncertainty band. The central curves are vastly different
between the two Monte Carlo approaches except for the first two bins.

In the right plot of \fig{fig:results-lowmA-A} we show the corresponding
curves for the modified approaches (\mcnlo{} with fixed shower scale,
m\powheg{}).  Evidently, the agreement among the results is
significantly improved. Also in this case, a restriction of the shower
scale in \powheg{} closes the large gap to \fnlo{} in the tail of the
distribution, such that the m\powheg{} curve nicely approaches \fnlo{}
at large \pt{}.  The \mcnlo{} uncertainty band is very similar to the
one of \AR{} and the central predictions of the Monte Carlos are almost
on top of each other.  Note that for such small Higgs boson masses, the
shower scale distribution as defined in \sct{sec:mcnlo} ranges down to
very low values ($\qsh{}\sim 1$\,GeV) which are not well suited for a
Monte Carlo approach, both from a physical and a technical point of
view. This can be avoided by using narrower distributions---a
$\delta$-function in our case---for $\qsh{}$, thus leading to a better
agreement with the other codes and a more reasonable size of the
uncertainty band.\footnote{In the default implementation of {\tt
    MadGraph5\_aMC@NLO}\,\cite{Alwall:2014hca}, there is a hard cut
  $\qsh{}\ge 3$\,GeV that avoids such low values; for this study,
  however, we have removed this cut.}




\clearpage
\section{Conclusions}
\label{sec:conclusions}
A reliable theoretical prediction of the Higgs transverse-momentum
distribution must include an estimate of the uncertainties associated
with the matching of fixed- and all-order results. The latter are
relevant to describe the large- and the low-$\pth$ parts of the
spectrum, respectively. The matching procedure is not unique; in this
paper, we studied three different methodological approaches and their
dependence on the respective matching scale.

In general, the gluon-fusion process is characterized by two external,
physical scales: the mass of the Higgs boson and the mass of the quark
running in the loop of the gluon-gluon-Higgs vertex. Since in the \sm{},
where the top quark is dominant, these two scales are quite close to
each other, the issue of setting the matching scale is less problematic.
In \bsm{} scenarios with enhanced bottom Yukawa coupling, however, the
proper choice of the matching scale has become a matter of debate.

In this paper we presented two distinct comparisons:
\begin{enumerate}
\item
The determination of a suitable central value for the matching scales
associated to the top, bottom, and interference terms, as proposed in
the \hmw{}\,\cite{Harlander:2014uea} and \bv{}\,\cite{Bagnaschi:2015qta}
approaches, is compared qualitatively and quantitatively.
\fig{fig:scales-compare} summarizes the main outcome as a function of
the Higgs mass, both for a \cp-even and a \cp-odd Higgs boson.
\item
The predictions of three resummation codes to the problem of the
matching (\AR, \mcnlo{} and \powheg{}), all at \nlo\ \qcd{} accuracy
concerning their normalization, but different in their logarithmic
accuracy, are compared at the level of their central values and of the
respective uncertainty bands, obtained through variation of the matching
parameter by a factor of two around a central value. The latter has
been fixed following the prescriptions of point $i)$. In this comparison
we considered the production of a Higgs boson in the framework of
the \sm{} and a number of \thdm{} scenarios which are representative of
the different possible interplays between top- and bottom-quark effects
in the gluon-fusion scattering amplitude.
\end{enumerate}

In these comparisons, it is important to keep in mind that resummation
considerably improves the theoretical prediction in the low-$\pth$
region, while one cannot expect the resummed (and matched) result to be
in any way superior to the fixed-order prediction at intermediate and
high-$\pth$. We showed that the discrepancies among the codes outside
the low-$\pt$ region can be greatly reduced by a modification of
the treatment of formally subleading logarithmic terms.

In the low-$\pt$ region, we find very compatible results among the three
codes in all scenarios, independent of whether one uses \bv\ or \hmw\
scales. Even in cases where the central scales suggested by the two
methods differ relatively strongly, the respective $\pth$ distributions
are consistent with each other once their error band is taken into
account. For the interference term, this reflects the complementary
reasoning behind the \bv\ and \hmw\ methods which allows them to diverge
as long as the sensitivity of the $\pth$ distribution on the matching
scale is small.

Let us summarize the outcome of our study as follows:
\begin{itemize}
\item
The matching uncertainty on the prediction of the Higgs transverse
momentum distribution amounts to one up to several tens of percent,
depending on the specific value of \pt{} and on the model under
study. This uncertainty should be combined with the usual perturbative
uncertainty estimated via renormalization and factorization scale
variations.
\item
As far as small transverse momenta are concerned, we find reasonable
agreement among the codes irrespective of the specific matching scale
choice (\bv{} or \hmw{}), although the central \AR{} result shows a
generally much softer spectrum as compared to the central prediction of
the Monte Carlo codes.  The agreement among the three codes is in part
caused by an increase of their error band towards very low \pt{}
(essentially in the first 5\,GeV bin of the distribution), which is most
apparent though for the \AR{} result. Note, however, that the cross
section is typically very small in this region. Let us recall that these
results have been obtained with the {\tt Pythia8} parton shower;
variations with the parton shower are left for future studies.

\item
In the intermediate \pt{} range, the estimate of the width of the
matching uncertainty bands shows differences in the three approaches
under study, which are more pronounced in the bottom dominated
scenarios.  In this last case the size and shape of the uncertainty
bands strongly depends on the matching formulation, as it has been
explicitly shown in the \mcnlo{} case with the use of different
probability functions to extract the value of the shower scale.

\item
In the large \pt{} range, where all the codes have only \lo{} accuracy,
sizable differences appear in their predictions, where the \powheg{}
result is systematically larger than the \mcnlo{} and \AR{} curves,
which in turn are compatible with the \fnlo{} one. One source of this
discrepancy has been identified in the different treatment of high-\pt{}
multiple parton emissions beyond the first one, still allowed in
\powheg{} and suppressed in the other two cases.  We expect the precise
size of the discrepancy to depend on the details of the parton shower;
quantitative studies will be deferred to a future publication.  The
codes compared in this note provide either a transition to the \lo{}
prediction at large $\pt$ (\AR{} and \mcnlo{}), which has a well defined
perturbative accuracy, but misses the effects of additional radiation, or
a prediction that includes multiple parton emissions at all transverse
momenta, but describes them by means of a \PS{} (\powheg{}), which is
not adequate at large $\pt$ because it is based on the soft/collinear
approximation. It was shown that a simple modification of the interface
between \powheg{} and the \PS{} (referred to as m\powheg{} in the text),
which restricts the shower scale of the remnant events, can be applied
to ensure a consistent matching to the \fnlo{} prediction in the tail of
the distribution, similar to the other two approaches.
\end{itemize}

The approaches studied here do not lead to {\it identical} results; not
only their central values, but also their matching-induced uncertainty
bands show characteristic differences.  Nevertheless, it is fair to say
that, in the low-$\pt$ region and including these error estimates, the
results obtained by following either the \bv\ or the \hmw\ approach, and
by employing either of the three codes are consistent with each
other. This allows the conclusion that any of these approaches leads to
a valid prediction for the low-$\pth$ distribution once the matching as
well as the perturbative uncertainties are taken into account. The
former should be estimated by a variation of the respective matching
scale around its central value, as given by \bv\ or \hmw, the latter by
a variation of the renormalization and factorization scales.  In the
intermediate- and high-$\pt$ region, where large \PS{} effects become
quite doubtful, all approaches can be made compatible with the
\fnlo{} result by suitable minor modifications.

Let us point out that the new generation of Monte Carlo event
generators, merged
at \nlo{}\plus{}\PS{}\,\cite{Frederix:2012ps,Hoeche:2014lxa,Buschmann:2014sia}
or \nnlo{} \qcd{} accurate for quantities inclusive over the
radiation\,\cite{Hamilton:2013fea,Hoche:2014dla,Hamilton:2015nsa,Alioli:2013hqa,Alioli:2015toa},
with \nlo{} \qcd{} accuracy in the description of the Higgs transverse
momentum spectrum, offers a more accurate description of the $\pth$
spectrum, however only in the \sm{}-like scenario.  The evaluation of
the associated matching uncertainties is left to a future study.



\section*{Acknowledgments}
We would like to thank Stefano Frixione for helpful discussions and for
carefully reading the manuscript, Emanuele Re for confirming our
observations concerning \powheg{}, as well as Carlo Oleari and Paolo
Torrielli for several fruitful discussions. We thank Sabine Kraml and
the authors of \citere{Bernon:2014nxa} for providing the parameters of
the \lowma{} scenario.  This research was supported in part by an
Italian PRIN2010 grant, by a European Investment Bank EIBURS grant, by
the European Commission through the HiggsTools Initial Training Network
PITN-GA-2012-316704, the Swiss National Science Foundation (SNF) under
contract 200020-141360, by BMBF under contract 05H2012 and by the
European Union as part of the FP7 Marie Curie Initial Training Network
MCnetITN PITN-GA-2012-315877.



\clearpage
\appendix
\section{Numerical values of the matching scales}

In the appendix we include two tables with the values of the matching
scales as computed in the \hmw{} and \bv{} approaches, for both a scalar and a
pseudo-scalar Higgs boson. The top-quark pole mass has been set to
$172.5$ GeV, while the bottom-quark pole mass is assumed to be equal to
$4.75$ GeV. Both sets of scales are dependent on inputs at the hadron
level. For \hmw{} this is due to the fact that the scales are determined
using resummed hadronic cross sections, while in the \bv{} procedure,
hadronic physics enters in the merging of the two sets of scales
obtained separately for the $gg$ and $qg$ channels.  This implies the
existence of a (minimal) dependence on the center-of-mass energy and on
the PDF set used in the determination.  For our study we have chosen
$\sqrt{S}=13$ TeV and we have used the \textbf{{\tt MSTW2008nlo68cl}}
PDF set.

\begin{table}
\begin{center}
\begin{tabular}{|c|ccc|ccc|}
\hline
\multirow{2}{*}{$m_{h/H}$ [GeV]}  & \multicolumn{3}{c|}{\hmw{} [GeV]} & \multicolumn{3}{c|}{{\abbrev \bv{} [GeV]}} \\
& $Q_t$ & $Q_b$ & $Q_{int}$ & $w_t$ & $w_b$ & $w_{int}$\\
\hline
20 & 11.5 & 4 & 11 & 29 & 5 & 6\\
25 & 17 & 5.5 & 16 & 30 & 5 & 3\\
30 & 21 & 6.5 & 28.5 & 30 & 6 & 1\\
35 & 23 & 8 & 21.5 & 31 & 7 & 2\\
40 & 25.5 & 9 & 20.5 & 31 & 7 & 3\\
50 & 29 & 11 & 20.5 & 33 & 8 & 4\\
60 & 32.5 & 12.5 & 22 & 34 & 10 & 5\\
70 & 35.5 & 14 & 23.5 & 36 & 11 & 5\\
80 & 38 & 15.5 & 25 & 38 & 12 & 6\\
90 & 40 & 17 & 26.5 & 40 & 13 & 6\\
100 & 41 & 18 & 28 & 42 & 14 & 7\\
125 & 45 & 21 & 31 & 48 & 18 & 9\\
150 & 48 & 24 & 34 & 55 & 21 & 11\\
175 & 51 & 27 & 37 & 62 & 24 & 12\\
200 & 53 & 29 & 39 & 71 & 27 & 14\\
225 & 55 & 31 & 41 & 85 & 29 & 16\\
250 & 56 & 34 & 43 & 108 & 32 & 18\\
275 & 58 & 36 & 45 & 112 & 35 & 20\\
300 & 59 & 38 & 47 & 111 & 38 & 23\\
325 & 59 & 40 & 48 & 103 & 41 & 25\\
350 & 58 & 42 & 49 & 87 & 43 & 26\\
375 & 59 & 44 & 52 & 81 & 46 & 25\\
400 & 60 & 46 & 56 & 81 & 49 & 23\\
425 & 61 & 47 & 61 & 84 & 51 & 22\\
450 & 62 & 49 & 65 & 87 & 53 & 21\\
475 & 64 & 50 & 72 & 91 & 56 & 19\\
500 & 66 & 52 & 78 & 96 & 58 & 17\\
525 & 68 & 53 & -- & 100 & 61 & 14\\
550 & 70 & 55 & -- & 104 & 63 & 11\\
575 & 71 & 56 & 86 & 108 & 66 & 6\\
600 & 72 & 58 & 84 & 113 & 68 & 6\\
\hline
\end{tabular}
\end{center}
\caption[]{Table of the matching scales (in GeV) in the \hmw{} and \bv{} approach for a \cp-even Higgs boson.
A dash is used to indicate the case where the determination procedure of a scale has not been successful.}
\end{table}

\begin{table}
\begin{center}
\begin{tabular}{|c|ccc|ccc|}
\hline
\multirow{2}{*}{$m_A$ [GeV]}  & \multicolumn{3}{c|}{\hmw{} [GeV]} & \multicolumn{3}{c|}{\bv{} [GeV]} \\
& $Q_t$ & $Q_b$ & $Q_{int}$ & $w_t$ & $w_b$ & $w_{int}$\\
\hline
20 & 11.5 & 5 & 21.5 & 26 & 5 & 2\\
25 & 16.5 & 6.5 & 18 & 26 & 6 & 1\\
30 & 21 & 7.5 & 17 & 27 & 6 & 3\\
35 & 23 & 9 & 17 & 27 & 7 & 3\\
40 & 25 & 10 & 18 & 28 & 8 & 4\\
50 & 29 & 11.5 & 19.5 & 30 & 9 & 5\\
60 & 32.5 & 13.5 & 21 & 32 & 10 & 6\\
70 & 35 & 15 & 23 & 34 & 11 & 6\\
80 & 37.5 & 16.5 & 24.5 & 37 & 12 & 7\\
90 & 39 & 18 & 26 & 40 & 13 & 7\\
100 & 41 & 19 & 27 & 43 & 14 & 8\\
125 & 45 & 22 & 31 & 52 & 18 & 10\\
150 & 48 & 25 & 34 & 61 & 21 & 12\\
175 & 50 & 28 & 36 & 72 & 24 & 14\\
200 & 53 & 31 & 39 & 102 & 27 & 16\\
225 & 54 & 33 & 41 & 110 & 30 & 18\\
250 & 56 & 36 & 43 & 112 & 33 & 20\\
275 & 57 & 38 & 44 & 109 & 35 & 23\\
300 & 58 & 40 & 45 & 103 & 38 & 25\\
325 & 57 & 42 & 46 & 91 & 41 & 27\\
350 & 55 & 44 & 52 & 70 & 43 & 23\\
375 & 59 & 46 & -- & 80 & 46 & 18\\
400 & 61 & 48 & -- & 86 & 49 & 14\\
425 & 63 & 49 & -- & 92 & 51 & 10\\
450 & 66 & 51 & -- & 98 & 53 & 2\\
475 & 68 & 52 & -- & 104 & 55 & 9\\
500 & 70 & 54 & -- & 109 & 58 & 14\\
525 & 72 & 55 & -- & 115 & 61 & 19\\
550 & 73 & 57 & -- & 120 & 63 & 23\\
575 & 75 & 58 & -- & 126 & 65 & 28\\
600 & 76 & 60 & -- & 132 & 68 & 39\\
\hline
\end{tabular}
\end{center}
\caption[]{Table of the matching scales (in GeV) in the \hmw{} and \bv{} approach for a \cp{}-odd Higgs boson.
A dash is used to indicate the case where the determination procedure of a scale has not been successful.
}
\end{table}



\vfill
\newpage

\def\app#1#2#3{{\it Act.~Phys.~Pol.~}\jref{\bf B #1}{#2}{#3}}
\def\apa#1#2#3{{\it Act.~Phys.~Austr.~}\jref{\bf#1}{#2}{#3}}
\def\annphys#1#2#3{{\it Ann.~Phys.~}\jref{\bf #1}{#2}{#3}}
\def\cmp#1#2#3{{\it Comm.~Math.~Phys.~}\jref{\bf #1}{#2}{#3}}
\def\cpc#1#2#3{{\it Comp.~Phys.~Commun.~}\jref{\bf #1}{#2}{#3}}
\def\epjc#1#2#3{{\it Eur.\ Phys.\ J.\ }\jref{\bf C #1}{#2}{#3}}
\def\fortp#1#2#3{{\it Fortschr.~Phys.~}\jref{\bf#1}{#2}{#3}}
\def\ijmpc#1#2#3{{\it Int.~J.~Mod.~Phys.~}\jref{\bf C #1}{#2}{#3}}
\def\ijmpa#1#2#3{{\it Int.~J.~Mod.~Phys.~}\jref{\bf A #1}{#2}{#3}}
\def\jcp#1#2#3{{\it J.~Comp.~Phys.~}\jref{\bf #1}{#2}{#3}}
\def\jetp#1#2#3{{\it JETP~Lett.~}\jref{\bf #1}{#2}{#3}}
\def\jphysg#1#2#3{{\small\it J.~Phys.~G~}\jref{\bf #1}{#2}{#3}}
\def\jhep#1#2#3{{\small\it JHEP~}\jref{\bf #1}{#2}{#3}}
\def\mpl#1#2#3{{\it Mod.~Phys.~Lett.~}\jref{\bf A #1}{#2}{#3}}
\def\nima#1#2#3{{\it Nucl.~Inst.~Meth.~}\jref{\bf A #1}{#2}{#3}}
\def\npb#1#2#3{{\it Nucl.~Phys.~}\jref{\bf B #1}{#2}{#3}}
\def\nca#1#2#3{{\it Nuovo~Cim.~}\jref{\bf #1A}{#2}{#3}}
\def\plb#1#2#3{{\it Phys.~Lett.~}\jref{\bf B #1}{#2}{#3}}
\def\prc#1#2#3{{\it Phys.~Reports }\jref{\bf #1}{#2}{#3}}
\def\prd#1#2#3{{\it Phys.~Rev.~}\jref{\bf D #1}{#2}{#3}}
\def\pR#1#2#3{{\it Phys.~Rev.~}\jref{\bf #1}{#2}{#3}}
\def\prl#1#2#3{{\it Phys.~Rev.~Lett.~}\jref{\bf #1}{#2}{#3}}
\def\pr#1#2#3{{\it Phys.~Reports }\jref{\bf #1}{#2}{#3}}
\def\ptp#1#2#3{{\it Prog.~Theor.~Phys.~}\jref{\bf #1}{#2}{#3}}
\def\ppnp#1#2#3{{\it Prog.~Part.~Nucl.~Phys.~}\jref{\bf #1}{#2}{#3}}
\def\rmp#1#2#3{{\it Rev.~Mod.~Phys.~}\jref{\bf #1}{#2}{#3}}
\def\sovnp#1#2#3{{\it Sov.~J.~Nucl.~Phys.~}\jref{\bf #1}{#2}{#3}}
\def\sovus#1#2#3{{\it Sov.~Phys.~Usp.~}\jref{\bf #1}{#2}{#3}}
\def\tmf#1#2#3{{\it Teor.~Mat.~Fiz.~}\jref{\bf #1}{#2}{#3}}
\def\tmp#1#2#3{{\it Theor.~Math.~Phys.~}\jref{\bf #1}{#2}{#3}}
\def\yadfiz#1#2#3{{\it Yad.~Fiz.~}\jref{\bf #1}{#2}{#3}}
\def\zpc#1#2#3{{\it Z.~Phys.~}\jref{\bf C #1}{#2}{#3}}
\def\ibid#1#2#3{{ibid.~}\jref{\bf #1}{#2}{#3}}
\def\otherjournal#1#2#3#4{{\it #1}\jref{\bf #2}{#3}{#4}}
\newcommand{\jref}[3]{{\bf #1}, #3 (#2)}
\newcommand{\hepph}[1]{\href{http://arXiv.org/abs/hep-ph/#1}{{\tt hep-ph/#1}}}
\newcommand{\hepth}[1]{\href{http://arXiv.org/abs/hep-th/#1}{{\tt hep-th/#1}}}
\newcommand{\heplat}[1]{\href{http://arXiv.org/abs/hep-lat/#1}{{\tt hep-lat/#1}}}
\newcommand{\mathph}[1]{\href{http://arXiv.org/abs/math-ph/#1}{{\tt math-ph/#1}}}
\newcommand{\arxiv}[2]{\href{http://arXiv.org/abs/#1}{{\tt arXiv:#1}}}
\newcommand{\bibentry}[4]{#1, {\it #2}, #3\ifthenelse{\equal{#4}{}}{}{,
}#4.}


\end{document}